\documentclass[twocolumn]{aastex631}
\usepackage{amsmath}
\usepackage{graphicx}
\usepackage{placeins}
\usepackage{float}
\usepackage{txfonts}
\usepackage{hyperref}

\usepackage{lipsum}

\clearpage

\begin{document}

\title{Searching for gravitational wave optical counterparts with the Zwicky Transient Facility: summary of O4a}

\author[0000-0002-2184-6430]{Tomás Ahumada}
\altaffiliation{These authors contributed equally} 
\affiliation{Division of Physics, Mathematics and Astronomy, California Institute of Technology, Pasadena, CA 91125, USA}

\author[0000-0003-3768-7515]{Shreya Anand}
\altaffiliation{These authors contributed equally} 
\affiliation{Division of Physics, Mathematics and Astronomy, California Institute of Technology, Pasadena, CA 91125, USA}

\author[0000-0002-8262-2924]{Michael W. Coughlin}
\affiliation{School of Physics and Astronomy, University of Minnesota, Minneapolis, Minnesota 55455, USA}

\author[0000-0002-7672-0480]{Vaidehi Gupta}
\affiliation{Department of Physics, Indian Institute of Technology Kharagpur, West Bengal, 721302, India}
\affiliation{School of Physics and Astronomy, University of Minnesota, Minneapolis, Minnesota 55455, USA}

\author[0000-0002-5619-4938]{Mansi M. Kasliwal}
\affiliation{Division of Physics, Mathematics and Astronomy, California Institute of Technology, Pasadena, CA 91125, USA}

\author[0000-0003-2758-159X]{Viraj R. Karambelkar}
\affiliation{Division of Physics, Mathematics and Astronomy, California Institute of Technology, Pasadena, CA 91125, USA}

\author[0000-0003-2434-0387]{Robert D. Stein}
\affiliation{Division of Physics, Mathematics and Astronomy, California Institute of Technology, Pasadena, CA 91125, USA}

\author[0000-0003-3630-9440]{Gaurav Waratkar}
\affiliation{Department of Physics, Indian Institute of Technology Bombay, Powai, 400 076, India}

\author[0000-0002-7942-8477]{Vishwajeet Swain}
\affiliation{Department of Physics, Indian Institute of Technology Bombay, Powai, 400 076, India}

\author[0009-0003-6181-4526]{Theophile Jegou du Laz}
\affiliation{Division of Physics, Mathematics and Astronomy, California Institute of Technology, Pasadena, CA 91125, USA}

\author[0000-0002-8935-9882]{Akash Anumarlapudi}
\affiliation{Department of Physics, University of Wisconsin-Milwaukee, P.O. Box 413, Milwaukee, WI 53201, USA}

\author[0000-0002-8977-1498]{Igor Andreoni}
\altaffiliation{Neil Gehrels Fellow} 
\affiliation{Joint Space-Science Institute, University of Maryland, College Park, MD 20742, USA} 
\affil{Department of Astronomy, University of Maryland, College Park, MD 20742, USA}
\affiliation{Astrophysics Science Division, NASA Goddard Space Flight Center, MC 661, Greenbelt, MD 20771, USA}

\author[0000-0002-8255-5127]{Mattia Bulla}
\affiliation{Department of Physics and Earth Science, University of Ferrara, via Saragat 1, I-44122 Ferrara, Italy} 
\affiliation{INFN, Sezione di Ferrara, via Saragat 1, I-44122 Ferrara, Italy}
\affiliation{INAF, Osservatorio Astronomico d’Abruzzo, via Mentore Maggini snc, 64100 Teramo, Italy}

\author[0000-0002-6428-2700]{Gokul P. Srinivasaragavan}
\affiliation{Department of Astronomy, University of Maryland, College Park, MD 20742, USA}

\author[0009-0008-9546-2035]{Andrew Toivonen}
\affiliation{School of Physics and Astronomy, University of Minnesota, Minneapolis, Minnesota 55455, USA}

\author[0000-0002-9998-6732]{Avery Wold}
\affiliation{IPAC, California Institute of Technology, 1200 E. California Blvd, Pasadena, CA 91125, USA}

\author[0000-0001-8018-5348]{Eric C. Bellm}
\affil{DIRAC Institute, Department of Astronomy, University of Washington, 3910 15th Avenue NE, Seattle, WA 98195, USA}

\author[0000-0003-1673-970X]{S. Bradley Cenko}
\affiliation{Astrophysics Science Division, NASA Goddard Space Flight Center, MC 661, Greenbelt, MD 20771, USA} 
\affiliation{Joint Space-Science Institute, University of Maryland, College Park, MD 20742, USA} 

\author[0000-0001-6295-2881]{David L. Kaplan}
\affiliation{Department of Physics, University of Wisconsin-Milwaukee, P.O. Box 413, Milwaukee, WI 53201, USA}

\author[0000-0003-1546-6615]{Jesper Sollerman}
\affiliation{The Oskar Klein Centre, Department of Astronomy, Albanova, Stockholm University, SE-106 91 Stockholm, Sweden}

\author[0000-0002-6112-7609]{Varun Bhalerao}
\affiliation{Department of Physics, Indian Institute of Technology Bombay, Powai, 400 076, India}

\author[0000-0001-8472-1996]{Daniel Perley}	
\affiliation{Astrophysics Research Institute, Liverpool John Moores University, IC2, Liverpool Science Park, 146 Brownlow Hill, Liverpool L3 5RF, UK}

\author[0000-0003-3173-4691]{Anirudh Salgundi}
\affiliation{Department of Physics, Indian Institute of Technology Bombay, Powai, 400 076, India}

\author[0009-0005-8230-030X]{Aswin Suresh}
\affiliation{Department of Physics, Indian Institute of Technology Bombay, Powai, 400 076, India}

\author[0000-0002-0129-806X]{K-Ryan Hinds}
\affiliation{Astrophysics Research Institute, Liverpool John Moores University, IC2, Liverpool Science Park, 146 Brownlow Hill, Liverpool L3 5RF, UK}

\author[0000-0002-7788-628X]{Simeon Reusch}
\affiliation{Deutsches Elektronen-Synchrotron DESY, Platanenallee 6, 15738 Zeuthen, Germany}

\author[0000-0003-0280-7484]{Jannis Necker}
\affiliation{Deutsches Elektronen-Synchrotron DESY, Platanenallee 6, 15738 Zeuthen, Germany}

\author[0000-0002-6877-7655]{David O. Cook}
\affiliation{IPAC, California Institute of Technology, 1200 E. California Blvd, Pasadena, CA 91125, USA}

\author[0009-0008-8062-445X]{Natalya Pletskova}
\affiliation{Department of Physics, Drexel University, Philadelphia, PA 19104, USA}

\author[0000-0001-9898-5597]{Leo P. Singer}
\affiliation{Astrophysics Science Division, NASA Goddard Space Flight Center, MC 661, Greenbelt, MD 20771, USA}


\author[0000-0001-6595-2238]{Smaranika Banerjee}
\affiliation{The Oskar Klein Centre, Department of Astronomy, Albanova, Stockholm University, SE-106 91 Stockholm, Sweden}

\author[0000-0002-4843-345X]{Tyler Barna}
\affiliation{School of Physics and Astronomy, University of Minnesota, Minneapolis, Minnesota 55455, USA}

\author[0000-0001-7983-8698]{Christopher M. Copperwheat}
\affiliation{Astrophysics Research Institute, IC2, Liverpool Science Park, 146 Brownlow Hill, Liverpool L3 5RF, UK}


\author[0000-0002-7718-7884]{Brian Healy}
\affiliation{School of Physics and Astronomy, University of Minnesota, Minneapolis, Minnesota 55455, USA}

\author[0000-0002-9108-5059]{R. Weizmann Kiendrebeogo}
\affiliation{Laboratoire de Physique et de Chimie de l'Environnement, Université Joseph KI-ZERBO, Ouagadougou, Burkina Faso}
\affiliation{Artemis, Observatoire de la Côte d'Azur, Université Côte d'Azur, Boulevard de l'Observatoire, F-06304 Nice, France}
\affiliation{School of Physics and Astronomy, University of Minnesota, Minneapolis, Minnesota 55455, USA}


\author[0000-0003-0871-4641]{Harsh Kumar}
\affiliation{Department of Physics, Indian Institute of Technology Bombay, Powai-400076, India}
\affiliation{Center for Astrophysics, Harvard \& Smithsonian}

\author[0009-0008-6428-7668]{Ravi Kumar}
\affiliation{Department of Aerospace Engineering, Indian Institute of Technology Bombay, Powai, 400 076, India}


\author{Marianna Pezzella}
\affiliation{Embry-Riddle Aeronautical University. Daytona Beach, Florida, United States.}

\author[0000-0002-3498-2167]{Ana Sagués-Carracedo}
\affil{The Oskar Klein Centre, Department of Physics, Stockholm University, AlbaNova, SE-106 91 Stockholm, Sweden}

\author{Niharika Sravan}
\affiliation{Department of Physics, Drexel University, Philadelphia, PA 19104, USA}

\author[0000-0002-7777-216X]{Joshua S. Bloom}
\affil{Department of Astronomy, University of California, Berkeley, CA 94720, USA} \affil{Lawrence Berkeley National Laboratory, 1 Cyclotron Road, MS 50B-4206, Berkeley, CA 94720, USA}

\author[0000-0001-9152-6224]{Tracy X. Chen}
\affiliation{IPAC, California Institute of Technology, 1200 E. California Blvd, Pasadena, CA 91125, USA}

\author[0000-0002-3168-0139]{Matthew Graham}
\affiliation{California Institute of Technology, 1200 E. California Blvd, Pasadena, CA 91125, USA}

\author[0000-0003-3367-3415]{George Helou}
\affiliation{IPAC, California Institute of Technology, 1200 E. California Blvd, Pasadena, CA 91125, USA}
             
\author[0000-0003-2451-5482]{Russ R. Laher}
\affiliation{IPAC, California Institute of Technology, 1200 E. California Blvd, Pasadena, CA 91125, USA}

\author[0000-0003-2242-0244]{Ashish~A.~Mahabal}
\affiliation{Division of Physics, Mathematics and Astronomy, California Institute of Technology, Pasadena, CA 91125, USA}
\affiliation{Center for Data Driven Discovery, California Institute of Technology, Pasadena, CA 91125, USA}

\author[0000-0003-1227-3738]{Josiah Purdum}
\affiliation{Caltech Optical Observatories, California Institute of Technology, Pasadena, CA 91125, USA}


\author[0000-0003-3533-7183]{G. C. Anupama}
\affiliation{Indian Institute of Astrophysics, 2nd Block 100 Feet Rd, Koramangala Bangalore, 560 034, India}

\author[0000-0002-3927-5402]{Sudhanshu Barway}
\affiliation{Indian Institute of Astrophysics, 2nd Block 100 Feet Rd, Koramangala Bangalore, 560 034, India}

\author[0000-0001-7570-545X]{Judhajeet Basu}
\affiliation{Indian Institute of Astrophysics, 2nd Block 100 Feet Rd, Koramangala Bangalore, 560 034, India}

\author[0009-0005-2367-6999]{Dhananjay Raman}
\affiliation{Department of Computer Science and Engineering, Indian Institute of Technology Bombay, Powai, 400 076, India}

\author[0009-0003-9906-2745]{Tamojeet Roychowdhury}
\affiliation{Department of Electrical Engineering, Indian Institute of Technology Bombay, Powai, 400 076, India}

\begin{abstract}

During the first half of the fourth observing run (O4a) of the International Gravitational Wave Network (IGWN), the Zwicky Transient Facility (ZTF) conducted a systematic search for kilonova (KN) counterparts to binary neutron star (BNS) and neutron star--black hole (NSBH) merger candidates. Here, we present a comprehensive study of the five high-significance (False Alarm Rate less than 1 per year) BNS and NSBH candidates in O4a. Our follow-up campaigns relied on both target-of-opportunity observations (ToO) and re-weighting of the nominal survey schedule to maximize coverage. We describe the toolkit we have been developing, \texttt{Fritz}, an instance of \textit{SkyPortal},  instrumental in coordinating and managing our telescope scheduling, candidate vetting, and follow-up observations through a user-friendly interface. ZTF covered a total of 2841 deg$^2$ within the skymaps of the high-significance GW events, reaching a median depth of $g\approx$\,20.2\,mag. We circulated 15 candidates, but found no viable KN counterpart to any of the GW events. Based on the ZTF non-detections of the high-significance events in O4a, we used a Bayesian approach, \texttt{nimbus}, to quantify the posterior probability of KN model parameters that are consistent with our non-detections. Our analysis favors KNe with initial absolute magnitude fainter than $-$16\,mag. The joint posterior probability of a GW170817-like KN associated with all our O4a follow-ups was 64\%. Additionally, we use a survey simulation software, \texttt{simsurvey}, to determine that our combined filtered efficiency to detect a GW170817-like KN is 36\%, when considering the 5 confirmed astrophysical events in O3 (1 BNS and 4 NSBH events), along with our O4a follow-ups. Following \citet{Kasliwal2020kn}, we derived joint constraints on the underlying KN luminosity function based on our O3 and O4a follow-ups, determining that no more than 76\% of KNe fading at 1 mag day$^{-1}$ can peak at a magnitude brighter than $-$17.5\,mag. 

\end{abstract}



\section{Introduction}\label{sec:intro} 

The increased sensitivity of gravitational-wave detector networks have enabled unprecedented discoveries of compact binary mergers in the last decade. The International Gravitational Wave Network (IGWN) detected 102 binary black hole (BBH) mergers, 2 binary neutron star (BNS) mergers and  4 neutron star--black hole (NSBH) mergers between 2015 and 2020 during the first three observing runs \citep{GWTC3}. The growing population of BBH mergers have challenged the existence of both the upper and lower black hole mass gaps \citep{LVCGW190814, LVCGW190521}, and have revealed a unique population of low-spin black holes \citep{Tiwari2018}. The second observing run of IGWN marked the discovery of GW170817, the very first GW signal from a binary neutron star merger system \citep{AbEA2017b}, with its short gamma-ray burst (GRB) counterpart \citep{AbEA2017c, GoVe2017}, panchromatic afterglow \citep{Haggard2017, Hallinan2017, Margutti2017, Troja2017, Mooley2018170817, Pozanenko2018,Makhathini2021,Balasubramanian2022,Mooley2022}, and optical/IR kilonova (KN) \citep{Coulter2017,  Drout2017, Evans2017, Kasen2017,KaNa17, LiGo2017, SoHo2017, Valenti2017, Utsumi2017, Arcavi2018, KaKa2019}. IGWN's third observing run yielded another BNS merger \citep{LVCGW190425} along with the first ever detections of neutron star--black hole mergers \citep{LVCGW190814, LVCGW200105-GW200115, GWTC3}, though no electromagnetic counterpart was found for any of these events. 

Many collaborations such as the Zwicky Transient Facility (ZTF; \citealt{Bellm2019, Graham2019,Dekany2019}), Electromagnetic counterparts of Gravitational wave sources at the Very Large Telescope (ENGRAVE; \citealt{Engrave2020}), Global Rapid Advanced Network Devoted to the Multi-messenger Addicts (GRANDMA; \citealt{Grandma2020}), Gravitational-wave Optical Transient Observer (GOTO; \citealt{Goto2020}), All Sky Automated Survey for SuperNovae (ASAS-SN; \citealt{ShPr2014}), Asteroid Terrestrial Last Alert System (ATLAS; \citealt{ToDe2018}), Panoramic Survey Telescope and Rapid Response System (Pan-STARRS; \citealt{ChMa2016}), MASTER-Net \citep{LiGo2017}, Searches after Gravitational Waves Using ARizona Observatories (SAGUARO; \citealt{Lundquist2019}), Gravitational-wave Electromagnetic Counterpart Korean Observatory (GECKO; \citealt{Paek2024gecko}), the Dark Energy Survey Gravitational Wave Collaboration (DES-GW; \citealt{SoHo2017}), Global Relay of Observatories Watching Transients Happen (GROWTH\footnote{\url{http://growth.caltech.edu/}}), Burst Optical Observer and Transient Exploring System (BOOTES; \citealt{2023FrASS..10.2887H}), KM3Net{\footnote{\url{https://www.km3net.org/}}} and VINROUGE{\footnote{\url{https://www.star.le.ac.uk/nrt3/VINROUGE/}}} undertook targeted efforts during IGWN's third observing run (O3) to identify any associated electromagnetic counterparts. However, despite extensive tiling and galaxy-targeted searches, no EM counterparts were found \citep{CoAh2019, Goldstein2019S190426c, S190814growth, Andreoni2019S190510g, Kasliwal2020kn, Grandma2020, Vieira2020GW190814, Kilpatrick2021GW190814, Alexander2021GW190814, Wet2021GW190814, Thakur2021GW190814,Tucker2022GW190814,Rastinejad2022gwsearch,Dobie2022GW190814}. 
Amongst the 6 BNS and 9 NSBH merger candidates announced in O3, only 1 BNS merger (GW190425) and 4 NSBH merger candidates (GW190426, GW190814, GW200105, and GW200115) passed the False Alarm Rate (FAR) threshold for inclusion in the Gravitational Wave Transient Catalog (GWTC-3; \citealt{GWTC3}) as high-confidence signals, rendering the remainder of the candidates as subthreshold astrophysical events or noise sources. Nevertheless, the dearth of BNS mergers during O3 revised the projected astrophysical rate of BNS mergers to 50--440 $\mathrm{Gpc^{-3} yr^{-1}}$ \citep{AbEA2023a}, assuming uniform mass and spin distributions, and that the merger rate is constant in comoving volume out to z=0.15.

IGWN's fourth observing run (O4) commenced on May 24, 2023 and paused for a commissioning break on January 15, 2024, marking the end of the first half of the observing run (O4a). Based on the sensitivity of the LIGO and Virgo detectors, observing scenarios studies \citep{Kiendrebeogo23} predicted that 36$^{+49}_{-22}$ BNS and 6$^{+11}_{-5}$ NSBH mergers would be detected at the public alert release threshold during the first year of O4, which is consistent with the number of potential NS merger candidates (including those of low significance, there are 27 events with \texttt{HasNS} $> 0.5$ and FAR better than 1 per week) released thus far during O4a (lasting 8 months). These estimates included the Virgo detector as a part of the GW network, whose sensitivity was projected to be between 40--80\,Mpc. Virgo has now joined the O4 run since April 2024 at a rough sensitivity of $\approx$\,50\,Mpc. The rates are driven by the LIGO interferometers, and the inclusion of Virgo does not affect the predicted rates dramatically; however, it results in better localized NS mergers.


The Zwicky Transient Facility, mounted on the Samuel Oschin 48-inch Telescope at Palomar Observatory, is a public-private project that routinely acquires 30\,s images in the $g$-, $r-$ and $i$-band, covering the entire available northern night-sky every two nights. Due to its cadence, ZTF has one of the most complete records of the contemporary dynamic sky. This capability enables the detection of transients at the early stages of their active phase. The use of ZTF for GRB and GW optical counterparts searches, over thousands of square degrees \citep{Kasliwal2020kn,Ahumada2022grb} has allowed for the discovery of rare GRB afterglows: the shortest burst associated to a collapsar \citep{Ahumada2021sgrb}, an orphan afterglow during O3 \citep{perley2024orphan}, and the afterglow of one of the brightest GRBs \citep{Gokul2024grb}.
We used ZTF (more details in \S~\ref{sec:ZTF}) to conduct wide-field tiling searches of 5 high-significance GW candidates (S230518h, S230529ay, S230627c, S230731an, and S231113bw) aiming to detect an EM counterpart. For completeness, we also include 5 other (lower significance) GW candidates for which ZTF has coverage, in the Appendix (see \S~\ref{sec:ap_gwevents}). 

In this paper, we start in \S~\ref{sec:ZTF} describing how ZTF is used to perform searches for EM counterparts to GW sources during O4a. We outline the triggering mechanisms for ZTF in \S~\ref{sec:triggering}. In \S~\ref{sec:pipelines} we give a description of the analysis pipelines and candidate filtering criteria, alongside the new and improved software toolkit for enabling counterpart discovery. In \S~\ref{sec:triggers} we provide details of the GW events we triggered ZTF on, and in \S~\ref{sec:discussion} we determine the efficiency of our efforts, and derive constraints to the KN luminosity function. We finalize the paper with conclusions in \S~\ref{sec:conclusion}.

\section{Zwicky Transient Facility Follow-up}\label{sec:ZTF} 


 In this section, we describe the ZTF triggering criteria for GW events during O4a. We start by describing the IGWN public data products that were used to evaluate the relevance of an event, and we continue describing the ZTF triggering criteria and the methods used to trigger and schedule ZTF observations.

\subsection{GW metrics}

The strain data of the GW events is analyzed in real time by different pipelines. Some pipelines such as GSTLal \citep{gstlal}, PyCBC Live \citep{pycbclive}, the Multi-Band Template Analysis (MBTA; \citealt{mtba}), and the Summed Parallel Infinite Impulse Response (SPIIR; \citealt{spiir}) match the signal to a template bank of compact binaries coalescences (CBCs), while others, such as cWB \citep{cwb} and oLIB \citep{olib}, search for bursts of power in the GW spectra. These pipelines include the FAR of the event in their public data products, as well as an initial 3D localization map produced by BAYESTAR \citep{SiPr2016bayestar}. In addition to this, pipelines searching for CBCs release metrics related to the template matching results, indicating the probability of a merger to have a BBH, BNS, NSBH, or Terrestrial origin in the initial GCN announcement (\texttt{$p_{BBH}$}, \texttt{$p_{BNS}$}, \texttt{$p_{NSBH}$}, and \texttt{$p_{Terrestrial}$} respectively). This first online pipeline analysis is followed by a machine-learning-based inference \citep{embright}, that sheds light onto whether at least one NS was part of the binary (\texttt{HasNS}), whether the merger is likely to leave a non-zero remnant behind (\texttt{HasRemnant}), or if it involves an object in the 3--5 \(M_\odot\) mass gap (\texttt{HasMassGap}). 

\subsection{Triggering criteria}

During O3, ZTF conducted a search for optical counterparts for \textit{all} observable BNS, NSBH, and MassGap events \citet[\S 2]{Kasliwal2020kn}. These criteria resulted in 13 campaigns, spanning GW events with FARs between 10$^{-25}$ -- 24 year$^{-1}$. The offline GW analysis post-O3 confirmed only five of these candidates as likely CBCs (GW190425, GW190426, GW190814, GW200105, and GW200115), while retracting all other events \citep{AbEA2023a}. 
During O4, we decided to take the FAR and other low-latency  GW parameters into consideration at the time of triggering ZTF observations. Given that the FAR depends on the template bank of each pipeline, there are usually discrepancies between the different pipelines that have to be considered case by case. Generally, the ZTF trigger criteria prioritized events with FAR $<$ 1 year$^{-1}$ and one of the following: \texttt{HasNS} $>0.1$, \texttt{pBNS} $>0.1$, or \texttt{pNSBH} $>0.1$ to avoid BBHs and terrestrial events. These criteria were intended to address the substantial volume of low-significance events, rather than serving as rigid criteria. During O4a, there were 150 events with \texttt{pBNS} $>0.1$ or \texttt{pNSBH} $>0.1$ (for a comprehensive list see Table \ref{tab:O4_summary_all}). However, only 5 of these had false alarm rates less than 1 year$^{-1}$. We used ZTF to follow-up all 5 of them (see Table~\ref{tab:O4_summary} and \S~\ref{sec:triggers}). 

\subsection{ZTF strategies}

In O4, ZTF developed two observing strategies for GW events that were identified as interesting (FAR $<$ 1 year$^{-1}$, and \texttt{HasNS}$>0.1$ or \texttt{p$_{BNS}>0.1$} or \texttt{p$_{NSBH}>0.1$}). The first strategy relied on interrupting the nightly schedule of ZTF through a Target of Opportunity (ToO) trigger, in order to cover the GW region with exposures longer than the nominal 30\,s survey exposures. This strategy allowed us to conduct 300\,s observations, and was limited to high confidence and well localized events. Our nominal ToO strategy covers the skymaps in multiple filters during night 0, night 1, night 2, and night 7. To prepare for O4, ZTF developed a set of deep reference images of the ZTF grid, which allowed for robust image subtraction of our deeper ToO observations. The median limiting magnitude of these deeper references is 23.0\,mag for the $i$-band, and 23.5\,mag for $g$- and $r$-bands. 


The second strategy relies on the deliberate rearrangement of the ZTF fields that are part of the regular survey operations. The nightly ZTF schedule is optimized for the discovery and characterization of the dynamic optical sky, while systematically observing different areas of the sky \citep{Bellm2019schedule}. During O3, we relied on serendipitous ZTF coverage of GW skymaps for low significance or poorly localized events. However, for O4, we developed an alternative approach, referred to as ``re-weighting" that makes use of the nominal 30\,s exposures of the ZTF public survey and constructs a re-weighted schedule, prioritizing the ZTF fields that overlap with the GW localization area. This strategy conducts observations during the first and second night after a trigger.

\subsection{Triggering ZTF observations}\label{sec:triggering}

\begin{figure}
    \centering
    \includegraphics[width=0.5\textwidth]{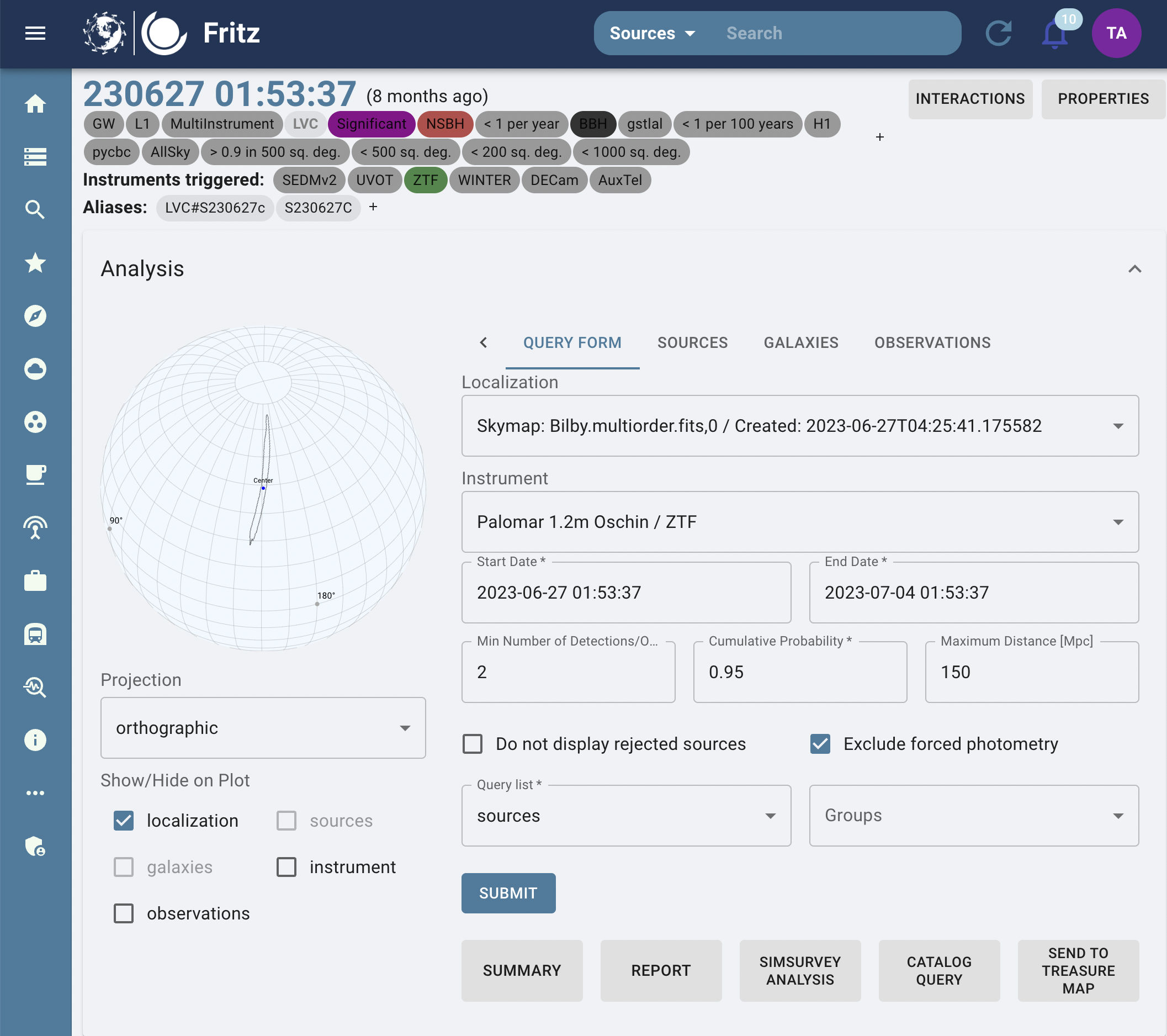}
    \caption{\label{fig:gcn-page}The \texttt{Fritz} page for a GW event displays information in tags located below the date of the event. In the Properties tab, it presents information originally available in the GCN. The page exhibits the most up-to-date information available, as well as the history of changes circulated through GCNs. }
    
\end{figure}

The scheduling of ZTF observations to tile and cover GW error regions can be done through multiple avenues, and the bulk of our triggers were managed through \texttt{Fritz}, an instance of \textit{SkyPortal} \citep{vanderwalt2019SkyPortal,Coughlin2023SkyPortal}.  \textit{SkyPortal} combines the functionalities of two separate tools: the GROWTH Marshal \citep{kasliwal2019growthmarshal} and the GROWTH Target of Opportunity Marshal \citep{CoAh2019}, while providing additional functionalities that further automate the EMGW follow-up process. While the GROWTH Marshal offered the ability to save candidates from different discovery streams and assign follow-up, the GROWTH ToO Marshal allowed for the interaction with skymaps. As a result, \textit{SkyPortal} provides a user-friendly tool that facilitates the management and exploration of astronomical data, allowing one to schedule observations and easily retrieve data associated to a skymap. Particularly, \texttt{Fritz} is optimized to interact with ZTF, as it retrieves data from the ZTF database \texttt{Kowalski} \citep{Duev2019}, displays light-curves and spectra of ZTF objects, and enables interaction with multi-messenger events, such as GWs, among other key features. \texttt{Fritz} continuously listens to the GCN stream of alerts \citep{newGCN} and generates an interactive GCN event page for each new alert, including for GWs, GRBs, and neutrino alerts (see Fig.\ref{fig:gcn-page}). 
Information intrinsic to each GCN, such as \texttt{$p_{BNS}$} or \texttt{HasNS}, is readily accessible through this page.
Additionally, \texttt{Fritz} facilitates the management and execution of ZTF observation plans (as well as for other facilities, such as DECam, WINTER, Palomar Gattini IR, and the GROWTH-India Telescope). As a new event comes in and is added to \texttt{Fritz}, a default ZTF observing plan is created with \texttt{gwemopt}, a schedule optimizer originally developed to handle GW skymaps \citep{CoTo2018}. The default \texttt{gwemopt} plan consists of three visits per field, each lasting 300s, in a $g$-, $r$-, $g$-band sequence. However, this default strategy can be modified by requesting a new observing plan with adjusted exposure times and filter sequences, or by targeting a subsection of the GW skymap. For each observing plan, \texttt{Fritz} additionally displays the tiling of the region in a dynamic skymap, and a summary of the plan including the duration of the observations, the areal coverage, and the probability enclosed. The finalized plan can be submitted to the ZTF queue through \texttt{Fritz}.

For events that required a re-organization and re-weighting of the nominal 30\,s ZTF observations, the procedure requires communication with the ZTF scheduler \citep{Bellm2019schedule}\footnote{\url{https://github.com/ZwickyTransientFacility/ztf_sim}}. This was accomplished by sending fields and their integrated probabilities from the GW skymap to the ZTF scheduler through an integrated API in \texttt{Fritz}. Once the fields are received, the scheduler assigns 30\,s epochs in $g$-, $r$-, and $i$-bands to the highest probability fields. 

Additionally, we developed an open source Simple Nodal Interface for Planning Electromagnetic Reconnaissance of Gravitational Waves (\texttt{SniperGW})\footnote{\url{https://github.com/robertdstein/snipergw}}, a programmatic avenue to access the ZTF scheduler, as a back-up that can be run on a laptop. \texttt{SniperGW} directly downloads maps from \texttt{GraceDB}, uses \texttt{gwemopt} to generate the schedules, and communicates directly with the ZTF scheduler via API. This serves as an ``offline" method for us to submit schedules in real-time in case the \texttt{Fritz} database is down, and also allows more flexibility to customize schedules if needed.

\section{Analysis Pipelines} \label{sec:pipelines}

The ZTF pipeline \citep{Masci2019}, running at the Infrared Processing and Analysis Center (IPAC \footnote{\url{https://www.ipac.caltech.edu/}}), reduces, calibrates and performs image subtraction in near real time. Any 5$\sigma$ flux deviation from the reference image issues an \textit{alert} \citep{Patterson2019}, containing metadata of the transient, including its light-curve history, real-bogus score \citep{Duev2019}, and cross-matches with PanSTARRS \citep{ChMa2016}, among other useful quantities. These \textit{alerts} are issued to brokers all around the globe, such as ALeRCE \citep{Forster2016alerce}, AMPEL \citep{ampel_19}, ANTARES \citep{saha2014antares}, Lasair \citep{Smith2019Lasair}, Fink \citep{Moller2020fink}, and Pitt-Google \footnote{\url{https://pitt-broker.readthedocs.io/en/latest/}}, where users can manage and filter the alerts in order to recover their transients of interest. 

\subsection{Transient searches: automatic filtering} \label{sec:filtering1}

Throughout O4a, we relied on four methods to select transients from the ZTF stream: \texttt{Fritz}, \texttt{nuztf}, \texttt{emgwcave}, and the ZTF REaltime Search and Triggering (\texttt{ZTFReST}; \citealt{Andreoni2021ztfrest}). Some of these tools were used during O3, and build on developments following the past IGWN run. Each tool developed a unique alert filtering scheme, however, they have a common core:

\begin{itemize}
    \item  \textbf{In the GW skymap:} The candidate is required to be inside the 95\% contour of the latest and most up-to-date GW skymap.
\item  \textbf{Positive subtraction:} We focus on sources that have brightened and have a positive residual after image subtraction.
\item  \textbf{Real astrophysical sources:} ZTF has developed a machine learning (ML) model to identify sources that are created by ghosts or artifacts in the CCDs. The model was trained with known ZTF artifacts and it relies on a deep convolutional neural network \citep{Duev2019}. Generally, sources with Real-Bogus score $> 0.3$ are considered to be of astrophysical origin.
\item \textbf{Avoid known point sources:} To avoid contamination from stars, we enforce transients to be greater than 3 arcsec\ from any point source in the PS1 catalog based on \cite{TaMi2018}. 
\item  \textbf{Minimum of two detections:} To reject slow moving solar system objects and cosmic rays, we enforce a minimum of two detections separated by at least 15\,min. 
\item  \textbf{Far from a bright star:} It is well known that bright sources produce artifacts and ghosts, thus we require a minimum distance of $>$ 20 arcsec\, from sources with $m_\textrm{AB} < 15$ mag. 
\item  \textbf{First detection after the GW event:} KNe and relativistic afterglows are only expected after the merger, thus we filter out sources with activity previous to the GW event. 
\end{itemize}

The majority of the analysis was carried out on \texttt{Fritz}: from planning the observing strategy, to the selection of candidates, and the orchestration of their follow-up. For the selection of candidates, we set in place \textit{two} \texttt{MongoDB} filters to interact with \texttt{Kowalski}, the ZTF database, via \texttt{Fritz}. Both filters followed the points established above, and while the \texttt{EM+GW} filter aims to serve as a thorough census of all the extragalactic sources spatially and temporally consistent with a GW event, the \texttt{EM+GW PtAu} filter was designed to recover transients within 150 kpc of projected distance from a galaxy, either in the Census of the Local Universe (CLU; \citealt{Cook2019clu}) or in the NASA/IPAC Extragalactic Database - Local Volume Sample (NED-LVS; \citealt{Cook2023ned}) catalogs. A major development in O4a is the flexible candidate searches in different skymaps. We used to rely on offline cross-matching for each candidate, in order to determine at what credible level within the GW skymap each candidate was discovered. Now, the searches can be customized through \texttt{Fritz}, by selecting a skymap, a credible level, and a detection date, in order to retrieve the candidates that meet the selected criteria. This new feature allows us to easily determine which ZTF sources are inside a skymap, and it has been used to revise candidates when a newly updated GW skymap is circulated (see Fig. \ref{fig:gcn-filtering}).  


\texttt{Fritz} was intended to provide a stable and reliable way to access, filter, visualize, and interact with ZTF data. It was optimized to cater to multiple science cases with a trade-off in flexibility. Although alert filters can easily be modified, real-time fine-tuning adjustments are difficult to implement. For this reason, we have other software stacks that enable independent queries to \texttt{AMPEL} and \textit{Kowalski}, the ZTF databases. Having multiple tools analyzing the ZTF data stream allows us to be meticulously thorough, to increase our completeness, and to understand how the different alert filters affect our results. In this section we describe complementary methods used to filter ZTF alerts.

Firstly, we conducted an independent search using the \texttt{nuztf}\footnote{\url{https://github.com/desy-multimessenger/nuztf}} \texttt{python} package \citep{nuztf}, originally developed for the ZTF Neutrino Follow-Up Program (see \citealt{stein_23_nu} for further details). \texttt{nuztf} uses the \texttt{AMPEL} framework to conduct candidate filtering \citep{ampel_19}, and uses the \texttt{AMPEL} broker data archive to retrieve ZTF data at very low latency \citep{ampel_19}. \texttt{AMPEL} provides a direct healpix API query that can return candidates within a given skymap. We perform cuts similar to those listed above to select candidates, and then perform automated cross-matching with various multi-wavelength catalogues to flag likely variable AGN or stars. \texttt{nuztf} can export candidates to \texttt{Fritz}, as well as produce summary PDFs for quick candidate scanning. \texttt{nuztf} uses ZTF observation logs from IPAC to calculate survey coverage of a skymap, accounting for chip gaps and any processing failures in each of the 64 ZTF quadrants.

The \textit{Kowalski} database was queried independently through \texttt{emgwcave}\footnote{\url{https://github.com/virajkaram/emgwcave}}, a python-based script that retrieves candidates based on the cuts similar to the ones described above. \textit{emgwcave} offers an extra layer of flexibility, as the queries can be easily modified. Similar to the \texttt{nuztf} searches, the candidates are cross-matched with multiple catalogs in order to identify AGNs and variable stars. The resulting outcomes are then exported to a PDF file and simultaneously pushed to \texttt{Fritz}.

Finally, we made use of the \texttt{ZTFReST} infrastructure \citep{Andreoni2021ztfrest}. This open-source code allows the exploration of ZTF data, and the flagging of fast fading transients. \texttt{ZTFReST} derives the evolution of a given transient based on the photometry in the ZTF alerts and forced photometry \citep{Yao2020}. The ranking of transients considers factors such as the galactic latitude, the cross-match to multi-wavelength catalogs, and the magnitude evolution. \texttt{ZTFReST} highlights transients through a user-friendly \texttt{Slack}-bot that enables the scanning of candidates. 

All candidates passing the automatic filter are submitted to the Transient Name Server (TNS\footnote{\url{https://www.wis-tns.org/}}).

\subsection{Transient Vetting: source by source } \label{sec:filtering2}

\begin{figure}
    \centering
    \includegraphics[width=0.5\textwidth]{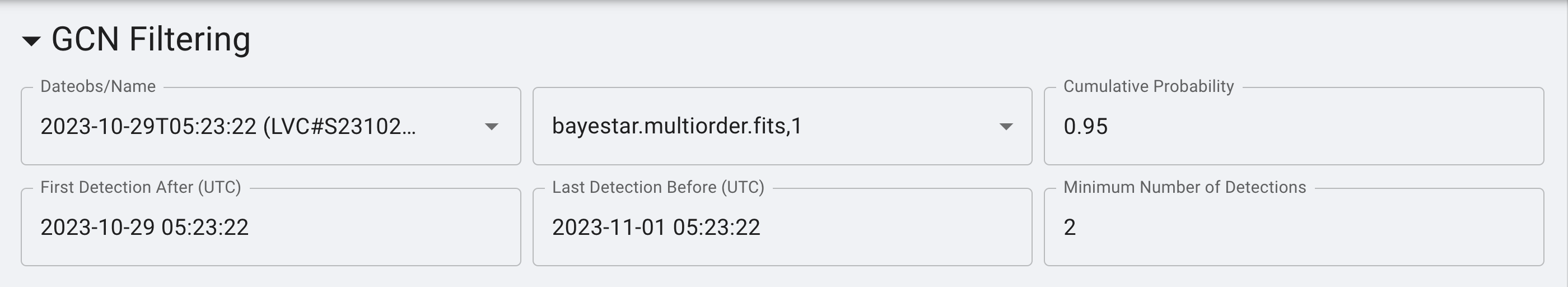}
    \caption{A snapshot illustrating the spatial and temporal constraints set on \texttt{Fritz} for transients selection. This feature is used to refine the candidate query, limiting it to a specific region (Cumulative Probability) on a skymap within a designated time-frame.}
    \label{fig:gcn-filtering}
\end{figure}

Once a transient passes either of the filters set in place (\texttt{EM+GW} or \texttt{EM+GW PtAu}), it can be easily retrieved through \texttt{Fritz} where we have implemented an efficient spatial filter through Healpix Alchemy \citep{singer2022healpix} that allow us to query transients in a given portion of a specific skymap. If a candidate passes any of the other offline filters (\texttt{nuztf}, \texttt{emgwcave}, or \texttt{ZTFReST}), it can easily be included in the main \texttt{Fritz} group and be analyzed using the \texttt{Fritz} capabilities. The \texttt{Fritz} interface allows one to easily modify the spatial query and retrieve ZTF transients at different credible levels, as seen in Fig. \ref{fig:gcn-filtering}. 

During O3, a key feature to discriminate candidates was the use of ZTF forced photometry (FP). Thanks to a number of modifications and improvements in the IPAC request and retrieval of FP products, \texttt{Fritz} has now integrated forced photometry capabilities. For each transient, there is the option to directly request FP from the \texttt{Fritz} source page, and additionally select the time window of interest, that could go back to the start of the survey. Similarly, \texttt{Fritz} has made use of the ATLAS FP service \citep{shingles2021atlasfp}, and it has implemented a similar system for data retrieval. For both services, the products include the flux information and its uncertainty. We set a threshold of 3$\sigma$ for detections and we take this information into account when ruling out sources. We also download the ATLAS images associated with the forced photometry for further inspection.


The \texttt{Fritz} alert filters can retrieve additional information for the candidates, as they are ingested in the \textit{Kowalski} database, they are also crossmatched with a number of surveys. Data from the Wide-Field Infrared Survey Explorer (WISE; \citealt{WrEi2010}) and Milliquas \citep{Flesch2023Milliquas} are retrieved and used to assess whether a source is associated to an active galactic nuclei (AGN): for WISE we use the W1$-$W2 $>$ 0.6 cut \citep{WrEi2010}, while for Milliquas we require a quasar probability $p_{QSO} < 0.8$. Since the WISE point spread function (PSF) is around 6 arcsec (compared to ZTF's 1 arcsec PSF), additional human vetting is required to ensure the association to an AGN.


\begin{figure}
    \centering
    \includegraphics[width=\linewidth]{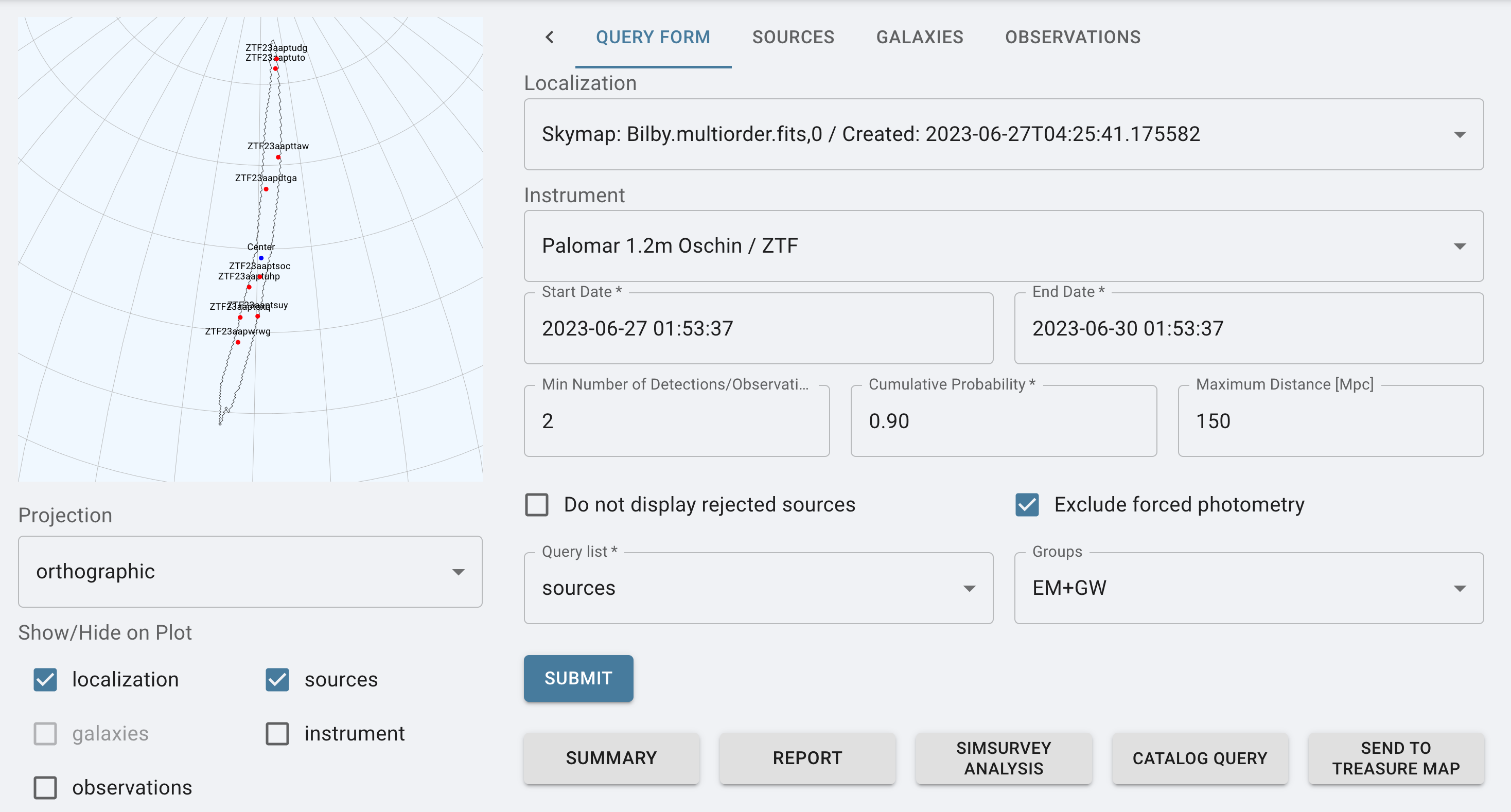}
    \caption{Snapshot of the GCN Analysis \texttt{Fritz} page. In this case, we display the sources within the 90\% localization of the GW event S230627c passing the EM+GW filter in the corresponding GW skymap.}
    \label{fig:sources-localization}
\end{figure}

\subsection{Transient Vetting: assigning follow-up} \label{sec:filtering4}

\begin{figure}
    \centering
    \includegraphics[width=\linewidth]{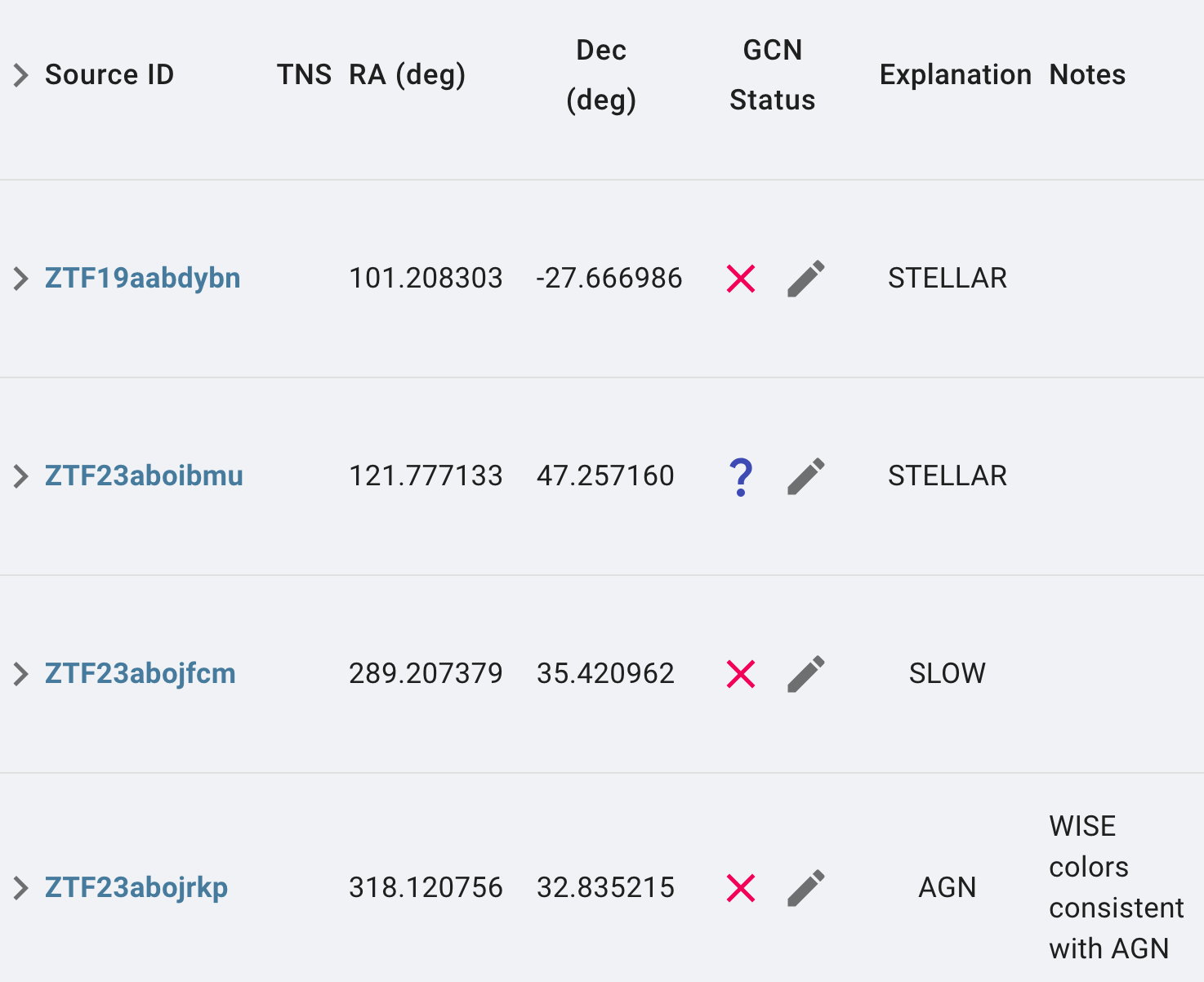}
    \caption{Snapshot of the GCN Analysis \texttt{Fritz} page showing the rejection criteria for candidates discovered within the 90\% localization of S231029k.}
    \label{fig:rejection-criteria}
\end{figure}

In many cases, the objects discovered in GW search campaigns require additional photometric and/or spectroscopic follow-up in order to discern the nature of the transients and determine whether they could be a viable EM counterpart. Objects passing the filtering criteria outlined in \S~\ref{sec:filtering1} and \ref{sec:filtering2} can be assigned external photometric and spectroscopic follow-up through \texttt{Fritz}. For example, we triggered the Spectral Energy Distribution Machine (SEDM; \citealt{BlNe2018,Rigault2019, Kim2022SEDM}) for both spectroscopy and imaging and Las Cumbres Observatory (LCO; \citealt{Brown2013LCO}) for imaging during our O4a GW search campaigns through \texttt{Fritz}. We triggered several other external photometric and spectroscopic facilities to photometrically monitor and classify transients found during our GW search campaigns; these facilities are described in Sections~\ref{subsec:imaging} and \ref{subsec:spectroscopy}.

After retrieving promising candidates within the GW localization (see Figure~\ref{fig:sources-localization}), we used in-built \texttt{Fritz} functionality to track the status of each candidate, a novelty during O4a. 
For each candidate, we can either highlight it, mark it as ambiguous, reject it, or flag it as a source that still needs to be vetted (see Figure~\ref{fig:rejection-criteria}). We can choose a reason for selection or rejection from a dropdown menu spanning the following categories:
\begin{itemize}
    \item \texttt{Local/Far} - based on the photometric/spectroscopic redshift of a potentially associated host galaxy, a candidate appears to be consistent with the GW distance, or too far to be associated with it.
    \item \texttt{New/Old} - based on either alerts or forced photometry, a candidate that is temporally consistent with the GW event (i.e. the first alerts occur after the GW trigger time) or has a history of previous detections.
    \item \texttt{Red} - based on either alerts or forced photometry a candidate exhibits red colors in its light curve ($g-r > 0.3$\,mag), as expected for a KN.
    \item \texttt{Fast/Slow} - based on either alerts or forced photometry, a candidate's light curve evolves more rapidly or slowly than 0.3 mag day$^{-1}$ (the minimum decay rate expected for a KN-like transient; \citealt{Andreoni2020fasttransient}).
    \item \texttt{Rock} - based on examination of image cutouts or light curve, a candidate is characterized as a moving object.
    \item \texttt{Stellar} - a star lies within 2 arcsec from the candidate position and/or the light curve has stellar-like variability.
    \item \texttt{AGN} - a candidate's host galaxy exhibits WISE colors consistent with an AGN, it shows photometric variability, and/or it is spectroscopically classified as AGN.
    \item \texttt{Bogus} - upon detailed examination of alert image point spread functions, a candidate appears to be an image artifact.
    \item \texttt{Specreject} - the spectrosopic classification of a candidate matches neither a GRB afterglow nor a kilonova.
\end{itemize}

Optionally, users can also leave a customized note on the candidate, providing additional information not captured in the dropdown menu. Since the selection/rejection tool is dynamic, users can update the status of a given candidate once additional information (such as forced photometry, or follow-up photometry/spectroscopy) has been obtained. One such example of the candidate selection/rejection tool is shown in Fig.~\ref{fig:rejection-criteria} for the GW event S231029k. 

\subsection{Transient Vetting: Dissemination of candidates} \label{sec:filtering5}

\begin{figure}
    \centering
    \includegraphics[width=0.5\textwidth]{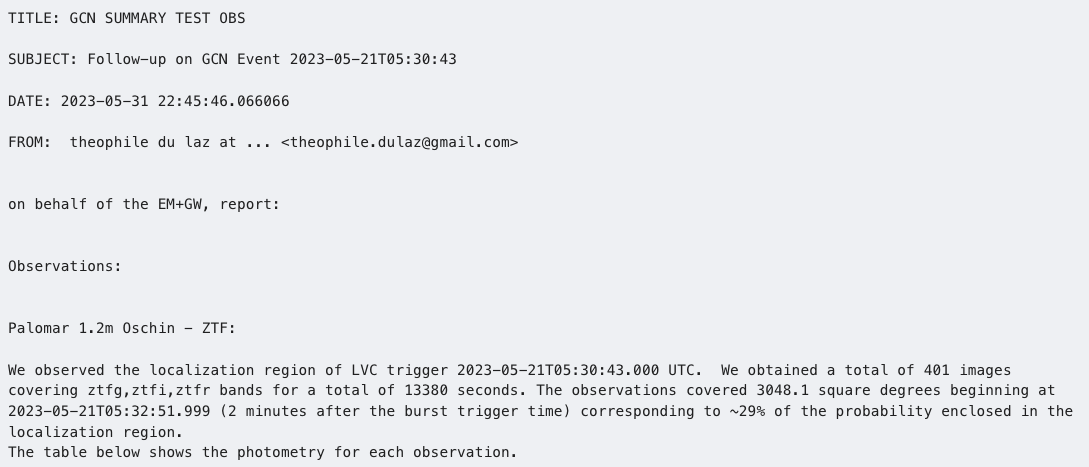}
    \includegraphics[width=0.5\textwidth]{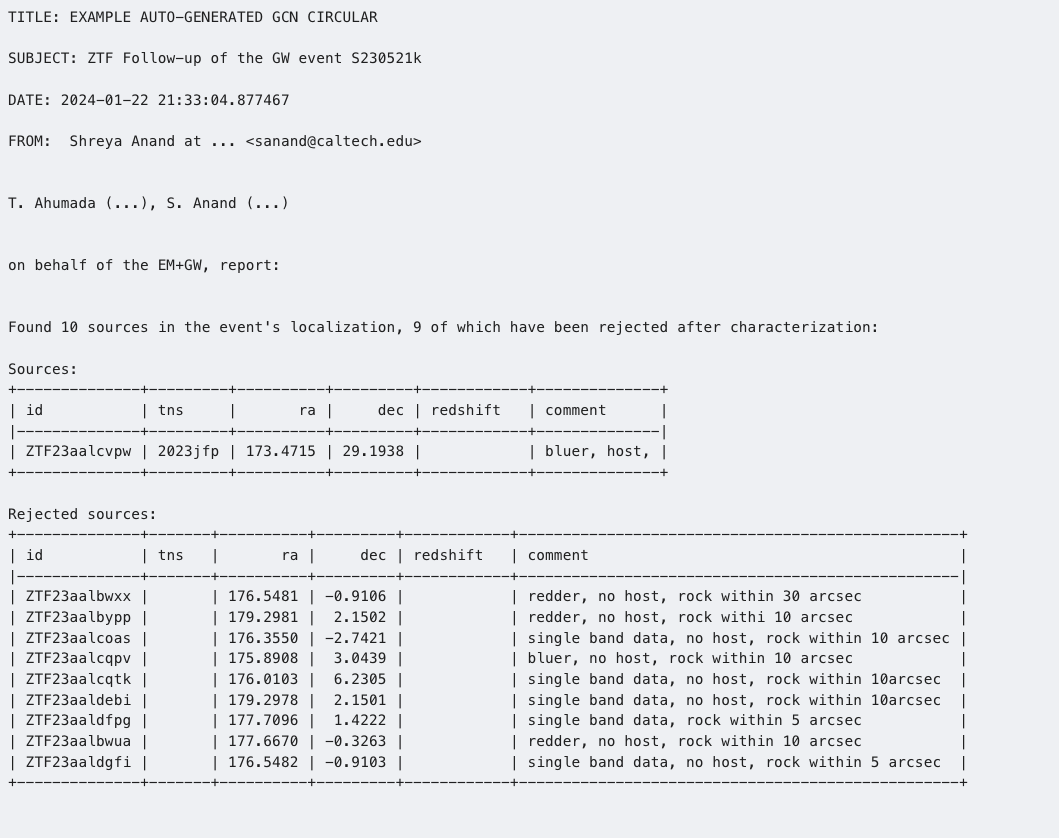}
    \caption{Two examples of auto-generated GCN circular text for the GW event S230521k. Top: a summary of the actual ZTF observations conducted. Bottom: selected and rejected candidates within the GW localization.}
    \label{fig:gcn-text}
\end{figure}

The last step is to disseminate the details of our observations and final candidate selection via GCN circular to the broader astronomical community. Based on the status of candidates marked in the selection/rejection tool, they will be automatically sorted into separate table and displayed in the content of the GCN circular. Furthermore, \texttt{Fritz} generates a summary of the conducted ZTF observations, with probability and areal coverage within the requested time window, along with a detailed table of the ZTF photometry. Examples of auto-generated GCN circular text summarizing ZTF observations as well as tables with highlighted and rejected candidates are shown in Fig.~\ref{fig:gcn-text}. This new, streamlined system for retrieving transients within the localization, tracking their status, and generating a GCN draft allowed for the timely circulation of interesting candidates discovered with ZTF to the rest of the multi-messenger astronomy (MMA) community. The ZTF fields and the coverage of the gravitational wave skymap is also made available through \texttt{Treasuremap} \citep{Wyatt2020tresuremap} to the community. 

\section{Summary of ZTF triggers}\label{sec:triggers}


In this section we describe the ZTF observations of 5 O4a GW events that had a probability of BNS or NSBH greater than 0.1 (see Table~\ref{tab:O4_summary}) and a FAR $<$ 1 year$^{-1}$. In Appendix \ref{apendixA} we describe the observations of 5 additional GW events with FAR greater than 1 year$^{-1}$. Of the events described in this section, only S230627c passed our criteria to trigger ToO observations. We obtained some serendipitous observations within the skymap of S230518h, but the updated skymap excluded the ZTF-observed regions. The remaining events (GW230529, S230731an, S231113bw) were observed using the re-weighting strategy (see Table~\ref{tab:O4_summary}).

\begin{table*}[tb]

\resizebox{\textwidth}{!}
{
\begin{tabular}{l|l|l|l|l|l|l|l|l|l|l|l|l}
{\bf Trigger }& {\bf Strategy }& {\bf  FAR } & {\texttt{$p_{BNS}$} }& {\texttt{$p_{NSBH}$} }& {\bf \texttt{HasNS} } & {\bf \texttt{HasRemnant} } & {\bf \texttt{HasMassGap} } & {\bf Distance } & {\bf Covered   }& {\bf Area covered }& {\bf g-band depth} & {\bf latency} \\
{\bf  }& {\bf  }& {\bf  [year$^{-1}$]} & {\bf  } & {\bf prob. } & {\bf prob. } & {\bf prob. } & {\bf prob. } & {\bf [Mpc]} & {\bf prob.  } &{\bf[deg$^2$]}& {\bf [AB mag]} & {\bf  [hr]} \\
\hline
S230518h  & No coverage          &  0.01  & 0.0  & 0.86 & 1.0  & 0.0  & 0.0  & 204  & --  & --   & --    & -- \\
GW230529 & Re-weighting  &  0.006 & 0.31 & 0.62 & 0.98 & 0.07 & 0.73 & 197  & 7\%  & 2425 & 20.6 & 10 \\
S230627c  & ToO           & 0.01   & 0.0  & 0.49 & 0.0  & 0.0  & 0.14 & 291  & 74\% & 72   & 21.03 & 2.2 \\
S230731an  & Re-weighting & 0.01   & 0.0  & 0.18 & 0.0  & 0.0  & 0.0  & 1001 & 3\%  & 43   & 18.7  & 12.4 \\
S231113bw & Re-weighting  & 0.42   & 0.0  & 0.17 & 0.0  & 0.0  & 0.02 & 1186 & 11\% & 301  & 21.17 & 7.7 \\
\end{tabular}
}
\caption{Summary of ZTF observations and GW properties of the 5 GW events selected and analyzed in this paper. We required their FAR to be less than 1 year$^{-1}$, and one of the following: $p_{BNS} > 0.1$, $p_{NSBH} > 0.1$, or \texttt{HasNS} $>$ 0.1. We quote other quantities intrinsic to the GW event, such as the mean distance to the merger, the \texttt{HasRemnant}, and the \texttt{HasMassGap} parameters. For each event we summarize the coverage, depth and latency for the ZTF observations. We include the events with FAR $>$ 1 year$^{-1}$ in Table \ref{tab:O4_summary_apendix} in the Appendix \ref{sec:ap_gwevents}. To determine the areal coverage and the enclosed skymap probability observed by ZTF, we require at least two ZTF observations in a given region. }
\label{tab:O4_summary}
\end{table*}

\subsection{S230518h}

The first event detected during O4a was during the engineering run, on May 18th, 2023 \citep{GCN2023ligo_S230518h}. This event was a highly significant event (FAR of one per 100\,years) and was originally classified as a likely NSBH (86\%) and its 90\% credible region spanned close to 460 deg$^2$.  The majority of the region was observable only from the Southern hemisphere, and ZTF covered $\sim$ 2\% of the initial region. However, IGWN circulated an updated localization 8 days after the event for which the ZTF coverage was negligible.

\subsection{GW230529}

\begin{figure*}[!t]
    \centering
    \includegraphics[width=0.9\textwidth]{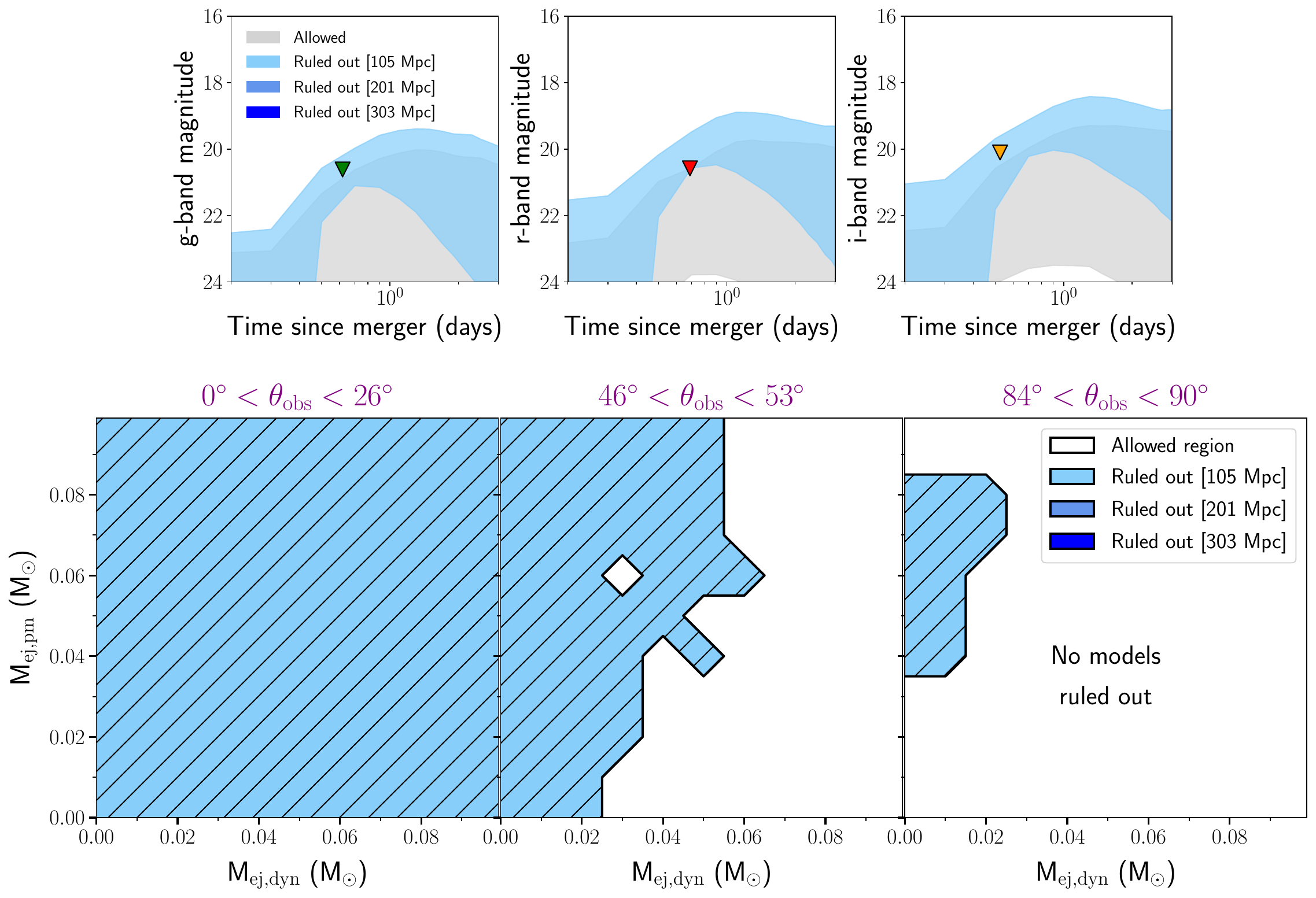}
    \caption{
    Constraints on KN model parameters based on the ZTF limiting magnitudes on GW230529. {\it Top panels}. the $g$ (left), $r$ (middle) and $i$ (right) band non-detections are shown together with NSBH KN models: the blue areas encompass light curves that are ruled out by the limits at three different distances (corresponding to median distances and $\pm 1\sigma$ distance uncertainties from LIGO), while those in grey encompass light curves that are compatible with the limits. These NSBH-specific models are computed with \texttt{POSSIS} \citep{Bulla2019, Anand2020} and have three free parameters: the mass of the lanthanide-rich dynamical ejecta ($M_{\rm ej,dyn}$), the mass of the post-merger disk-wind ejecta ($M_{\rm ej,pm}$) and the viewing angle ($\theta_{\rm jobs}$).  {\it Bottom panels}. Regions of the $M_{\rm ej,pm}$- $M_{\rm ej,dyn}$ parameter space that are ruled out at different distances and for different viewing angle ($\theta_{\rm jobs}$) ranges (from a face-on to a edge-on view of the system from left to right), assuming the KN fell within the ZTF footprint.}
\label{fig:models_GW230529}
\end{figure*}

GW230529 is a highly significant (FAR of 1 per 160\,yrs), single detector (LIGO Livingston) event \citep{GCN2023LigoS230529ay}. It was confirmed as an astrophysical event in April 5th, 2024 \cite{GW230529}. The 90\% credible region spans over 24000 deg$^2$, thus we did not trigger ToO observations and decided to re-weight the ZTF survey fields. The first observation started $\sim$ 10\,hours after the GW trigger and based on the first night, the median limiting magnitudes were $g=21.1$ and $r=21.0$ mag. Over three nights of observations, we covered 2425 deg$^2$, that translates to 7\% of the localization region. We originally found six candidates in this region; upon follow-up, none of them showed KN-like signatures and hence were rejected \citep{GCN2023ZTFS230529ay}. Details of the candidates are presented in Table~\ref{tab:candid}. Although our coverage is only 7\%, our limiting magnitudes allow us to set constraints in the properties of the KN, assuming the event was in the ZTF footprint (see Fig.\ref{fig:models_GW230529}). Specifically, we can rule out KNe with polar viewing angles ($0^\circ < \theta_{\mathrm{obs}} < 26\circ$) within the observed region, assuming a distance of 105\,Mpc (corresponding to the median$-1\sigma$ distance) for the NSBH merger (see Fig. \ref{fig:models_GW230529}). 

\subsection{S230627c}

\begin{figure}[!t]
    \centering
    \includegraphics[width=0.45\textwidth]{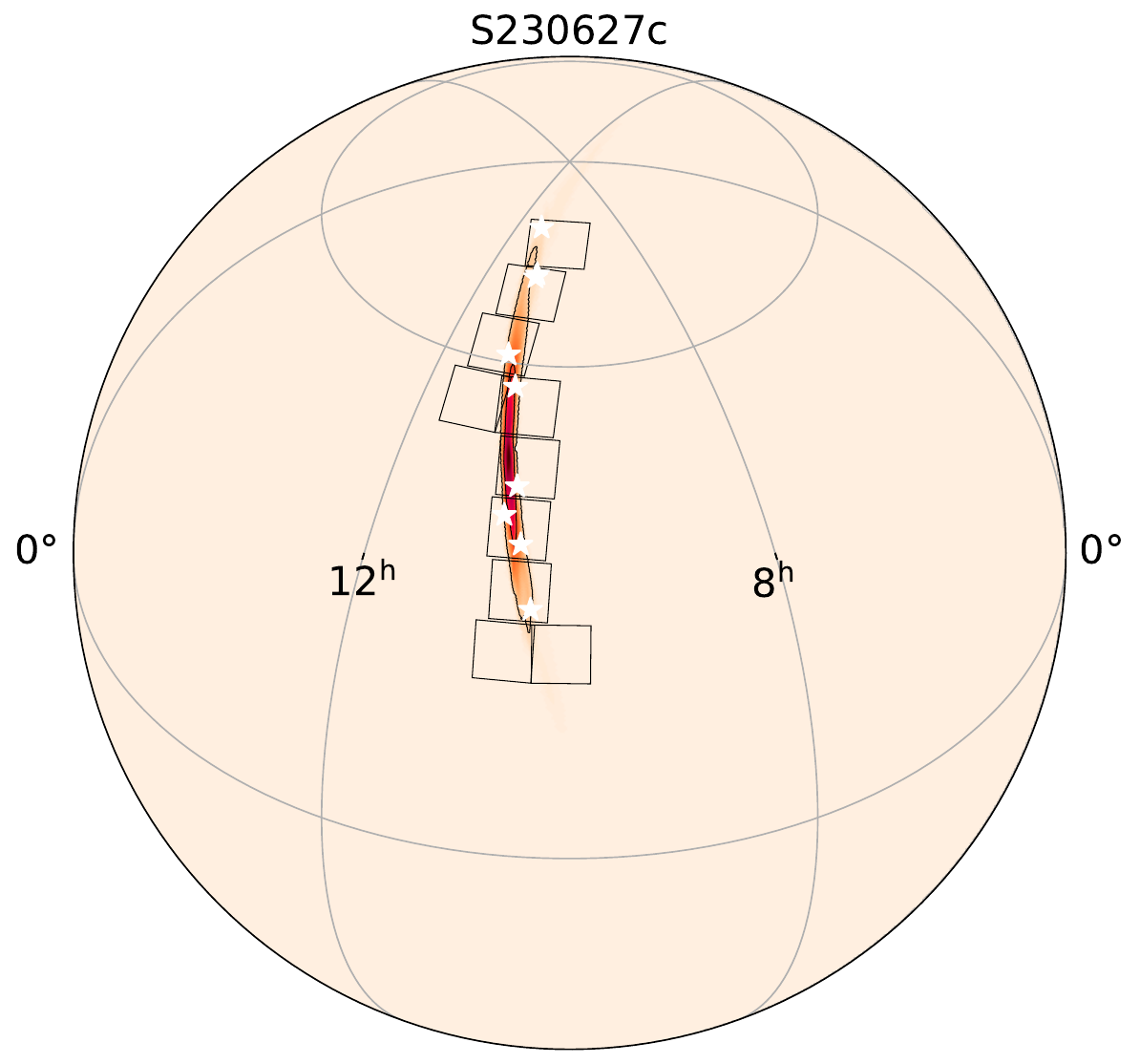}
    \caption{Localization of S230627c, overplotted with the ZTF coverage (black squares) and the 90\% probability contour. We show the candidates associated to this event as white stars in the localization region. The rest of the skymaps can be found in the Appendix \ref{sec:ap_gwevents}, in Figs. \ref{fig:map_S230529ay}-\ref{fig:map_S231029k}
    .}
    \label{fig:map_S230627c}
\end{figure}

S230627c, with a FAR of about 1 in 100 years, was classified by the \texttt{pycbc} \citep{pycbclive} pipeline as a likely NSBH ($\sim$50\%) or BBH ($\sim$50\%) with a relatively small localization: the 90\% of the probability spanned $\sim 82$ deg$^2$ \citep{GCN2023S230627cligo}. Even though the \texttt{GSTLAL} \citep{gstlal} pipeline classified this event as a BBH (100\%), we triggered a targeted search with ZTF. The observations started about 2.2 hours after the GW event and covered 74\% ($\sim 72$ deg$^2$) of the skymap (see Figure~\ref{fig:map_S230627c}). After an initial inspection of the candidates (Table~\ref{tab:candid}), we ran forced photometry on archival ZTF data, leading to 10 potential counterparts \citep{GCN2023S230627cztf1}. Further monitoring did not reveal color or magnitude evolution consistent with known KN models or an AT2017gfo-like transient.
Observations over the first night reached median magnitude limits of $g=21.0$ and $r=21.2$\,mag \citep{GCN2023S230627cztf2}.

\subsection{S230731an}

S230731an, had a FAR of a 1 per 100 years and the 90\% credible region of its initial localization covered 599 deg$^2$ \citep{GCN2023S230731anligo}. It was originally detected by the \texttt{pycbc} pipeline with a NSBH probability of 18\% (BBH probability of 81\%), while the \texttt{gstlal} pipeline classified it as a probable BBH (99\%). Due to its large inferred distance of 1001$\pm$242 Mpc, we decided to re-weight the ZTF fields. Due to weather, the ZTF coverage was $\sim$ 3\% (43 deg$^2$), reaching a depth of $g=18.7$\,mag, and no candidates were found in the region in a 72 hr window.

\subsection{S231113bw}
Detected by \texttt{pycbc}, this event had a relatively moderate FAR of about 1 per 2.35 years, and was initially classified as a likely BBH (79\%), or a NSBH (17\%) \citep{GCN2023S231113bwligo}. Offline analyses by IGWN later classified this event as a likely BBH (96\%), and lowered the probability of it being an NSBH to less than 1\%. The 90\% credible region spanned $\sim 1713$ deg$^2$, and although it was mostly a northern hemisphere event, the majority of the error region was in close proximity to the sun. We covered about 11\% of the skymap (301 deg$^2$), achieving a depth of $g=21.17$\,mag, and found no candidates that passed our filters \citep{GCN2023S231113bwztf}.

\label{sec:gw_trigger}

\section{Discussion}
\label{sec:discussion}

In this section, we quantify the efficiency of the ZTF searches during O4a, while also including in the analysis the confirmed astrophysical events from O3. We address this by taking both a Bayesian  (\S~\ref{subsec:nimbus}) and a frequentist approach (\S~\ref{subsec:simsurvey}). We use the ZTF observations to constrain the KN luminosity function under different assumptions. 

\subsection{nimbus} \label{subsec:nimbus} 
In our analysis of the events described above, we have utilised the hierarchical Bayesian framework \texttt{nimbus} \citep{Mohite_2022}. Briefly, \texttt{nimbus} uses a single ``average-band'' linear model (we will hereafter refer to this model as the \textit{Tophat} model) for the time evolution of the absolute magnitude using $M(t) = M_{0} + \alpha \ (t-t_0) $, where $M_{0}$ is the initial magnitude and $\alpha$ is the evolution rate, to determine the likelihood of obtaining the upper limits from ZTF observations given a model ($M_{0}$, $\alpha$).
The ``average-band'' model enables us to use ZTF observations across all bands. In order for \texttt{nimbus} to infer the intrinsic luminosity parameters, it requires information about the survey observations, which in this case includes the ZTF observation logs with the specific fields targeted, the Milky Way extinction values for each pointing, and a 3D GW skymap.

\texttt{nimbus} determines the posterior probability of a KN with a particular model (in this case, with a specific M$_{0}$ and $\alpha$) given the ZTF observations within the GW skymap. The framework self-consistently accounts for the probability of a GW event being of astrophysical origin (p$_{astro}$) and also factors in the ZTF coverage within the GW skymap. 
For every sample in the KN parameter space, \texttt{nimbus} calculates the likelihood of obtaining the observed limiting magnitude in the ZTF survey, given the model parameters for every field independently. 
For this, \texttt{nimbus} follows  \citet[\S~2.2]{Mohite_2022}. We have adopted a uniform distribution for the model priors, and flattened the \texttt{multiorder} skymap fits file for all the events to an \texttt{nside} of 256. 
Once the likelihoods have been determined of the observations for each event in all the corresponding ZTF fields, the overall posterior probability of the KN model parameters is determined as in \citet[Eq.~18]{Mohite_2022}.

\begin{figure}[ht]
  \centering
\includegraphics[width=0.53\textwidth]{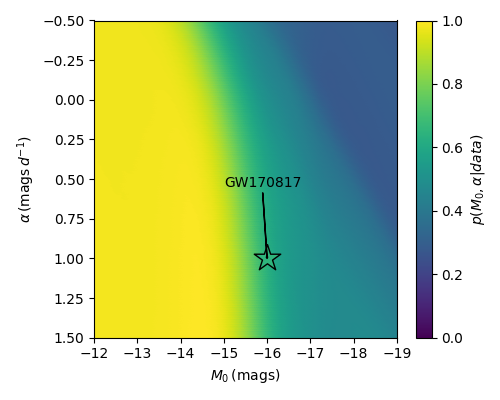}
\caption{ The \texttt{nimbus} results of the combined posterior probability for KN model parameters assuming the \textit{Tophat} model using events followed up by ZTF only during O4a. The x-axis shows the initial absolute magnitude $M_0$ of a model, while the y-axis shows its evolution rate $\alpha$. The color bar shows the posterior probability of each model, in the combined dataset, where yellow regions show the favored regions of parameter space given the non-detection of KNe from ZTF observations, and the bluer regions show less preferred combinations for initial $M_0$ and $\alpha$. We also mark the position of the average $r$-band decay rate for a GW170817-like KN over its first 3 days of evolution.}  \label{fig:nimbus_o4a}
\end{figure}

The combined posterior probability for KN model parameters using events followed up by ZTF during O4a is shown in Figure \ref{fig:nimbus_o4a}. Based on the ZTF observations of O4a events, \texttt{nimbus} shows a preference for models that are fainter than $M_0 = -16$ mag (at a credible level of 0.9), regardless of evolution rate. The yellow shaded regions in Fig.~\ref{fig:nimbus_o4a} correspond to portions of the KN parameter space that ZTF is unable to constrain based on event distances and ZTF upper limits. On the other hand, for fading KNe in the $-16 < M_{0} < -19$\,mag range, ZTF is partially sensitive, hence the posterior probability has some support for those models (at a credible level of 0.64). The bright KNe that show a rising behavior have the least preference in \texttt{nimbus}, with posterior probabilities less than 0.3. We note that the most constraining event is S230627c, as it has the best combination of coverage and depth, while for other events these numbers are more marginal. 

\subsection{simsurvey} \label{subsec:simsurvey}

\begin{figure}
    \includegraphics[width=0.47\textwidth]{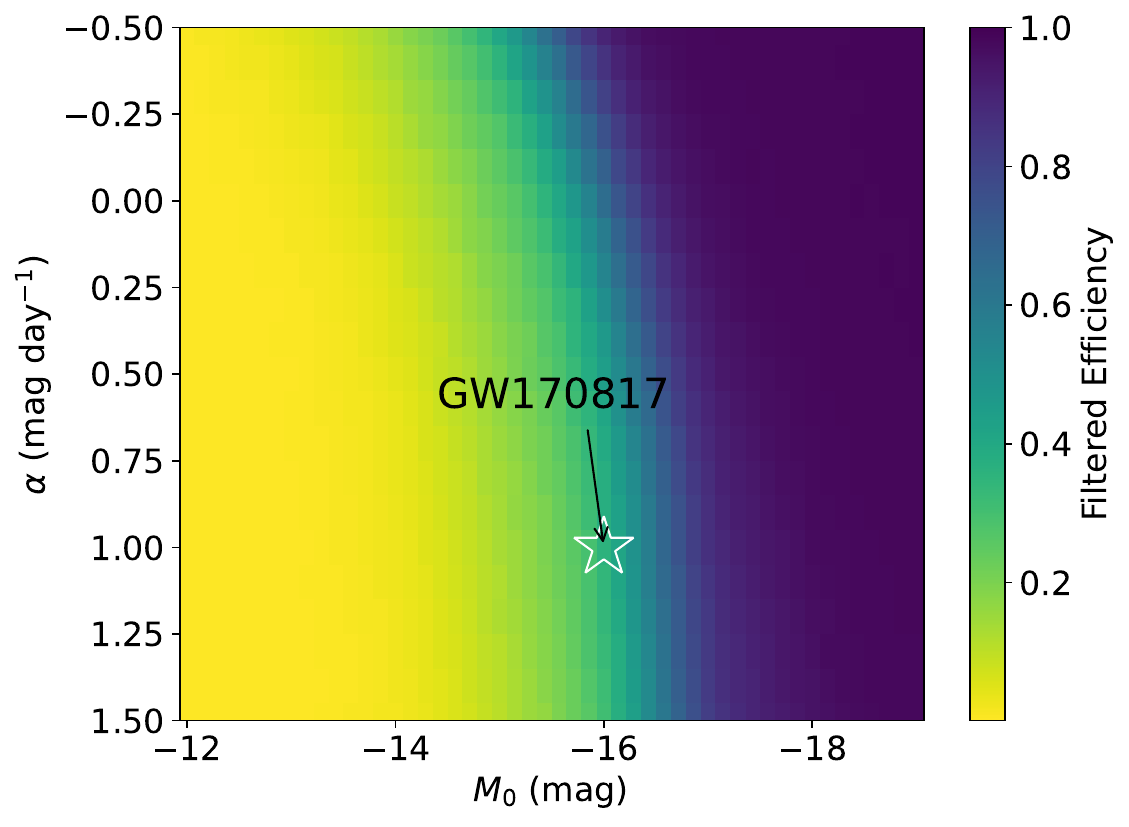}
    \caption{Filtered kilonova efficiency with \texttt{simsurvey} for the \textit{Tophat} model evolution. The filtering cuts we apply include a requirement of a minimum of two detections separated by 15 minutes at 5$\sigma$. The color bar shows the fraction of sources detected after the filtering versus the number of sources ingested in the GW volume for the O3 and O4a combined set of skymaps. Similar to Fig.~\ref{fig:nimbus_o4a}, we mark the position of a GW170817-like KN on this plot. For this dataset, GW170817 has 36\% of efficiency.}
    \label{fig:filt-efficiency}
\end{figure}

\begin{figure}
    \centering
    \includegraphics[width=0.5\textwidth]{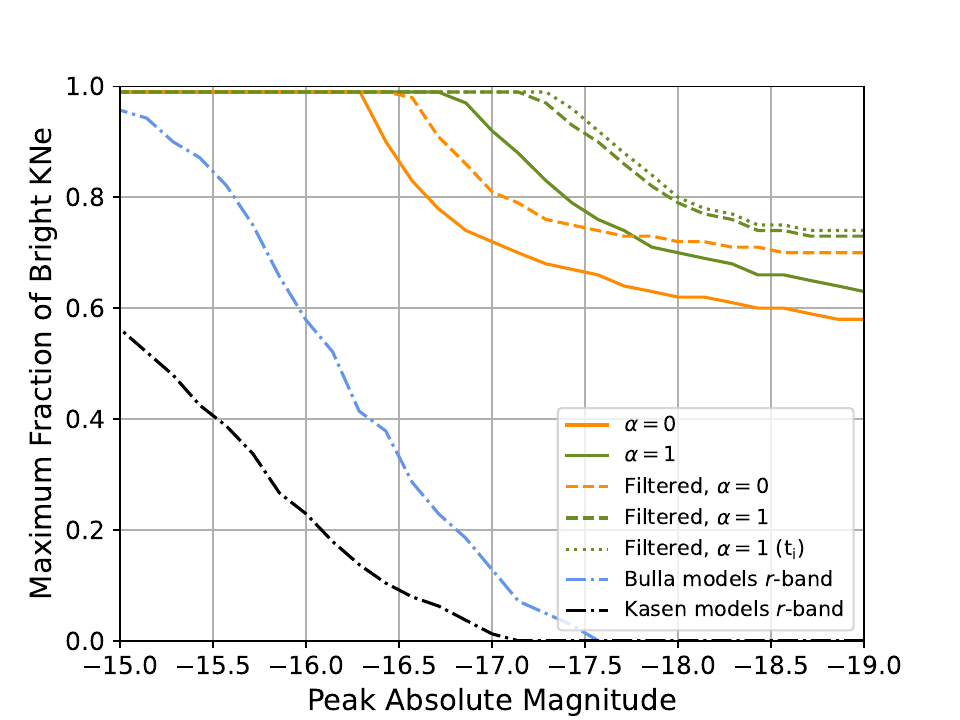}
    \caption{Kilonova luminosity function for events surviving O3, and high significance O4a events. We show in orange the models with flat evolution ($\alpha = 0$), and in green the fading models ($\alpha = 1$ mag day$^{-1}$. The solid lines show the unfiltered results, while the dashed lines show the results after selecting sources consistent with the ZTF filtering criteria (i.e. two detections). The green dotted line weights models with fading evolution passing the filtering criteria by the event's terrestrial probability (t$_{\mathrm{i}}$). The black and blue lines show the fraction of Kasen and Bulla models whose peak magnitudes fall within a particular luminosity bin. }
    \label{fig:lumfunc}
\end{figure}

Similarly to previous optical wide field of view (FoV) studies \citep{Ahumada2022grb,Kasliwal2020kn}, we make use of \texttt{simsurvey} to estimate the efficiency of the ZTF searches. The strategy that \texttt{simsurvey} takes starts with injecting KN-like light-curves in the GW localization volume, then uses the empirical ZTF coverage to measure the KN recovery rate (number of detected KNe divided by the number of injected KNe). We refer to this KN recovery rate as the KN efficiency. \texttt{simsurvey} also has filtering functionality, which we use to mimic our realistic candidate filtering criteria. In particular, for KNe to pass the filtering criteria in \texttt{simsurvey}, they must have at least two ZTF detections separated by 15 minutes above 5$\sigma$. We run separate simulations within the skymaps of each of the 5 GW events listed in Table~\ref{tab:O4_summary} as well as the five surviving O3 candidates for which we conducted ZTF follow-up (GW190425, GW190426, GW190814, GW200105 and GW200115). We chose to include GW190814 despite its ambiguous classification, since it remains unclear whether the merger was a BBH or NSBH. We inject three different sets of KN models into \texttt{simsurvey}:

\begin{enumerate}
    \item \textit{Tophat} - an empirical KN model parameterized by initial absolute magnitude ($M_{\rm 0}$) and evolution rate ($\alpha$). This same model was used in the \texttt{nimbus} framework.
    \item \textsc{POSSIS} - the 3-D, radiative transfer \textit{Bu2019lm} KN models described in \citet{Bulla2019} and  \citet{Diet2020}, parameterized by dynamical ejecta mass, disk wind ejecta mass, half-opening angle of the lanthanide-rich component, and viewing angle.
    \item \textit{Kasen} - 1-D, radiative transfer KN models described in \citet{Kasen2017}, parameterized by total ejecta mass, velocity, and lanthanide fraction (no viewing angle dependence).
    \item \textit{Banerjee} - 1-D radiative transfer KN model from \citep{Banerjee2022, Banerjee2023}, parameterized by the density, total ejecta mass, and lanthanide fraction (no viewing angle dependence).
\end{enumerate}

In Fig.~\ref{fig:filt-efficiency}, we plot the KN efficiency for the \textit{Tophat} model after applying the filtering criteria used in the ZTF searches (i.e. two detections). ZTF would detect a GW170817-like KN with M$_0 \approx -16.0$\,mag and $\alpha \approx 1.0$\,mag day$^{-1}$ passing the basic filtering criteria with 36\% efficiency. In contrast, during O3, our joint detection efficiency (i.e. one detection in \texttt{simsurvey}) for a GW170817-like KN was 93\% \citep{Kasliwal2020kn}. The lower joint efficiency for O3+O4a events compared to \citet{Kasliwal2020kn} can be attributed to the fact that many GW event candidates we followed up in O3 were retracted \citep{GWTC3}, and we assess efficiency using more realistic criteria of two detections in \texttt{simsurvey} rather than one. 
In the \texttt{simsurvey} simulations, we detect KNe brighter than M$_{0}=-17.5$\,mag with $>90\%$ efficiency, indicating that such bright KNe are unlikely to have existed in our dataset.

Next, we determine the efficiency with which we can recover GW170817-like KNe in our ZTF observations for more complex models: \textsc{POSSIS}, \textit{Kasen}, and \textit{Banerjee}.
Using the best-fit parameters of GW170817, we find that the filtered combined efficiency is 36\% and 35\% for the \textsc{POSSIS} and the \textit{Kasen} models respectively. The \textit{Banerjee} models, which assume a lanthanide fraction of $X_{\mathrm{lan}}=$0.1, are slightly more pessimistic, predicting a filtered combined efficiency of 20\%. We note that the proximity of results from KN models to the approximated \textit{Tophat} model efficiency of 36\% shows that the \textit{Tophat} model is a good initial approximation to the KN evolution. In particular, with the \textit{Tophat} model, we can recover GW170817-like KNe with $>$15\% efficiency only in the follow-ups of GW190425 and S230627c, indicating that our most successful EMGW follow-up campaigns with ZTF during O3 and O4a have been of those two events.


While \texttt{nimbus}, a Bayesian approach, and \texttt{simsurvey}, a frequentist approach, provide independent information about KNe given the ZTF observations, these frameworks are complementary to one another. \texttt{nimbus} provides insight into which KN model parameters are more or less favored, given the ZTF observations, while \texttt{simsurvey} allows us to assess the recovery efficiency of KNe with particular model parameters from the ZTF follow-ups. When comparing the two analyses, we note similar overall trends: bright KNe ($M\lesssim -17.5$\,mag) that exhibit rising behavior have the highest efficiencies in \texttt{simsurvey} and are the least preferred by \texttt{nimbus}, while faint, fast-fading KNe with the poorest detection efficiencies in \texttt{simsurvey} have the highest support in \texttt{nimbus} given the ZTF non-detections.

\subsection{Kilonova Luminosity Function Constraints}

Combining all of our EMGW follow-ups in O3 and O4a described above, we follow \citet{Kasliwal2020kn} in calculating the joint constraints on the KN luminosity function. The luminosity function is given by the following equation:

$$(1 - \rm{CL}) = \prod_{i=1}^{N} (1-f_{b}\cdot p_{i}\cdot(1-t_{i}))$$ \\
where CL is the confidence level, f$_{\rm b}$ is the maximum allowed fraction of KNe brighter than a given absolute magnitude, p$_{\rm i}$ is the probability of KN detection within a given GW event skymap, and t$_{\rm i}$ is the terrestrial probability, defined as $1 - (p_\mathrm{astro})$. We solve for f$_{\rm b}$ at 90\% confidence for each luminosity bin and plot the results in Figure~\ref{fig:lumfunc}. We include separate luminosity function curves corresponding to KNe with flat evolution and declining at 1 mag day$^{-1}$, with two tiers of criteria: KNe recovered with a single detection (solid lines), and KNe passing our filtering criteria of two 5$\sigma$ detections separated by 15\,minutes (dashed lines). In all of the curves except for the green dotted line, we set t$_{\rm i}$ to zero for all events, meaning that we assume that all of the events are astrophysical in those cases. 

For reference, we plot curves corresponding to the fraction of \textsc{POSSIS} (\textit{Bu2019lm}) and \textit{Kasen} models peaking at, or brighter than a particular luminosity bin (see Fig. \ref{fig:lumfunc}). The \textsc{POSSIS} models span $M_{\mathrm{ej, dyn}}=0.001-0.02 \rm M_{\odot}$, $M_{\mathrm{ej, wind}}=0.01-0.13 \rm M_{\odot}$, half-opening angles of the lanthanide rich component $\phi=15-75\deg$, and viewing angles $\theta=0-90\deg$; we exclude the \textsc{POSSIS} models with half-opening angles of $\phi=0 \deg$ and $\phi=90 \deg$. With our ZTF observations, we can place constraints on the luminosity function for fading KNe with $M\lesssim-16.5$\,mag, corresponding to $\sim$35\% of the \textsc{POSSIS} \textit{Bu2019lm} grid. The \textsc{POSSIS} models shown here are designed for KNe from BNS (and not NSBH) mergers. We note that though many of the events we followed up have a higher p$_{NSBH}$ than p$_{BNS}$, KNe from NSBH mergers are expected to be similar, but redder and fainter on average, compared to those from BNS mergers \citep{Anand2020}, and hence our ZTF observations would be much less sensitive to NSBH KNe. 

We also plot a subset of the \textit{Kasen} model grid consisting of total ejecta masses of 
$M_{\rm ej}=0.01-0.1 M_{\odot}$, velocities of $v_{\rm ej}=0.03-0.3$ c, and lanthanide fractions of $X_{\rm lan}=10^{-9} - 10^{-1}$, excluding the very faint KN models with low total ejecta masses (with $m_{\rm ej}<0.01 \rm M_{\odot}$). Approximately 10\% of the \textit{Kasen} grid KNe are brighter than $M\lesssim-16.5$\,mag, corresponding to the portion of the KN luminosity function our ZTF observations are sensitive to. Here, we choose to include a larger subset of the Bulla and Kasen grid models as compared to \citet{Kasliwal2020kn}; this choice is largely motivated by the fact that our limits are less constraining, and thus we cannot confidently exclude any portion of the KN model space.

We calculate a maximum fraction of 76\% for KNe (detected at least once by ZTF) brighter than $-$17.5\,mag and fading at 1 mag day$^{-1}$. If we take into account only KNe passing ZTF filtering criteria of two detections and fold in the event-by-event terrestrial probability, our maximum fraction of KNe brighter than $-$17.5\, mag and fading at 1 mag day$^{-1}$ becomes 92\%. At this point, our observations cannot constrain the maximum fraction of GW170817-like KNe (with $M_{\rm peak}=-16.5$\,mag, fading at 1 mag day$^{-1}$). Compared to the 40\% fraction found in \citet{Kasliwal2020kn} for objects brighter than $-18.0$\,mag with flat evolution and no filtering imposed, our constraints are slightly worse (we find a maximum fraction of 62\% for the same criteria). Out of the 13 GW events contributing to the luminosity function in \citet{Kasliwal2020kn}, only 5 survived to make it to GWTC-2 and GWTC-3. In addition to these events, we include 5 events from O4a; however amongst these events, we only triggered ToO observations on S230627c, achieving a skymap coverage $>$70\% (all other O4a events have $<$15\% skymap coverage). Thus many more GW events with $>$50\% ZTF coverage are required in O4b in order to place meaningful constraints on the maximum fraction of GW170817-like KNe.

\section{Conclusion}\label{sec:conclusion}

During the first half of IGWN's fourth observing run, O4a, we conducted GW follow-ups of five high significance GW events. In this work, we have reported our revised approach to triggering on GW events, novel \texttt{Fritz} machinery for rapidly vetting ZTF candidates found within GW skymaps, and our derived constraints on the properties of KNe.

One of the key developments during O4a is \texttt{Fritz}, a \texttt{SkyPortal} instance to manage ZTF data and coordinate follow-up. This new capability allowed us to receive the initial GW alert, create an observing plan for ZTF, trigger ZTF observations, display the sources on the GW maps, coordinate follow-up observations for telescopes, vet the candidates, and disseminate our results in an organized fashion. We complemented these searches with offline analyses (\texttt{nuzft}, \texttt{emgwcave}, and \texttt{ZTFReST}), to leave no stone un-turned in our counterpart searches.

In addition to the ZTF ToO observations, we set in place a novel approach to use the ZTF all-sky survey and observe the GW skymap regions by re-weighting the schedule to maximize the nightly coverage. In total, we conducted observations for 5 high-significance events, and used the re-weighting strategy for 4 of the cases. Only S230518h, S230529ay, S230627c, S230731an, and S231113bw were considered of high significance, as they all had FAR $<$\,1 year$^{-1}$, and $p_{BNS}\,>\,0.1$ or $p_{NSBH}\,>\,0.1$ or HasNS $>$ 0.1. We describe in Appendix \ref{sec:ap_gwevents} the results for the follow-up of additional events with a FAR $>$ 1 year$^{-1}$. In summary, we followed-up over 15 ZTF KN candidates and found no viable GW optical counterpart. 

Given the ZTF skymap coverages and limiting depths of these GW events, the lack of an associated KN counterpart is consistent with our non-detection analyses. For this, we used both Bayesian and frequentist frameworks. The Bayesian approach, \texttt{nimbus}, allows us to compare which combination of parameters are more likely to have been consistent with the non-detections during our O4a campaigns, and gives preference to KN models with starting absolute magnitudes fainter than $-$16 mag. Our frequentist approach used \texttt{simsurvey} to simulate sources in the GW skymap volumes, leading to an overall combined efficiency of 36\% for GW170817-like KNe in O3 and O4a. Both analyses show similar trends, with \texttt{nimbus} showing a preference for fainter models, and \texttt{simsurvey} exhibiting a high recovery efficiency for bright models, painting a cohesive picture between the two frameworks.

The combination of the ZTF observations during O3 and O4a allow us to set constraints on the KN luminosity function. We find that a maximum fraction of 76\% of all KNe can be brighter than $-$17.5\,mag. Our results are less constraining than the ones in \cite{Kasliwal2020kn}, mainly due to the number of high-significance events followed up and the ZTF skymap coverage for the events considered. By observing 9 (17) GW events with $>90\%$ (50\%) coverage to a sensitivity of M$_{\mathrm{peak}} > -16$\,mag, we would be able to set constraints on the maximum fraction of GW170817-like KNe at the 25\% level \citep{Kasliwal2020kn}.

New near-infrared (NIR) facilities, such as WINTER \citep{Lourie2020winter} and PRIME \citep{Kondo2023prime}, have recently joined the multi-messenger search campaigns. We expect that coordinated efforts in GW searches will lead to the use of these facilities to discriminate candidates based on their NIR evolution, and that they could conduct independent searches for GW events for skymaps in the $<$500 deg$^2$ regime. Such well-localized GW events are expected to be routinely detected in O5 \citep{Kiendrebeogo23}.
Upcoming wide field surveys, such as the Rubin observatory \citep{ivezic2019rubin}, ULTRASAT \citep{Shvartzvald2023ultrasat}, The Nancy Grace Roman Space Telescope, and UVEX \citep{Kulkarni2021uvex} will open a new window in the GW searches, surveying larger volumes and exploring the UV regime. Wide FoV surveys, such as ZTF, will continue to play a fundamental role in identifying fast fading counterparts that will likely have no previous history in these new data streams. 

One of the main challenges we faced during both O3 and O4a was the large localization areas associated with each of the events. 
We look forward to the second phase of O4, with the re-integration of Virgo to the network of interferometers at an increased sensitivity, which will reduce the sizes of IGWN sky localizations. New events discovered during O4b will likely improve our KN luminosity function constraints, while the verdict on whether the O4a events included in this analysis are recovered in offline GW analysis will also affect the results of this work.

The development of efficient tools to interface with ZTF, such as \texttt{Fritz}, has proven to be useful in the broader context of MMA during O4a, and will continue to be a valuable asset to our search efforts during O4b. \texttt{Fritz} has allowed multiple astronomers in the same team to analyze the ZTF data stream simultaneously, sharing notes and conclusions about the evolution or behavior of the candidates. \texttt{Fritz} also allows for the exploration of new observing strategies using \texttt{simsurvey}, as it can determine the KN recovery efficiency given a skymap, distance and latency. The ability to trigger automated follow-up of promising candidates within the Fritz interface itself, and generate ready-to-send GCNs summarizing our follow-up efforts are ways in which we have significantly reduced our latency in the GW follow-up process, increasing our chances of detecting the associated KN.

During O4b, we plan to include ZTF forced photometry throughout the candidate filtering stages, rather than post-facto. This is now possible because of the inclusion of forced photometry in the ZTF alert packets which are also accessible to the broader community. Additionally, new tools such as GWSkyNet may enhance our ability to target candidates that are less likely to be caused or influenced by detector noise by providing an independent metric that can be consistently interpreted across all candidates \citep{Cabero2020,Abbott2022,Raza2024}. GWSkyNet annotations are currently expected to be publicly available for LIGO-Virgo events in O4b on GraceDB\footnote{\url{https://emfollow.docs.ligo.org/userguide/content.html\#gwskynet-classification}}. 

Recent recommendations from the broader EM community \citep{WoU2024} underline the importance of prompt, public access to images and alerts, and not just the vetted counterpart candidates, from surveys conducting MMA search campaigns. Our frequent use of the re-weighting strategy during O4a has ensured immediate access to ZTF images and alerts from those GW follow-ups. This approach could be adopted by future surveys, such as the Vera C. Rubin Observatory's Legacy Survey of Space and Time survey. For both ToO and re-weighting follow-ups, we report our pointings (and limiting magnitudes) to the \texttt{TreasureMap} as soon as our observations have completed. Furthermore, the development of critical software infrastructure to streamline telescope coordination and efficiency is emphasized in the white paper. 

Following these recommendations, we highlight three specific areas where software infrastructure needs to be improved to boost multi-messenger discovery. First, joint querying of heterogeneous discovery streams in real-time (e.g., querying ZTF, WINTER, Rubin and LS4 simultaneously with kowalski) will enable both timely selection of the most promising multi-messenger candidates as well as timely rejection of the false positives. 
Second, a decentralized communications framework could facilitate active follow-up co-ordination between independent teams. This will enable optimal use of limited follow-up resources that are already the bottleneck in multi-messenger searches (e.g., communication between decentralized SkyPortal instances or similar softwares). Third, incorporating inclination angle constraints into the low-latency GW alert packets could help refine EM counterpart search strategies. For instance, one could tune the targeted depth in optical/IR bands or customize search strategies in radio/high-energy bands based on the expected emission from a KN model with GW inclination constraints applied.
Together, such improvements in software infrastructure would amplify the power of collaborative discovery.

Augmenting infrastructure used by the MMA community will make multi-messenger science more accessible to a diverse set of teams around the world. \texttt{Fritz} is an example of an open-source tool, catering to the needs of its users, designed to lower the entry barrier for astronomers into time-domain astronomy and MMA. It serves as an intuitive interface to analyze astronomical data, while exploiting the interactive nature of a number of surveys and online catalogs. We look forward to the infrastructure developments that will address the challenges raised by the MMA community, as they will foster a more inclusive approach to enabling MMA discoveries. 

%



\section{Acknowledgements}

We thank Fabio Ragosta, Nidhal Guessoum, and Albert K.H. Kong for the useful comments and suggestions. 
M.M.K., S.A. and T.A. acknowledge generous support from the David and Lucile Packard Foundation.
We acknowledge the support from the National Science Foundation GROWTH PIRE grant No. 1545949. 
M.W.C, B.~F.~H., A.T., and T.B. acknowledge support from the National Science Foundation with grant numbers PHY-2308862 and PHY-2117997.
This work used Expanse at the San Diego Supercomputer Cluster through allocation AST200029 -- ``Towards a complete catalog of variable sources to support efficient searches for compact binary mergers and their products'' from the Advanced Cyberinfrastructure Coordination Ecosystem: Services \& Support (ACCESS) program, which is supported by National Science Foundation grants \#2138259, \#2138286, \#2138307, \#2137603, and \#2138296.
C.M.C. acknowledges support from UKRI with grant numbers ST/X005933/1 and ST/W001934/1.
G.C.A. thanks the Indian National Science Academy for support under the INSA Senior Scientist Programme.

Based on observations obtained with the Samuel Oschin Telescope 48-inch and the 60-inch Telescope at the Palomar Observatory as part of the Zwicky Transient Facility project. ZTF is supported by the National Science Foundation under Grants No. AST-1440341 and AST-2034437 and a collaboration including current partners Caltech, IPAC, the Oskar Klein Center at Stockholm University, the University of Maryland, University of California, Berkeley , the University of Wisconsin at Milwaukee, University of Warwick, Ruhr University, Cornell University, Northwestern University and Drexel University. Operations are conducted by COO, IPAC, and UW.

SED Machine is based upon work supported by the National Science Foundation under Grant No. 1106171. 

The ZTF forced-photometry service was funded under the Heising-Simons Foundation grant \#12540303 (PI: Graham).

The Gordon and Betty Moore Foundation, through both the Data-Driven Investigator Program and a dedicated grant, provided critical funding for SkyPortal.

This research has made use of the NASA/IPAC Extragalactic Database (NED), which is funded by the National Aeronautics and Space Administration and operated by the California Institute of Technology.

The Liverpool Telescope is operated on the island of La Palma by Liverpool John Moores University in the Spanish Observatorio del Roque de los Muchachos of the Instituto de Astrofisica de Canarias with financial support from the UK Science and Technology Facilities Council.

This work relied on the use of HTCondor via the IGWN Computing Grid hosted at the LIGO Caltech computing clusters.


\FloatBarrier
\bibliography{main}
\bibliographystyle{aasjournal}

\appendix\label{apendixA}
\section{Observing and Data Reduction Details for Follow-up Observations}
\subsection{Photometric Follow-up} \label{subsec:imaging}

\begin{figure*}[h]
    \centering
\includegraphics[width=0.3\textwidth]{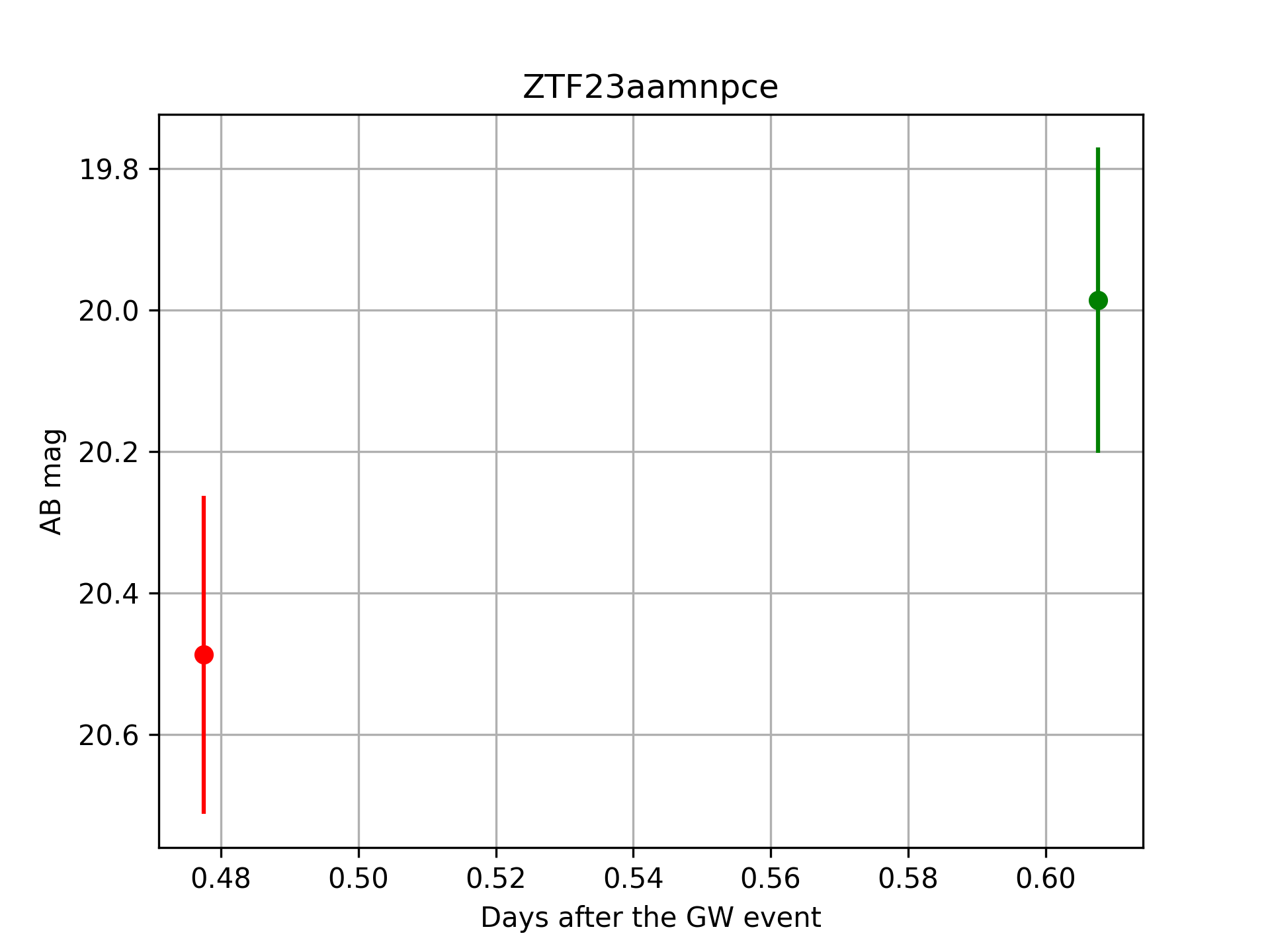}
\includegraphics[width=0.3\textwidth]{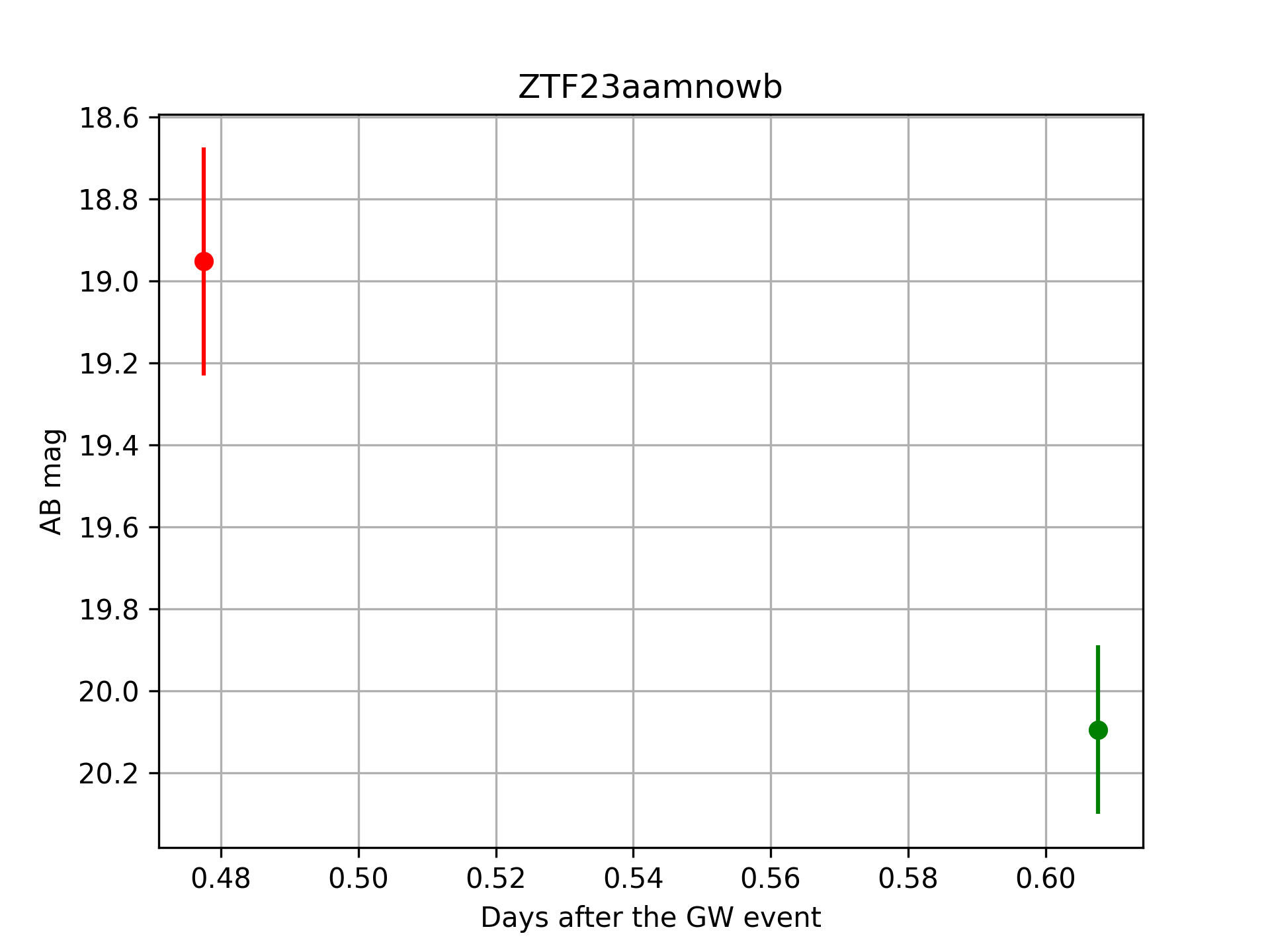}
\includegraphics[width=0.3\textwidth]{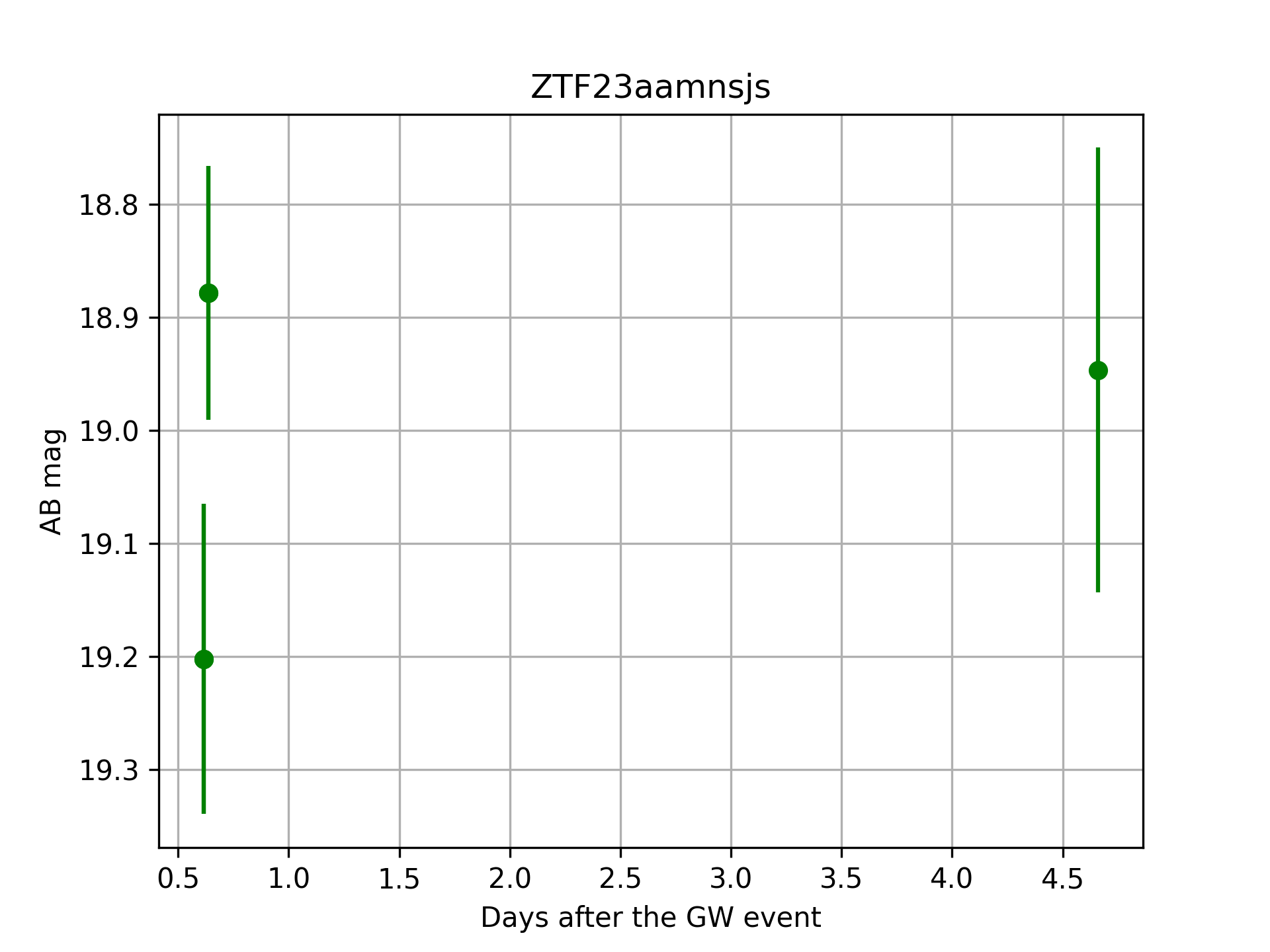}
\includegraphics[width=0.3\textwidth]{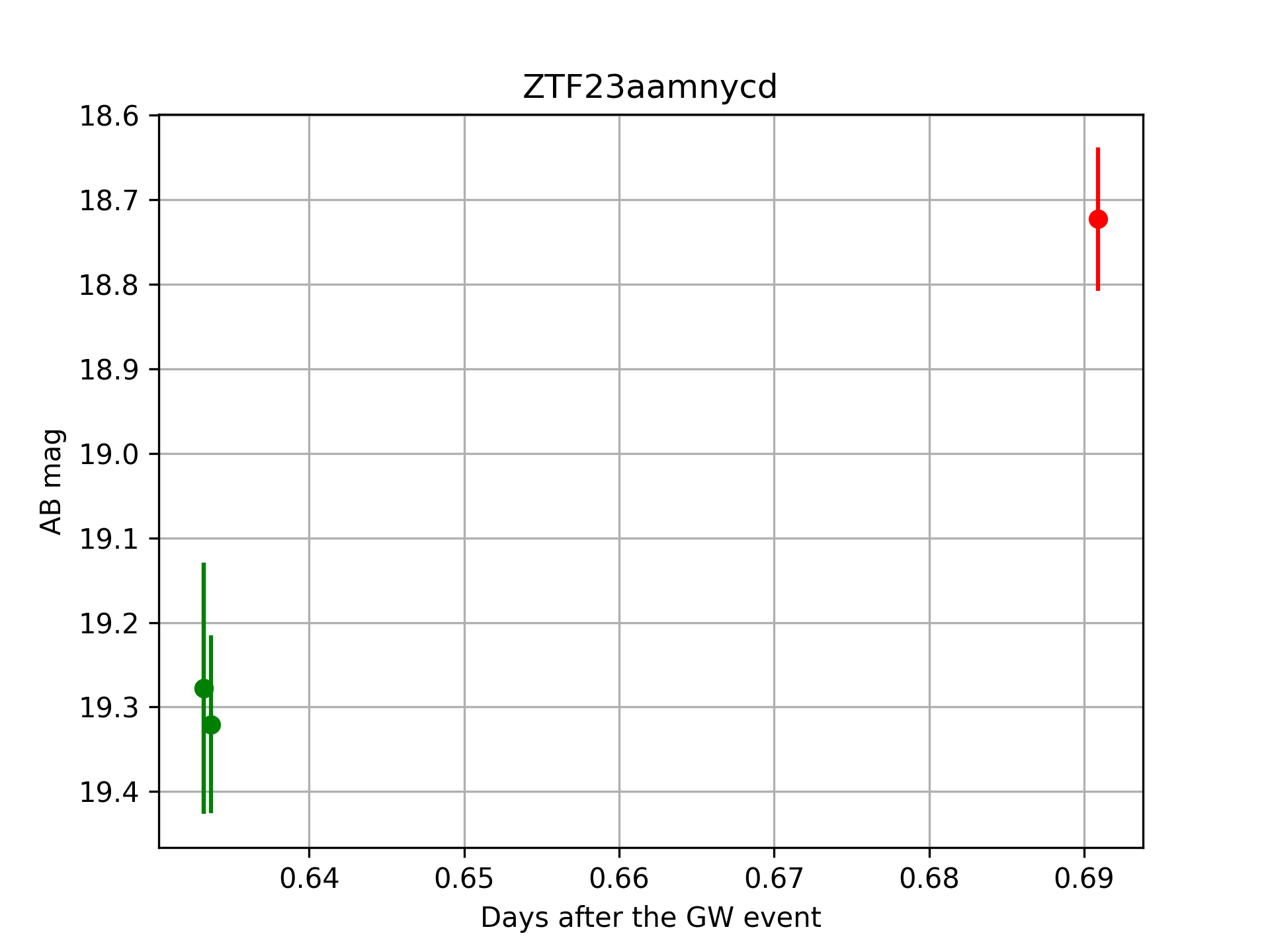}
\includegraphics[width=0.3\textwidth]{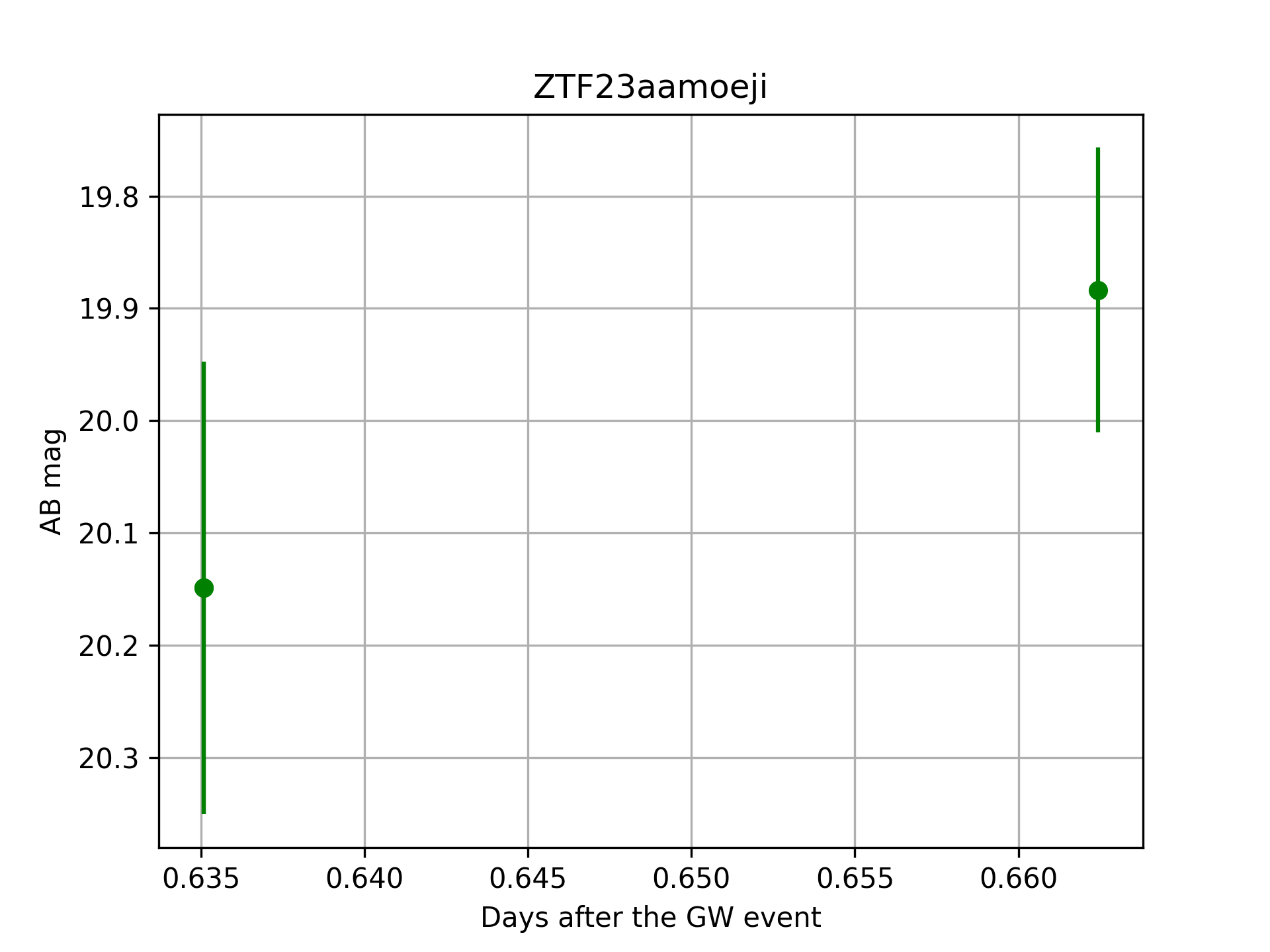}
\includegraphics[width=0.3\textwidth]{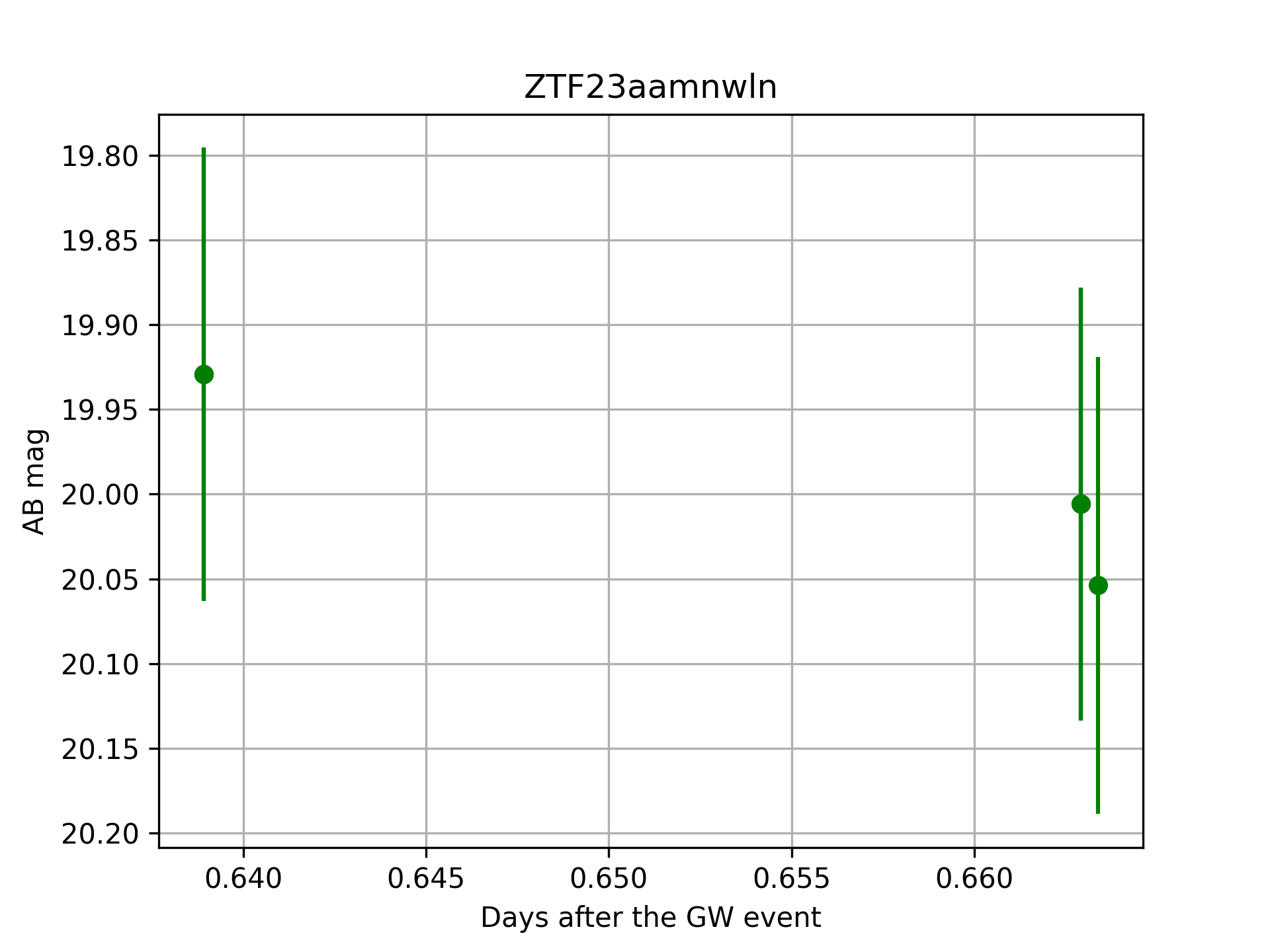}
\includegraphics[width=0.3\textwidth]{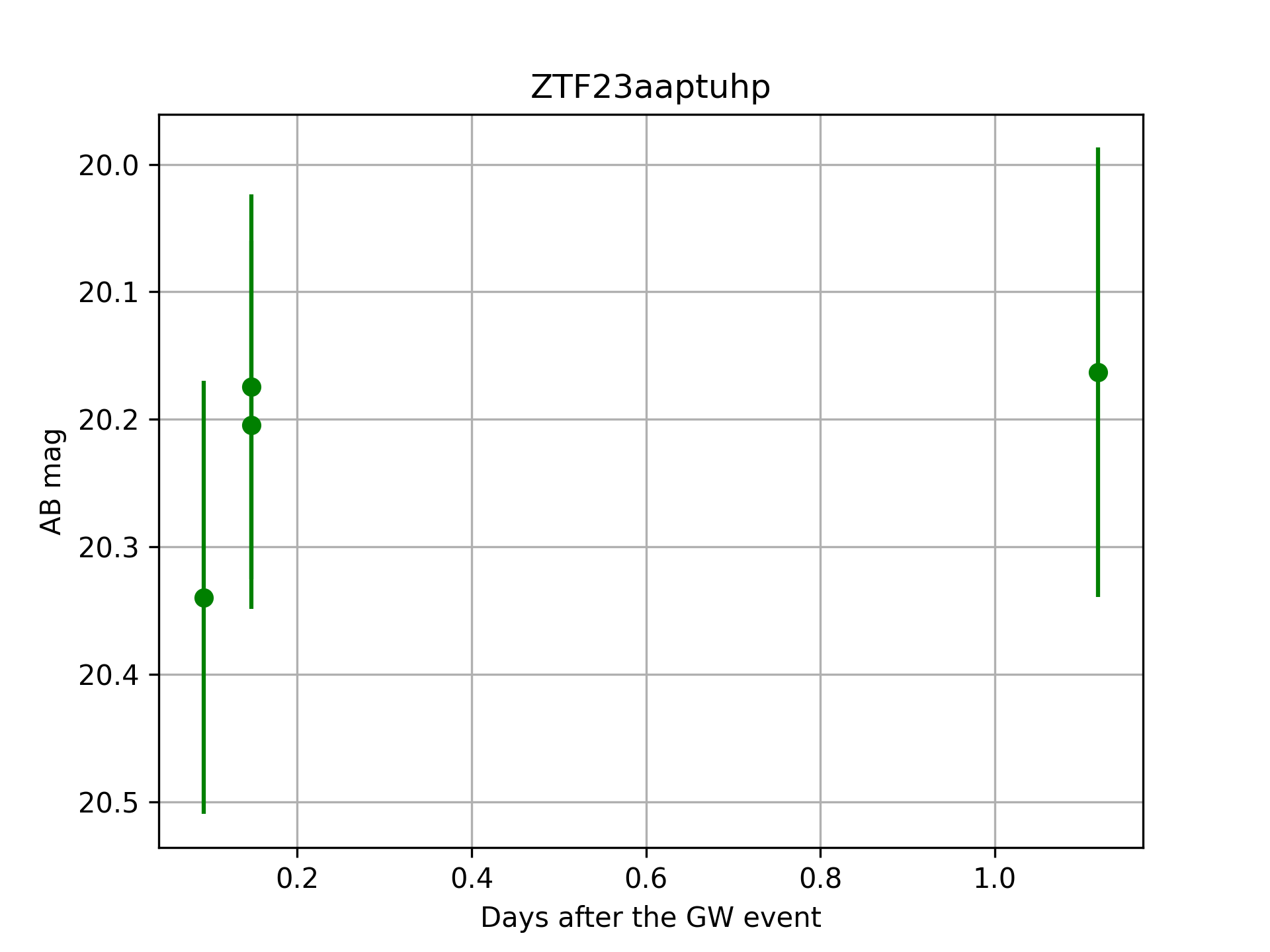}
\includegraphics[width=0.3\textwidth]{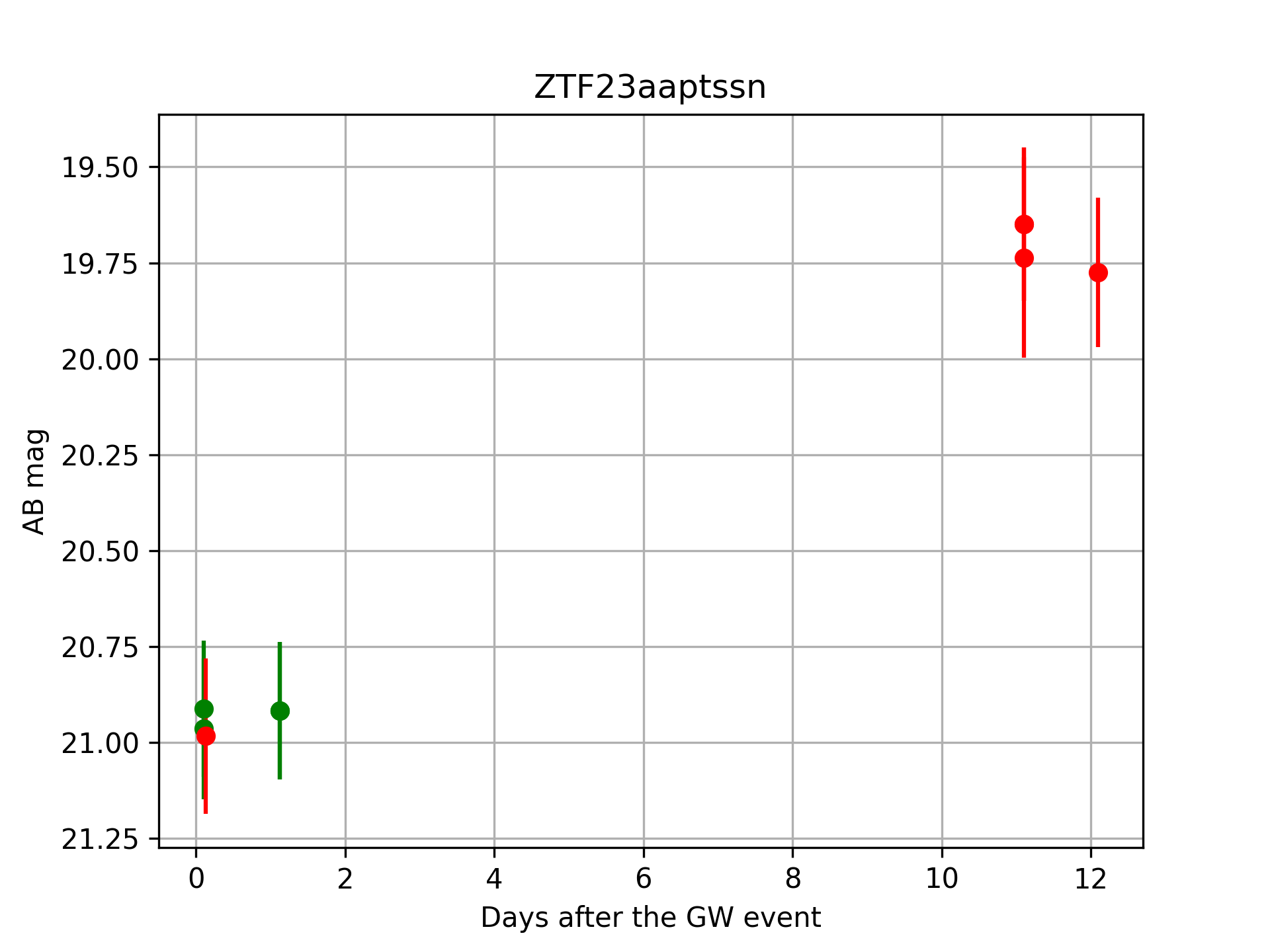}
\includegraphics[width=0.3\textwidth]{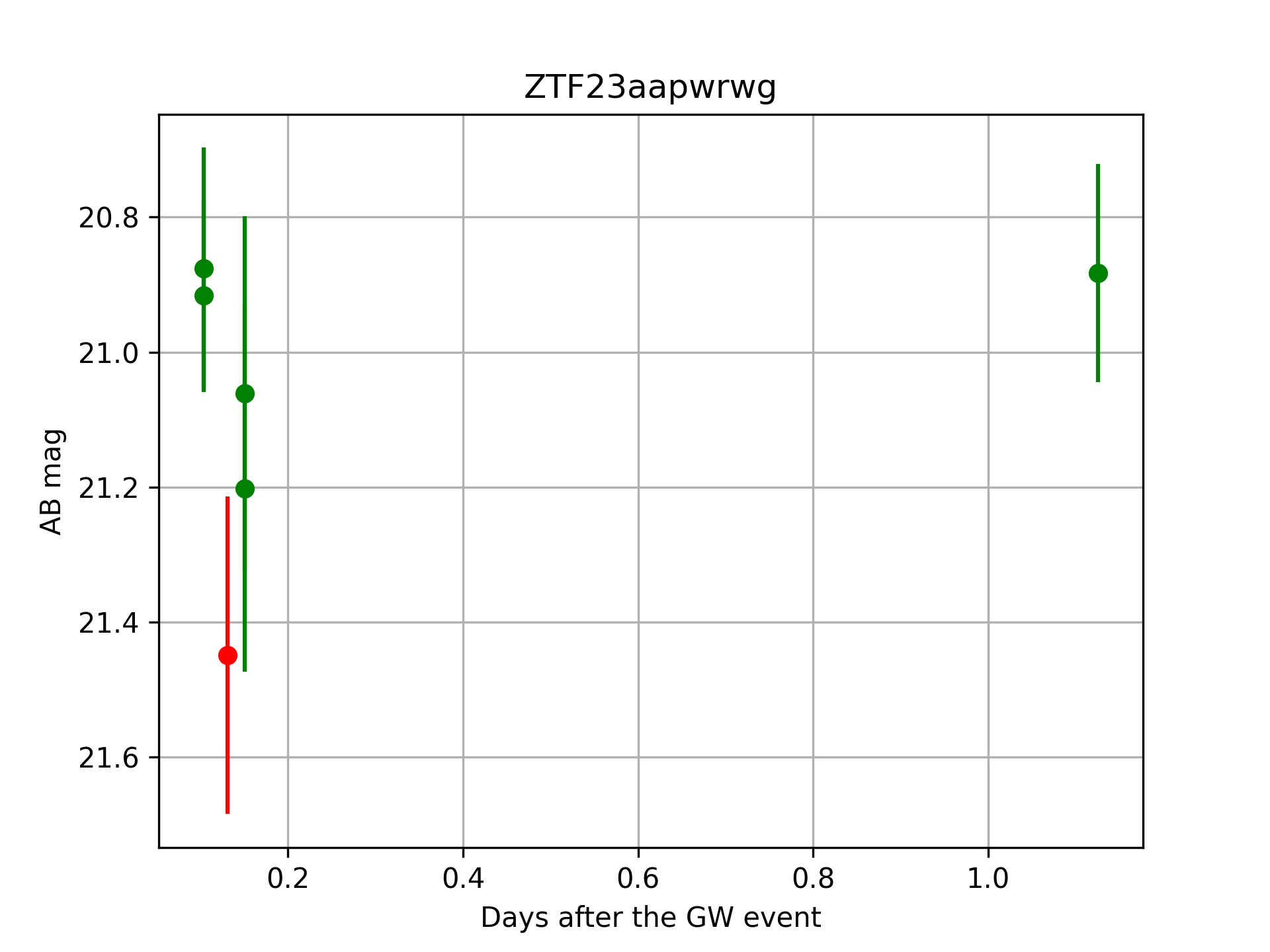}
\includegraphics[width=0.3\textwidth]{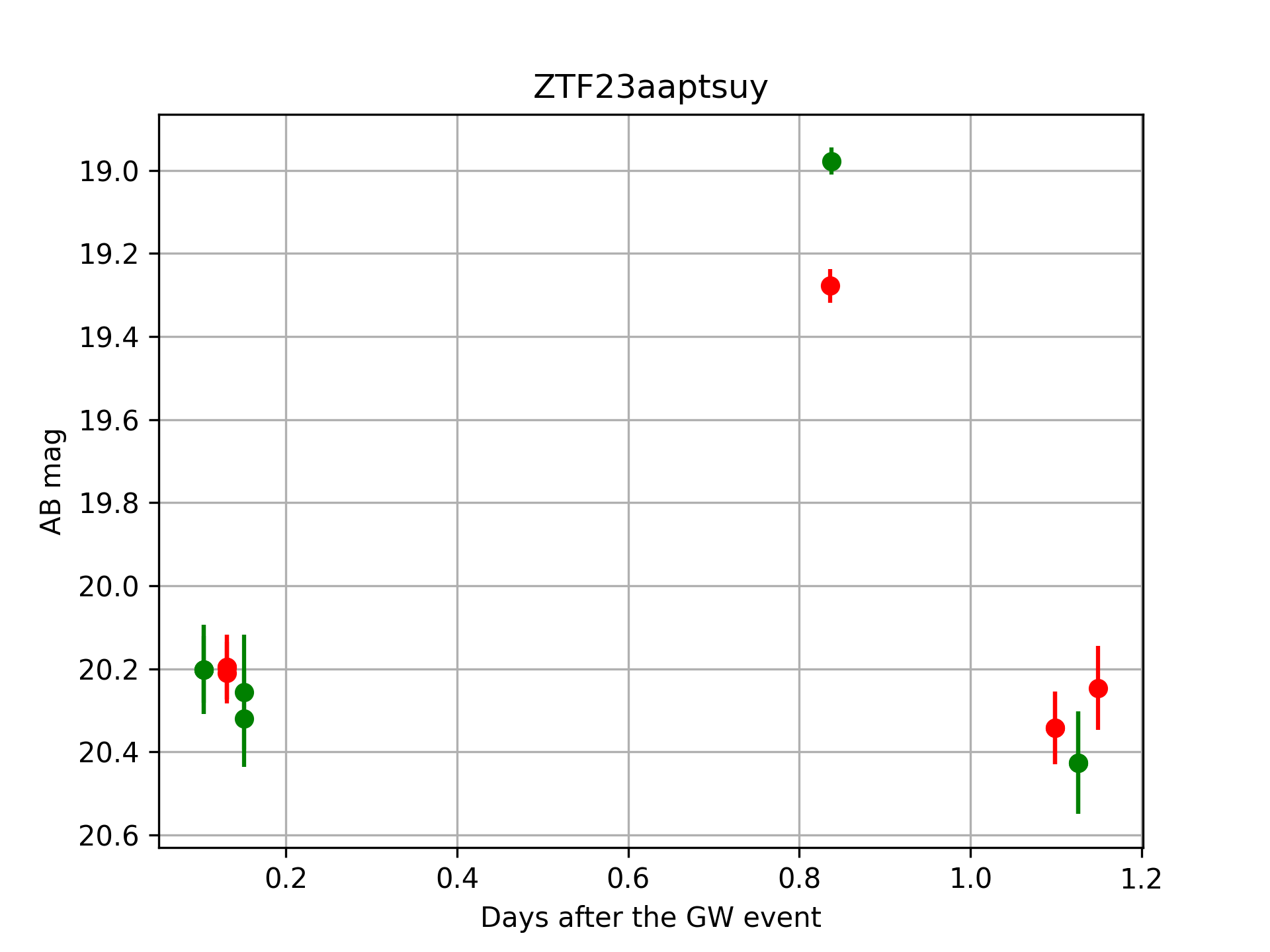}
\includegraphics[width=0.3\textwidth]{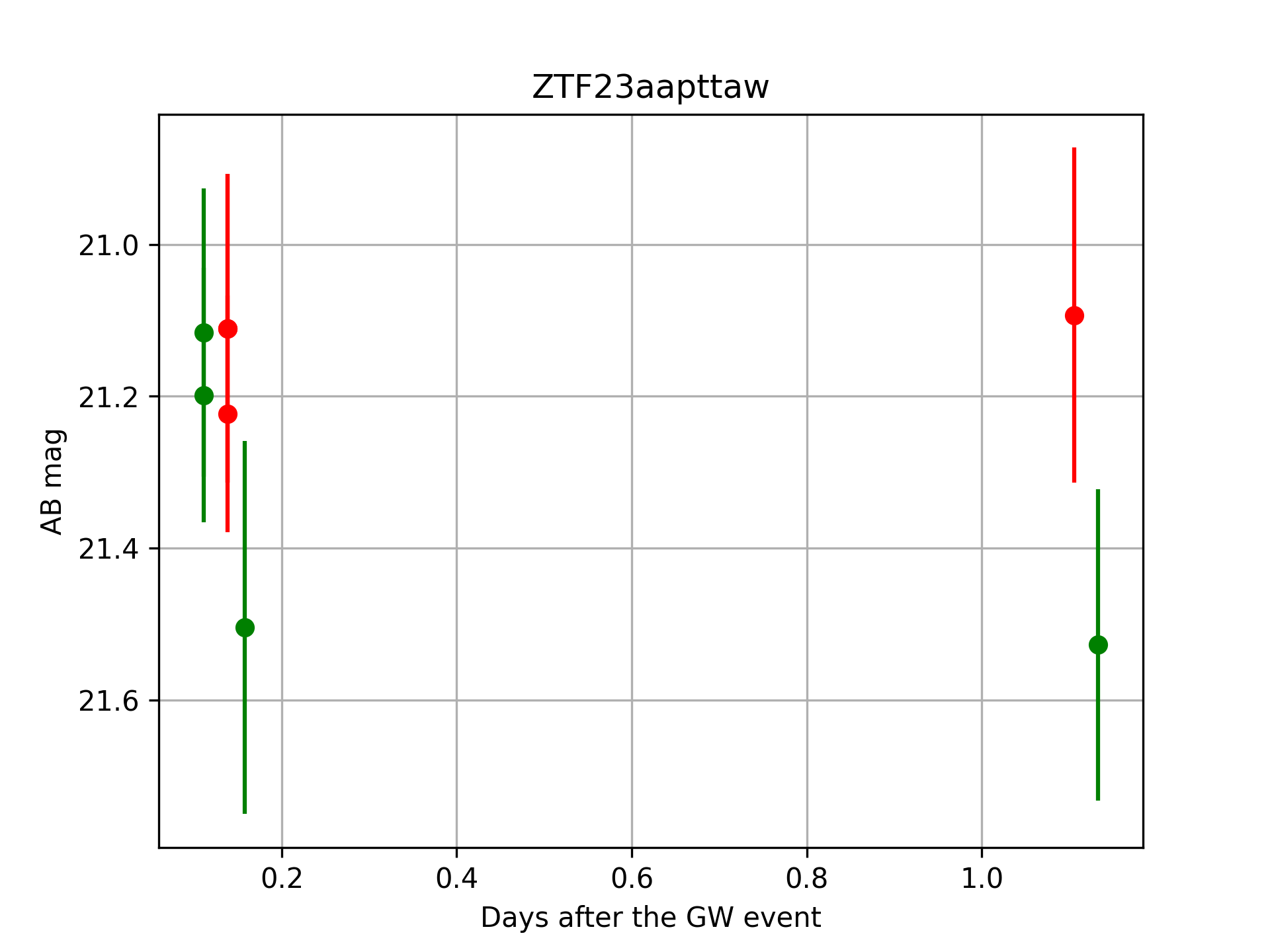}
\includegraphics[width=0.3\textwidth]{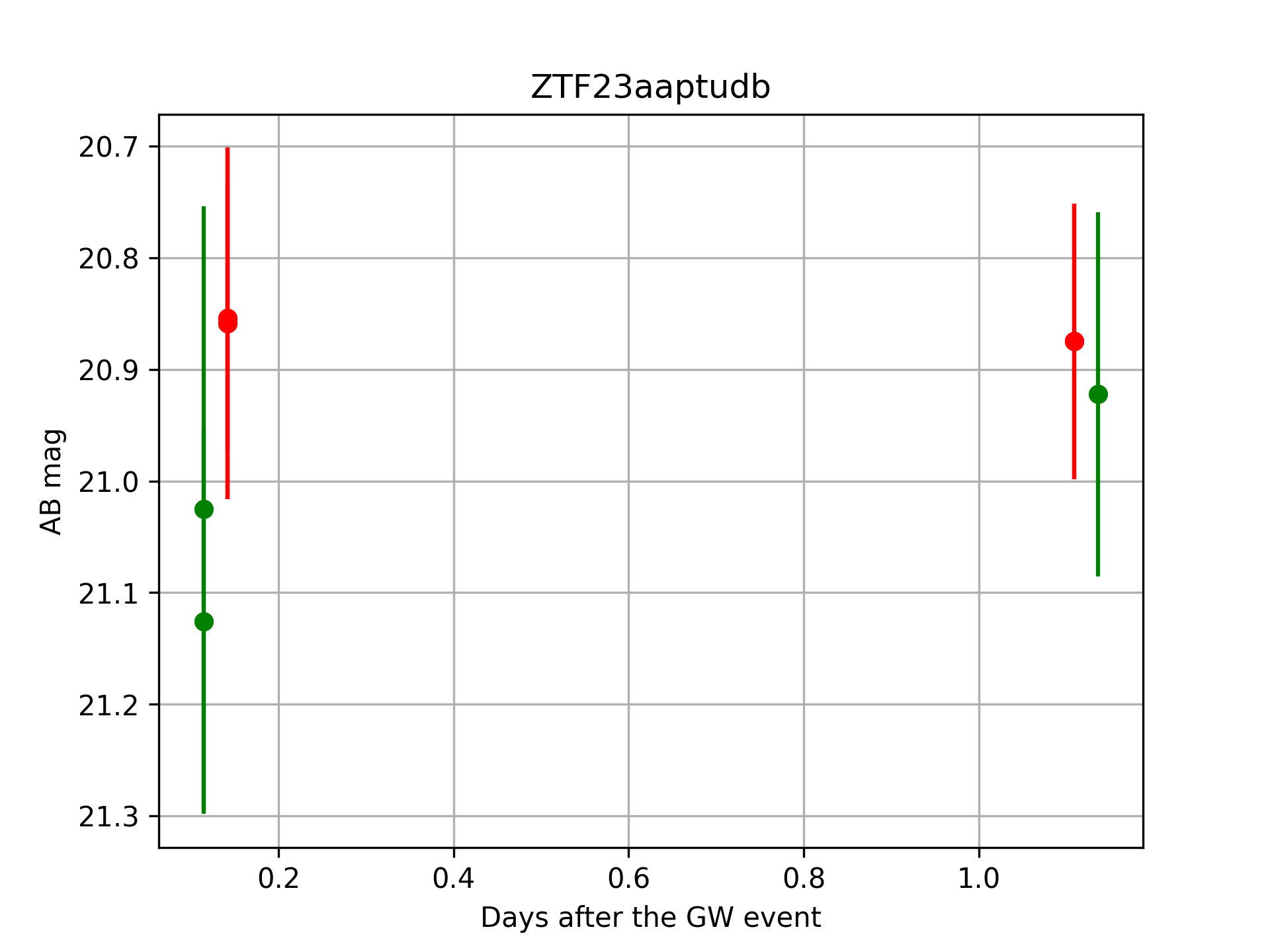}
\includegraphics[width=0.3\textwidth]{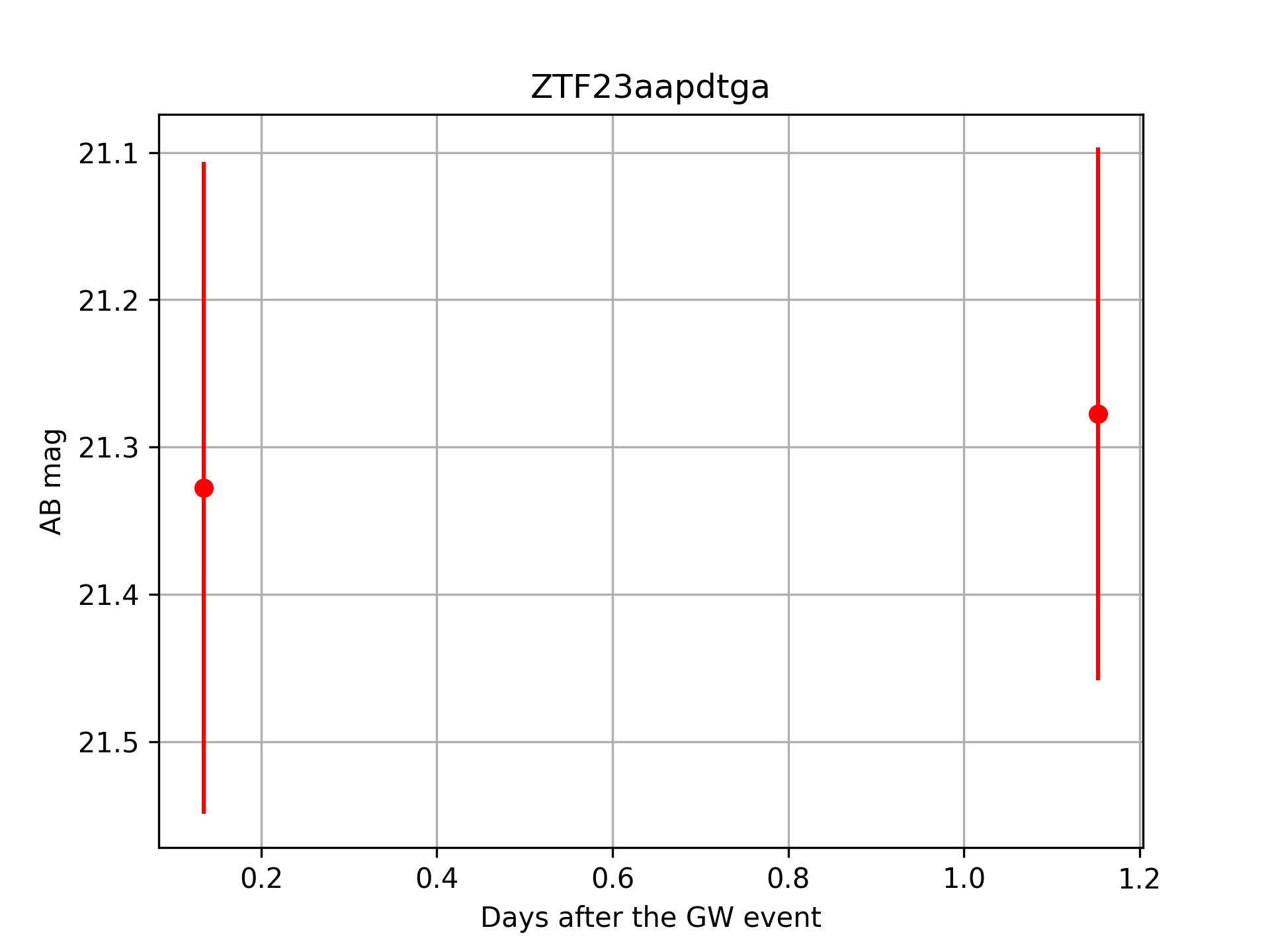}
\includegraphics[width=0.3\textwidth]{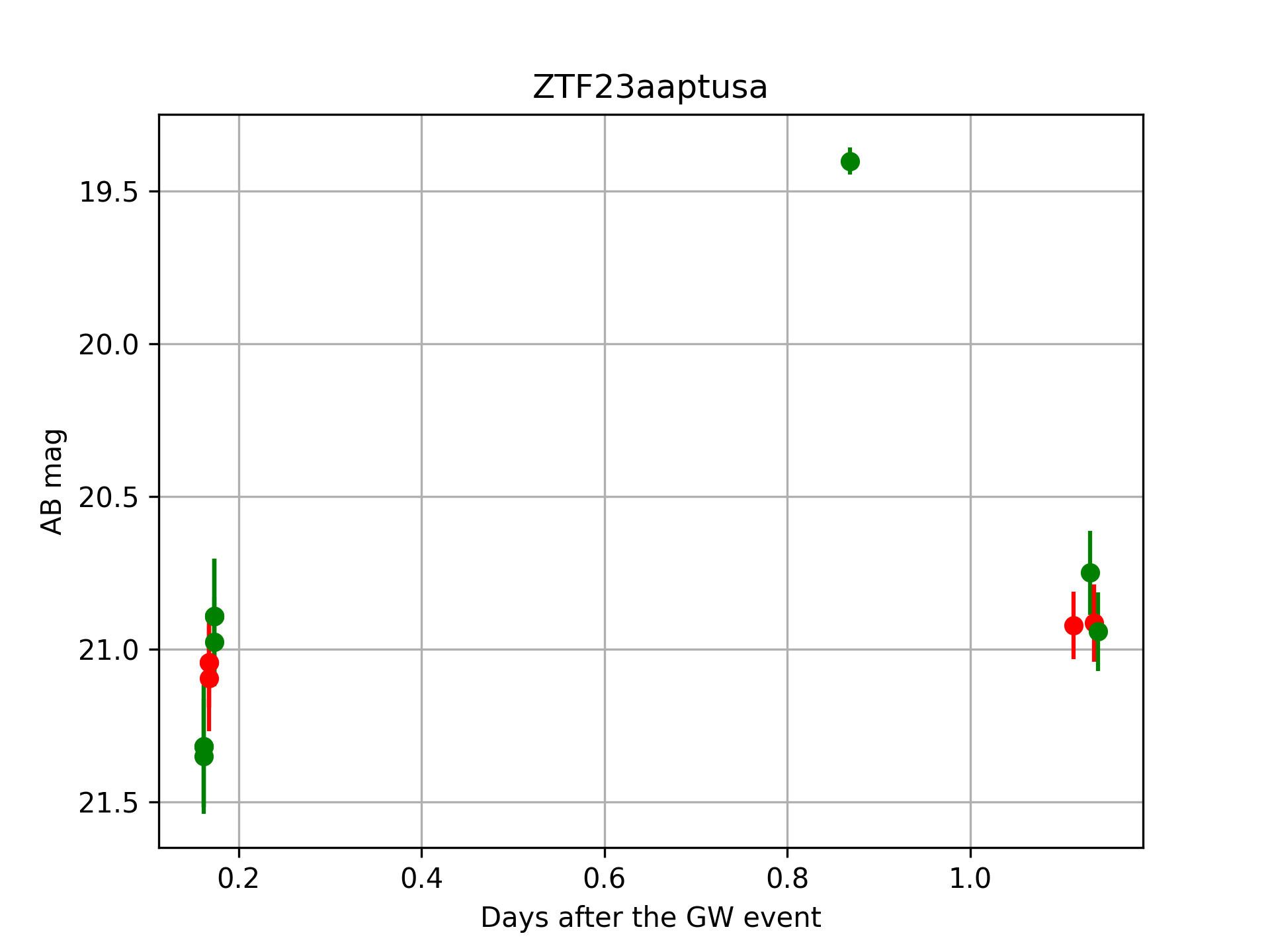}
\includegraphics[width=0.3\textwidth]{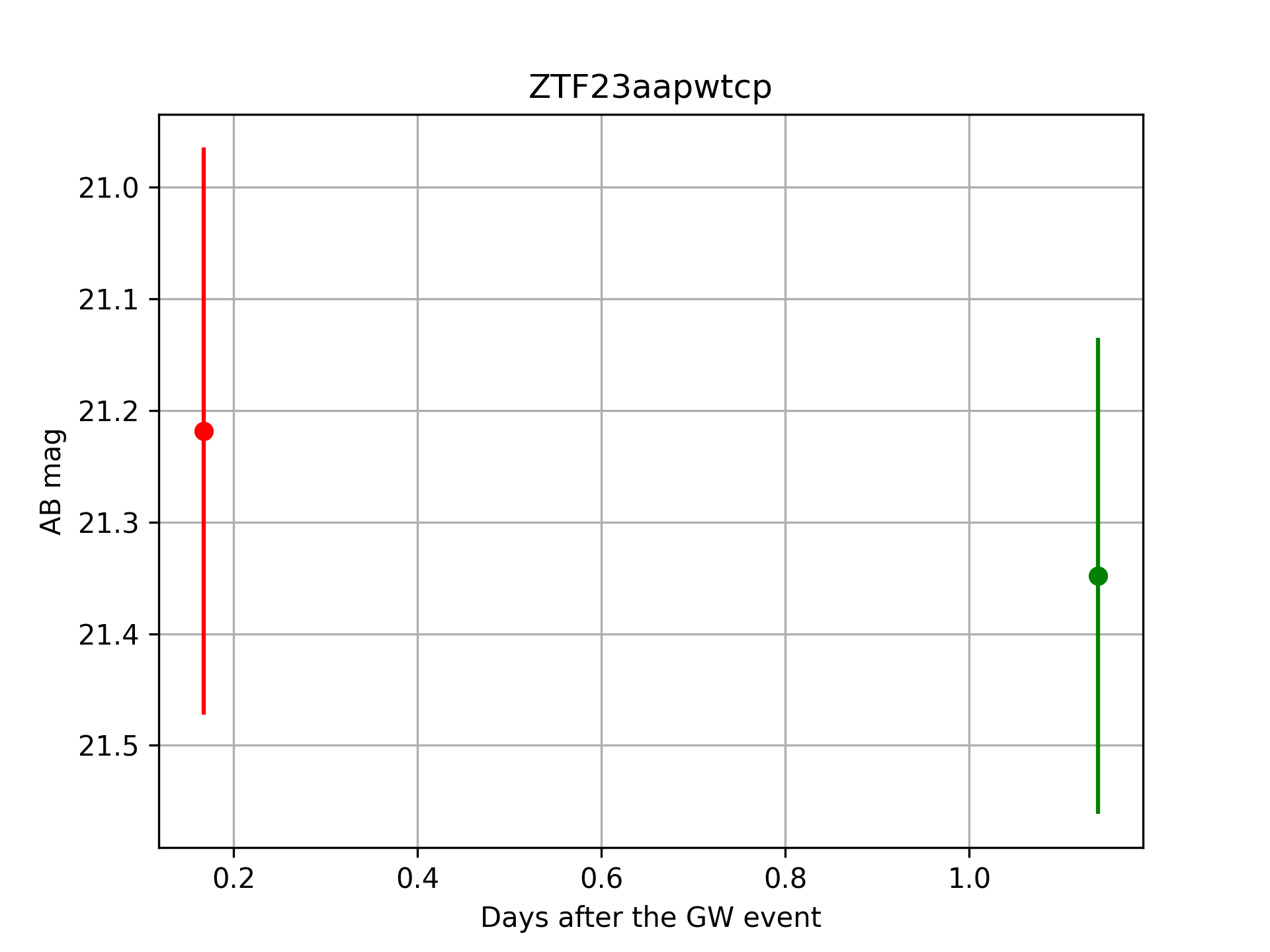}
    \caption{Light-curves for ZTF candidates found during O4a. These candidates correspond to the high-significant events S230529ay and S230627c. }
    \label{fig:LC_1}
\end{figure*}

\begin{figure*}[h]
    \centering
\includegraphics[width=0.3\textwidth]{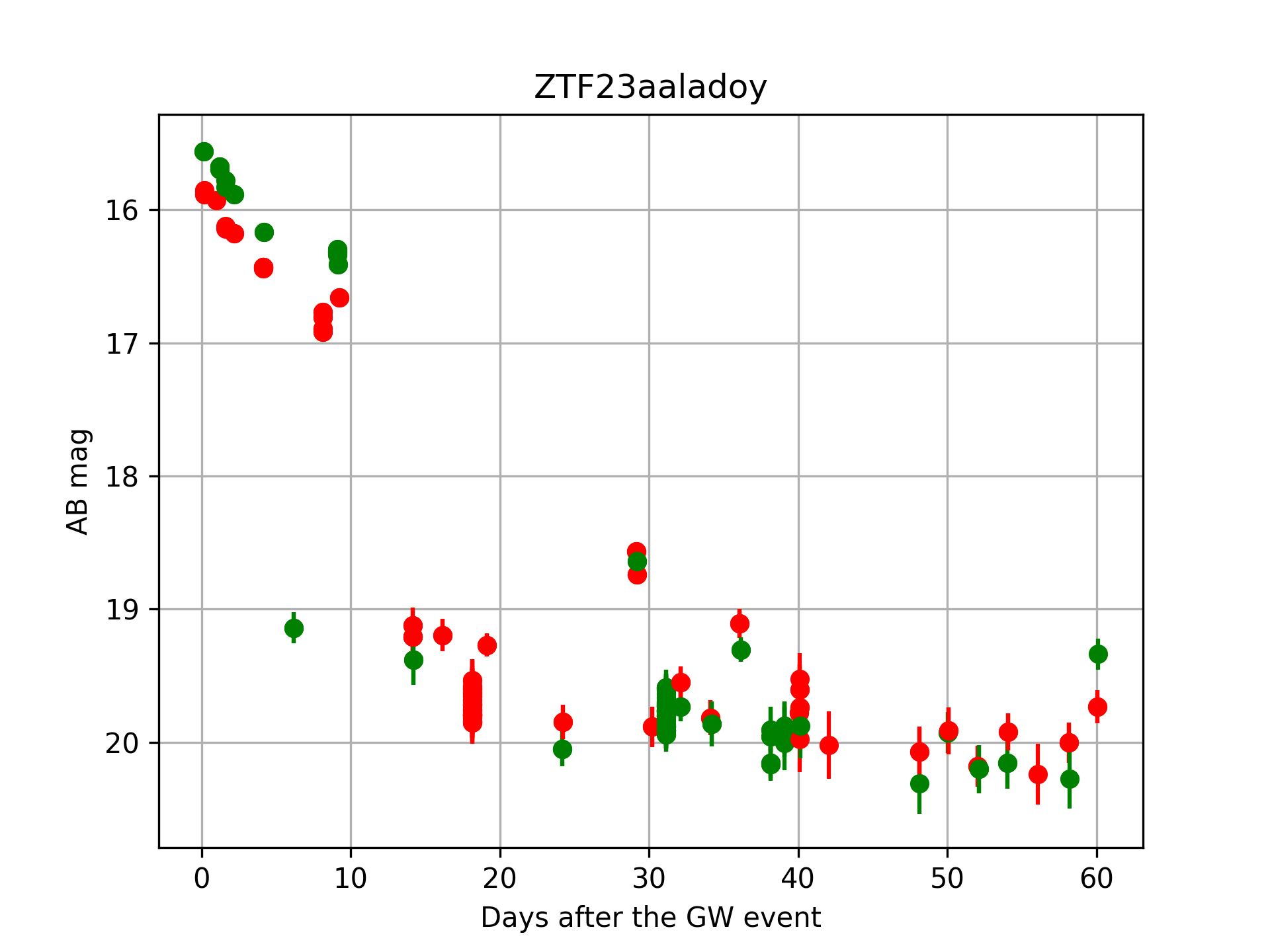}
\includegraphics[width=0.3\textwidth]{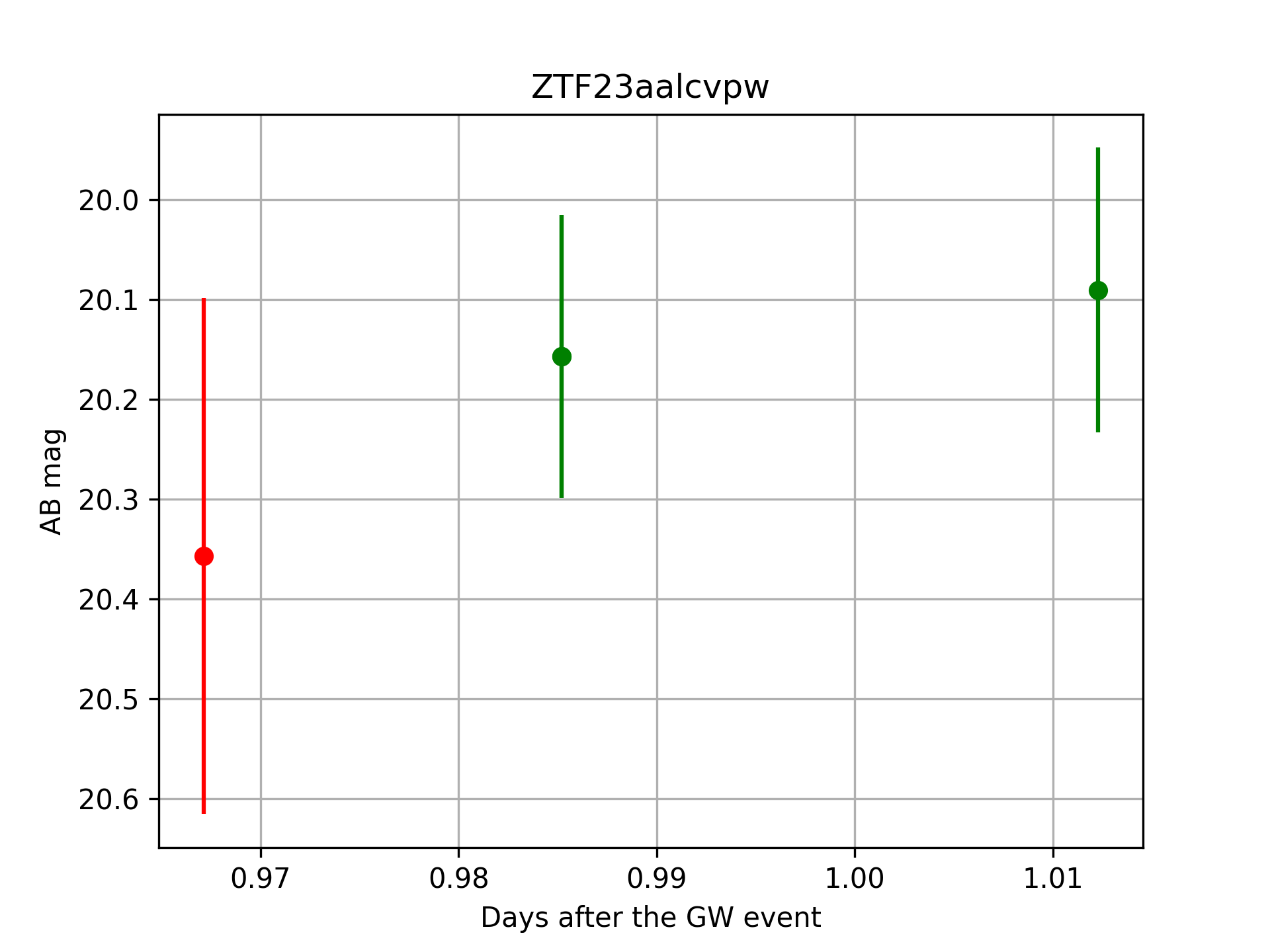}
\includegraphics[width=0.3\textwidth]{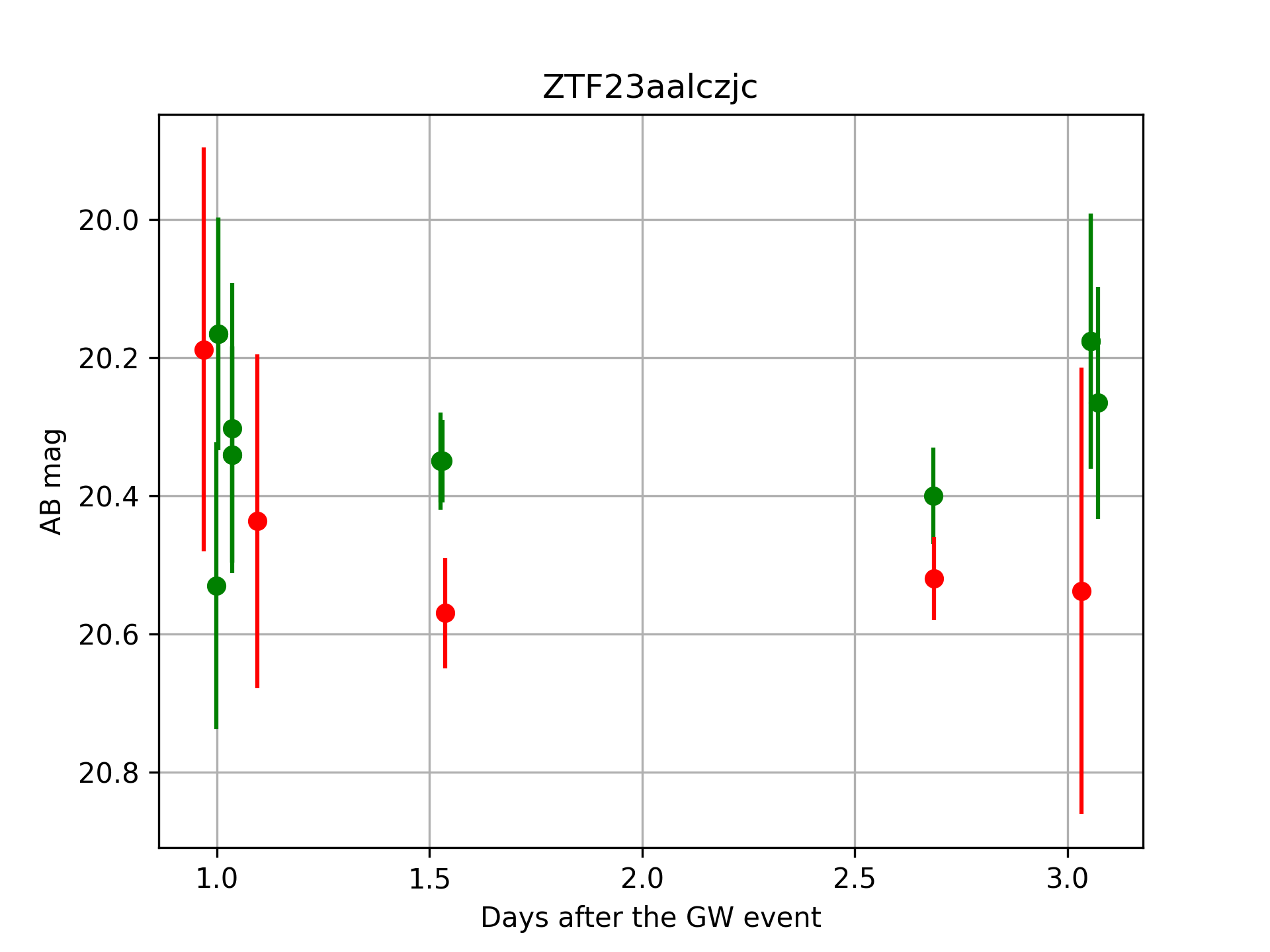}
\includegraphics[width=0.3\textwidth]{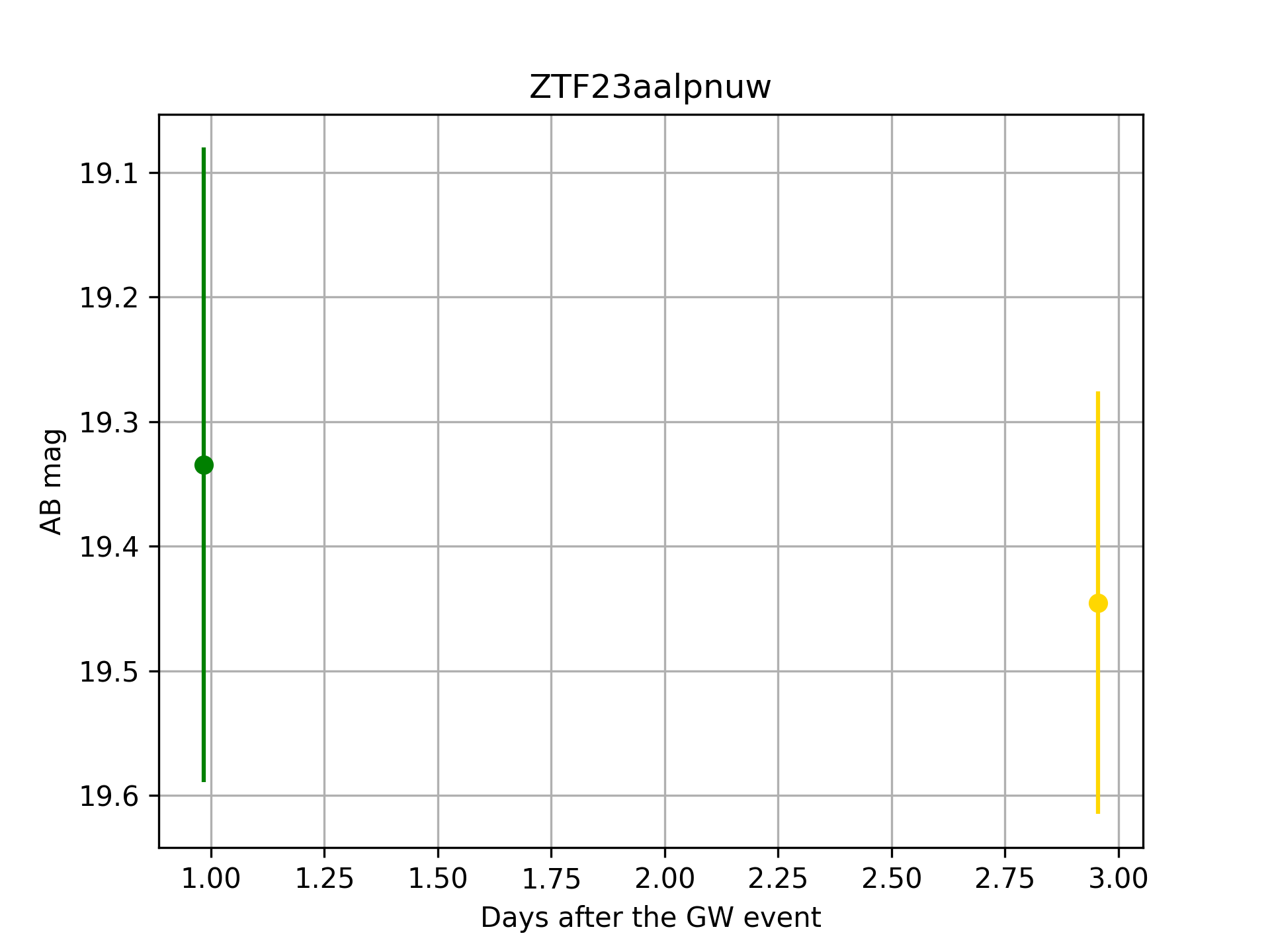}
\includegraphics[width=0.3\textwidth]{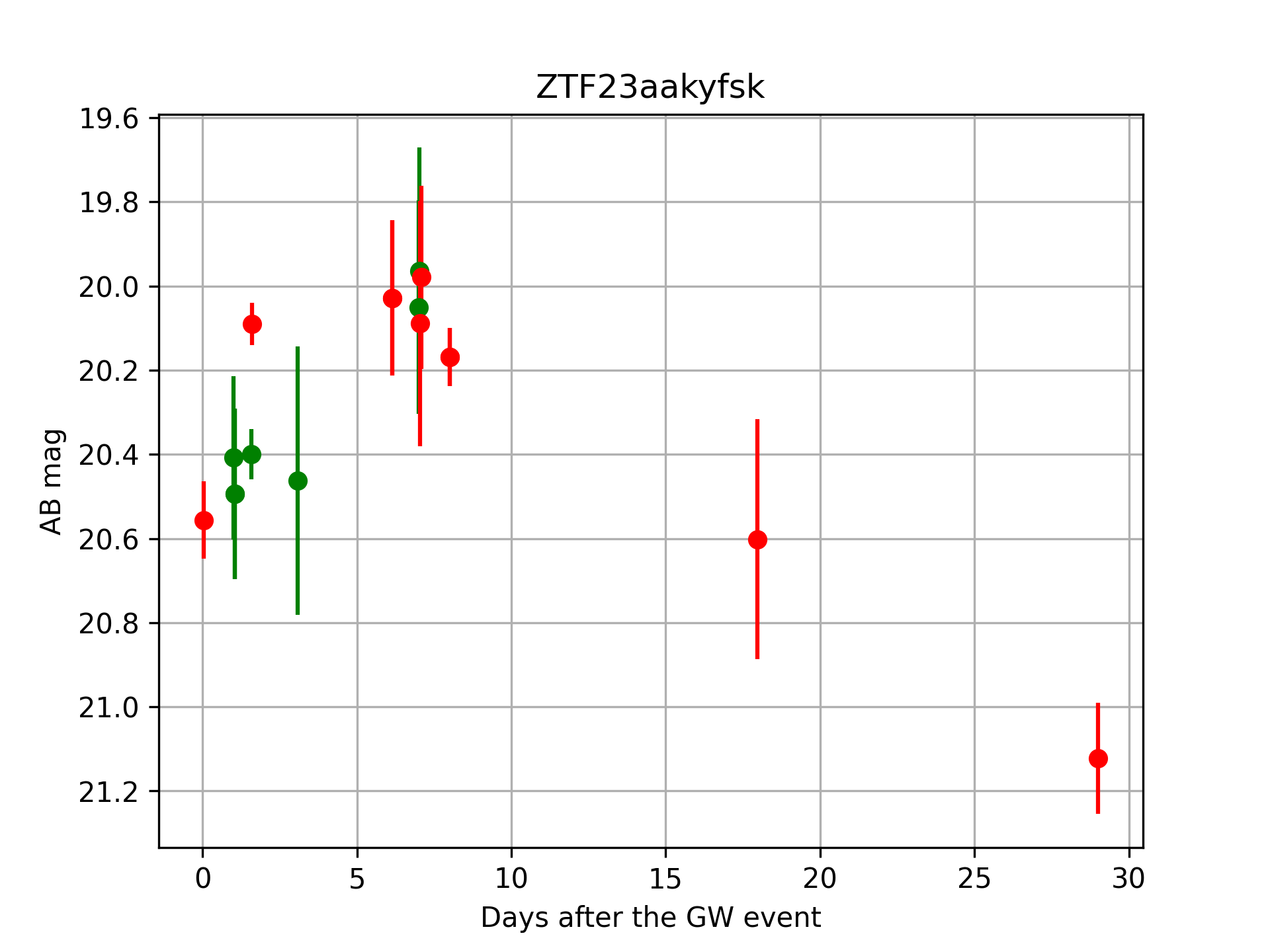}
\includegraphics[width=0.3\textwidth]{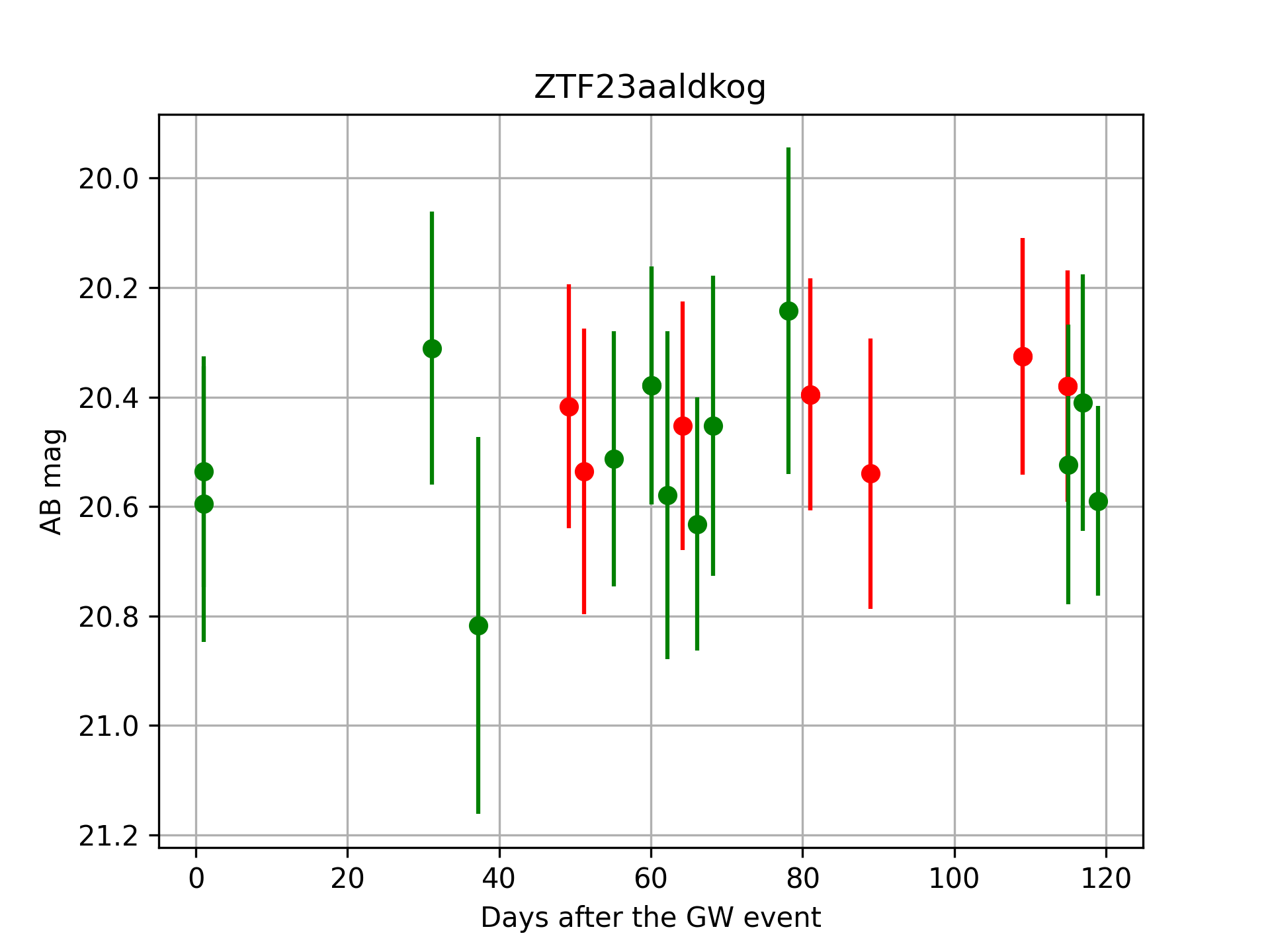}
\includegraphics[width=0.3\textwidth]{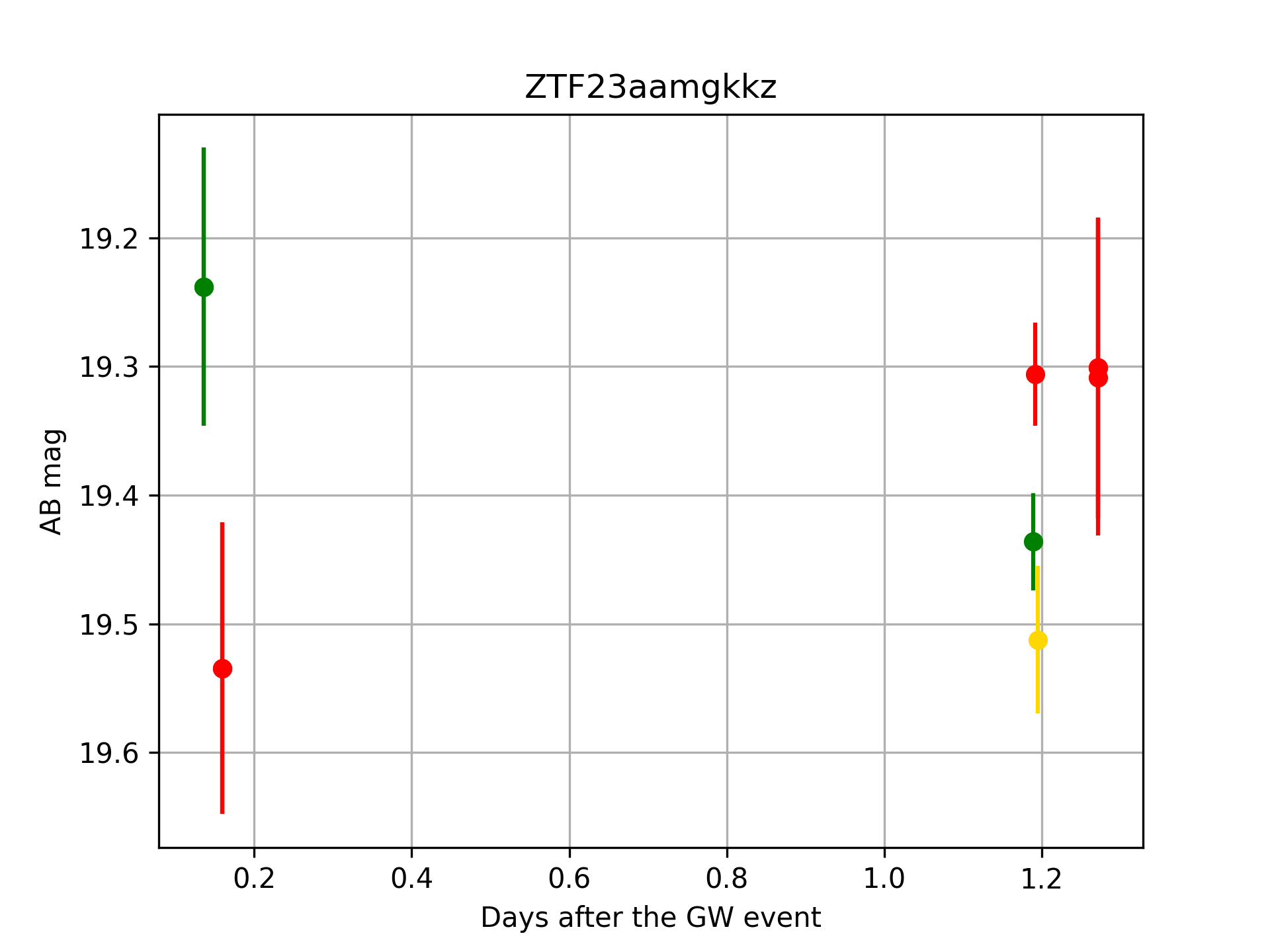}
\includegraphics[width=0.3\textwidth]{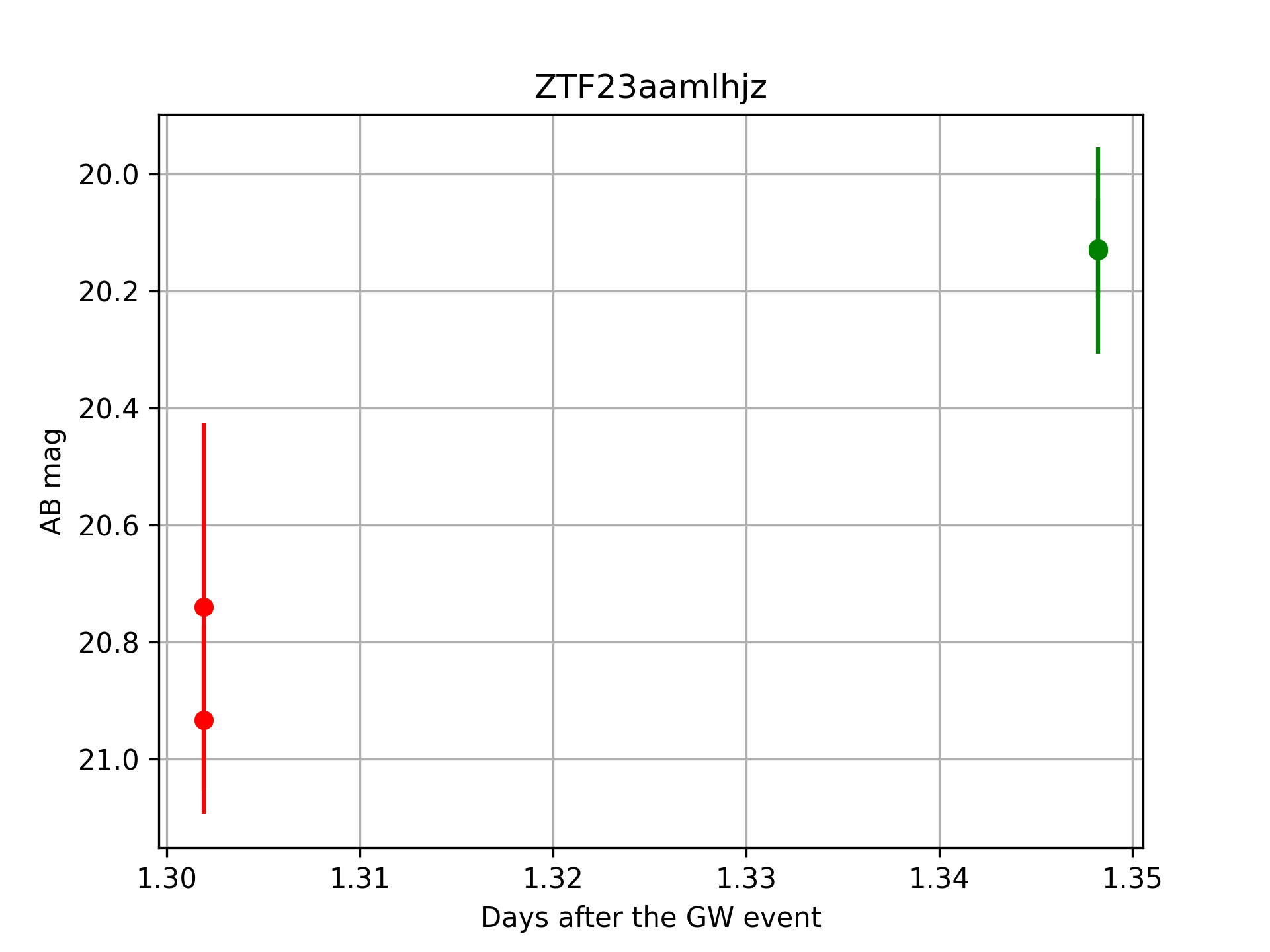}
\includegraphics[width=0.3\textwidth]{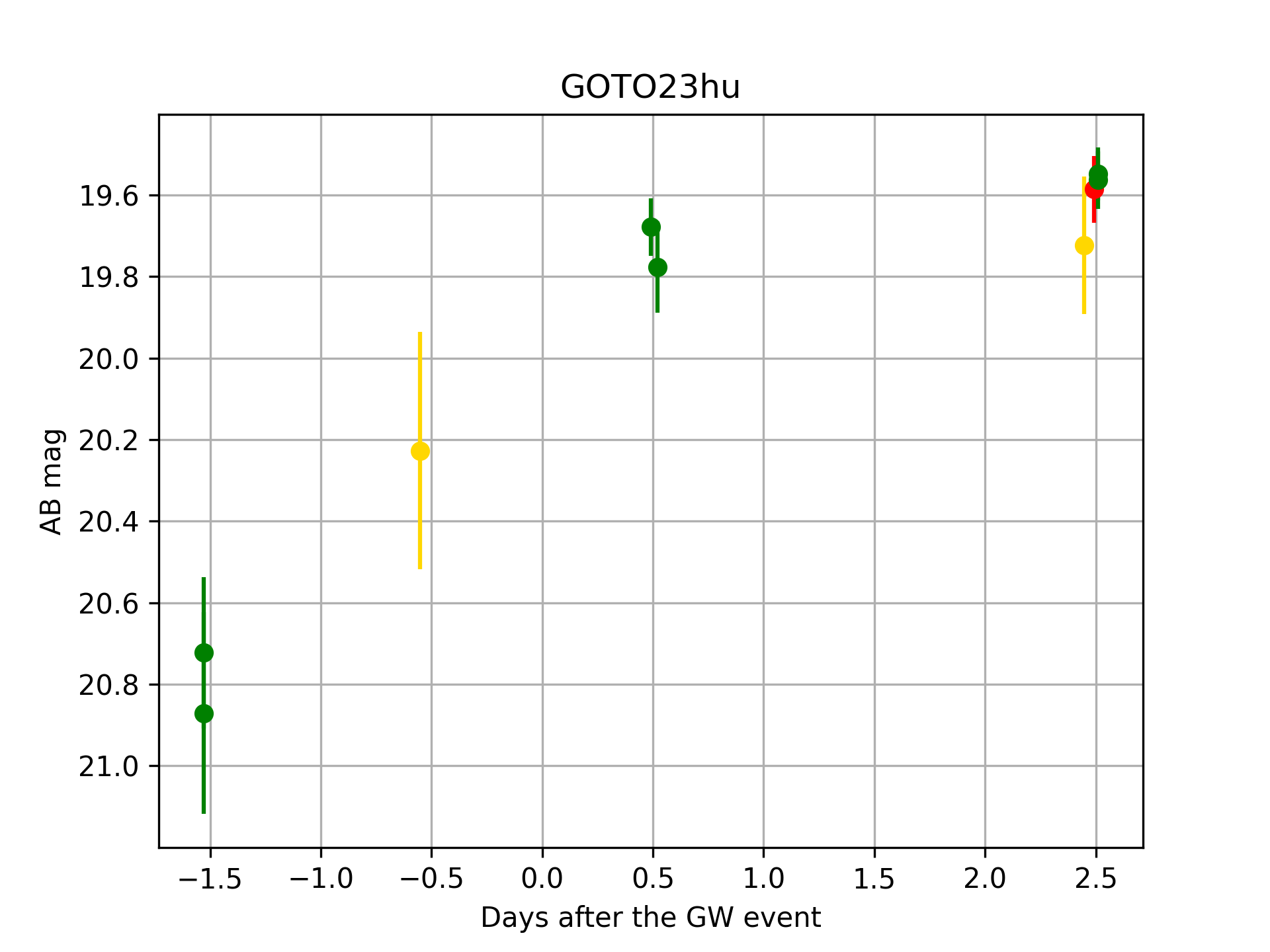}
\includegraphics[width=0.3\textwidth]{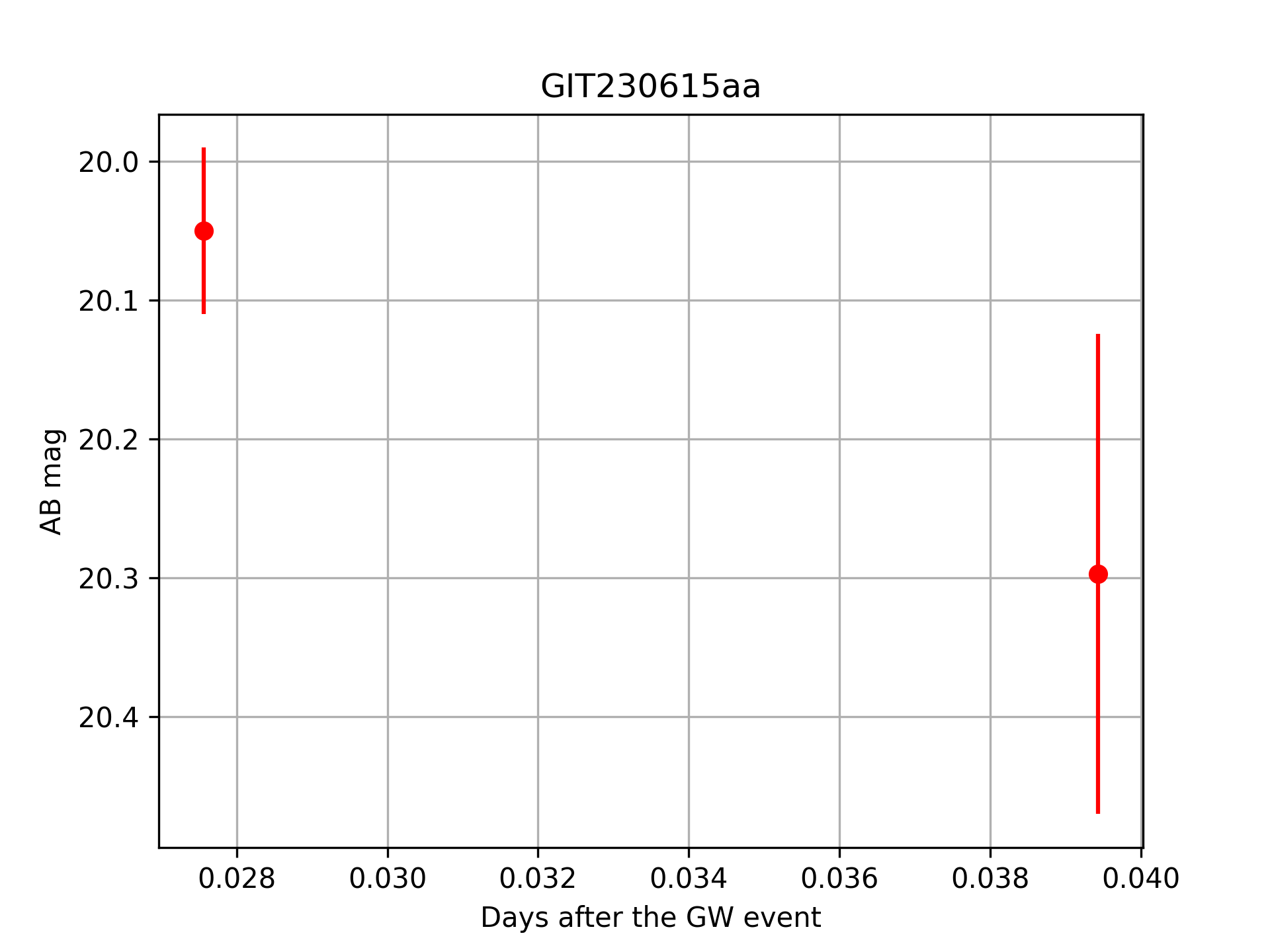}
\includegraphics[width=0.3\textwidth]{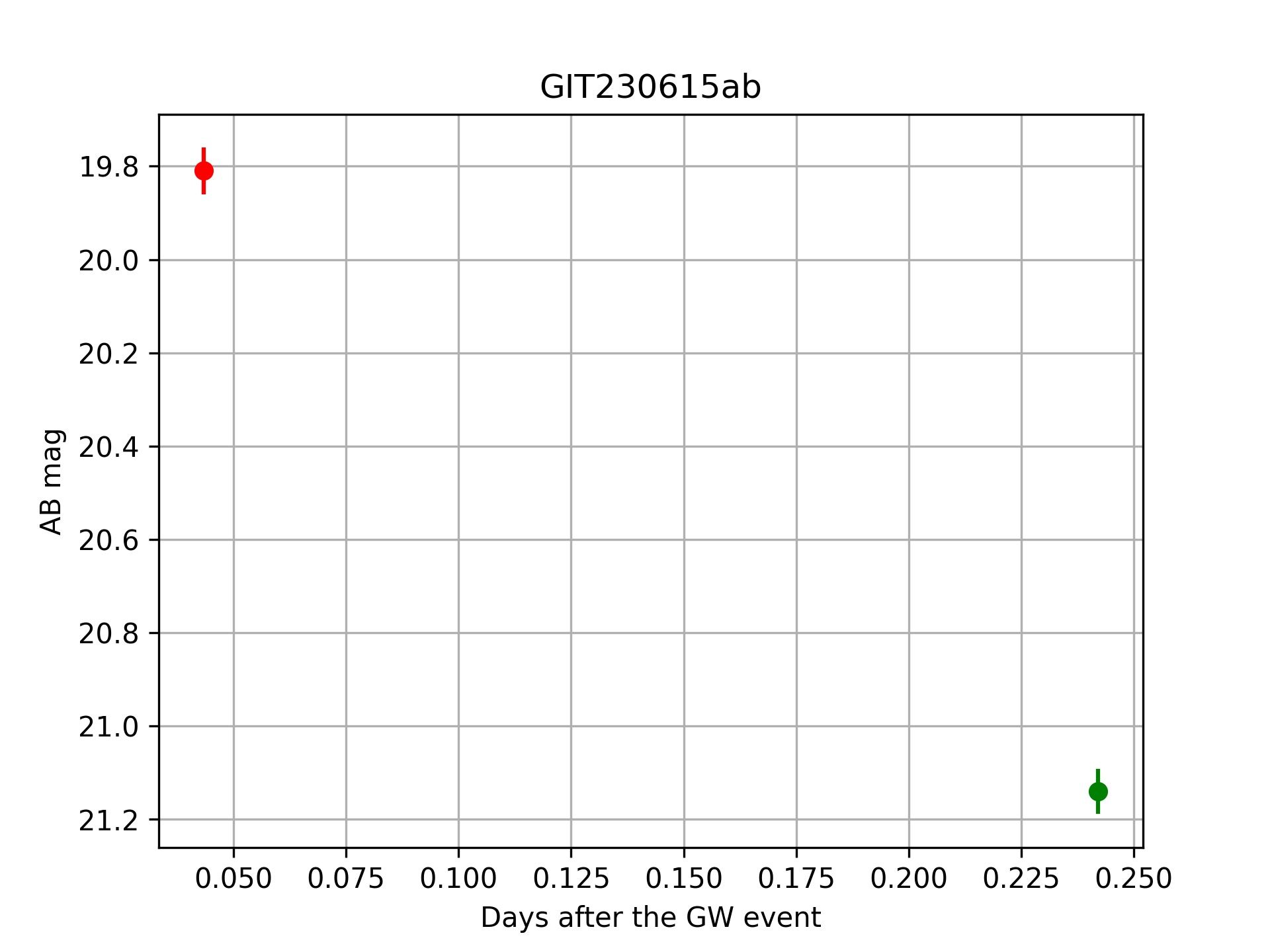}
\includegraphics[width=0.3\textwidth]{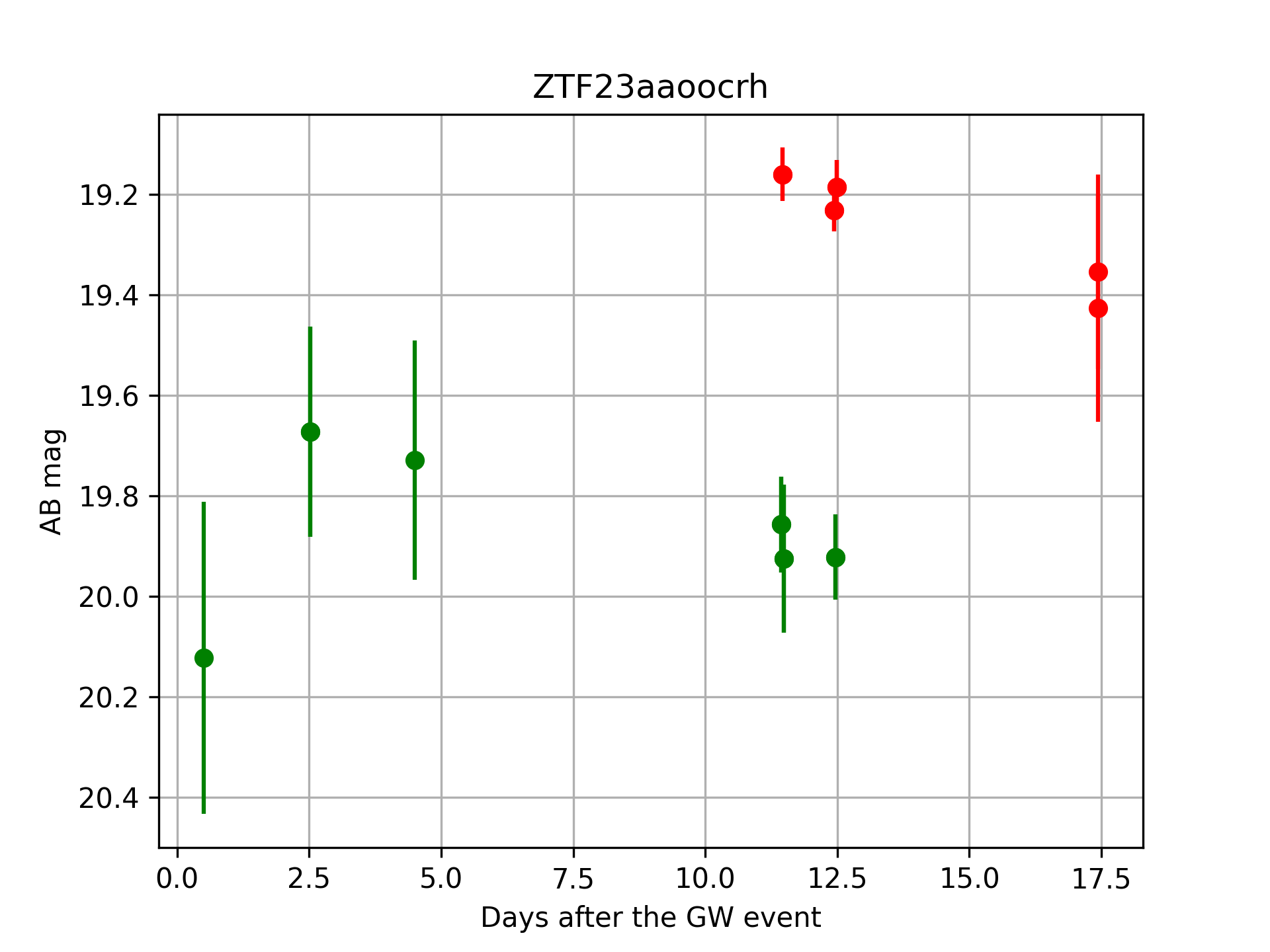}
\includegraphics[width=0.3\textwidth]{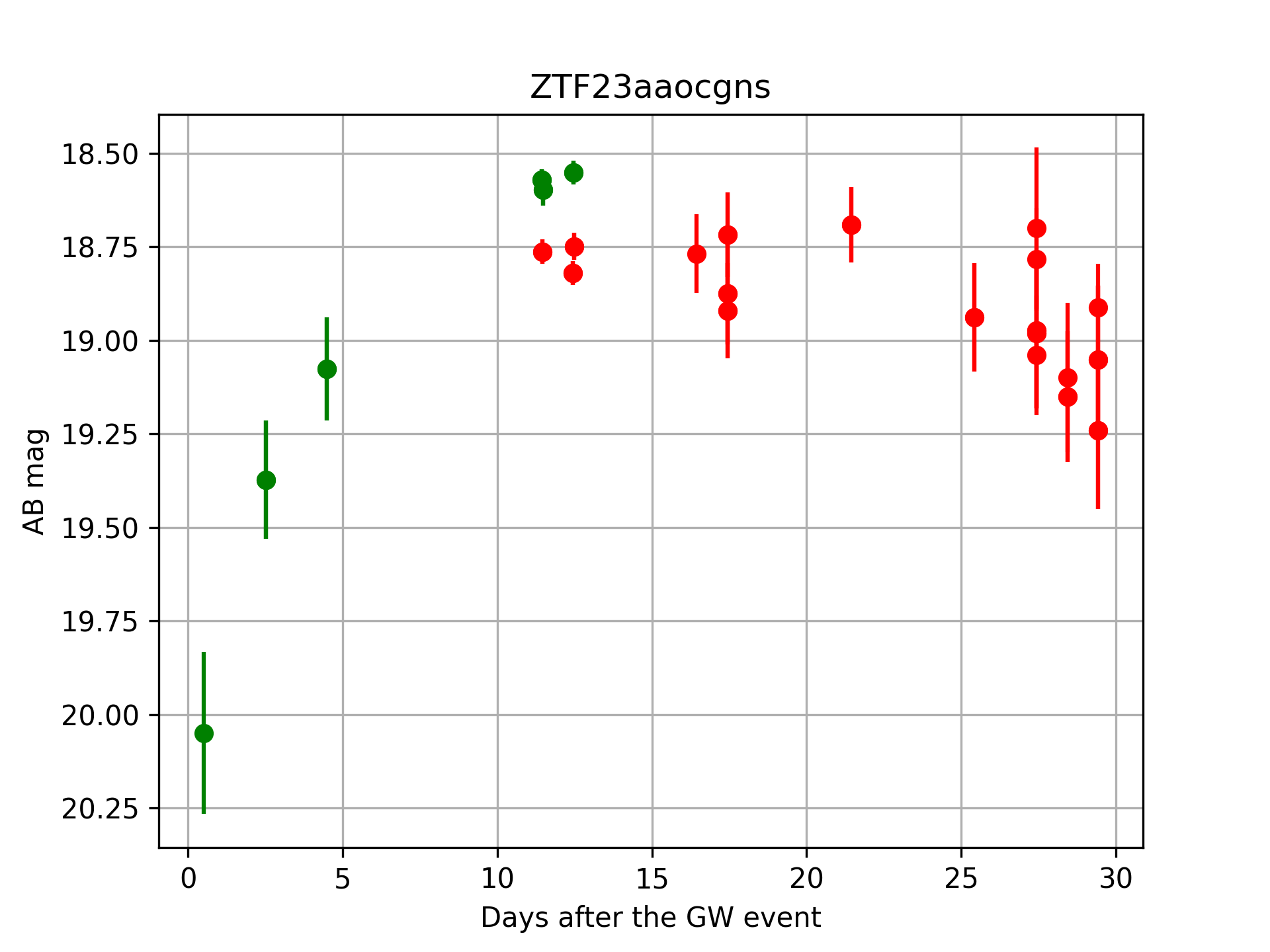}
\includegraphics[width=0.3\textwidth]{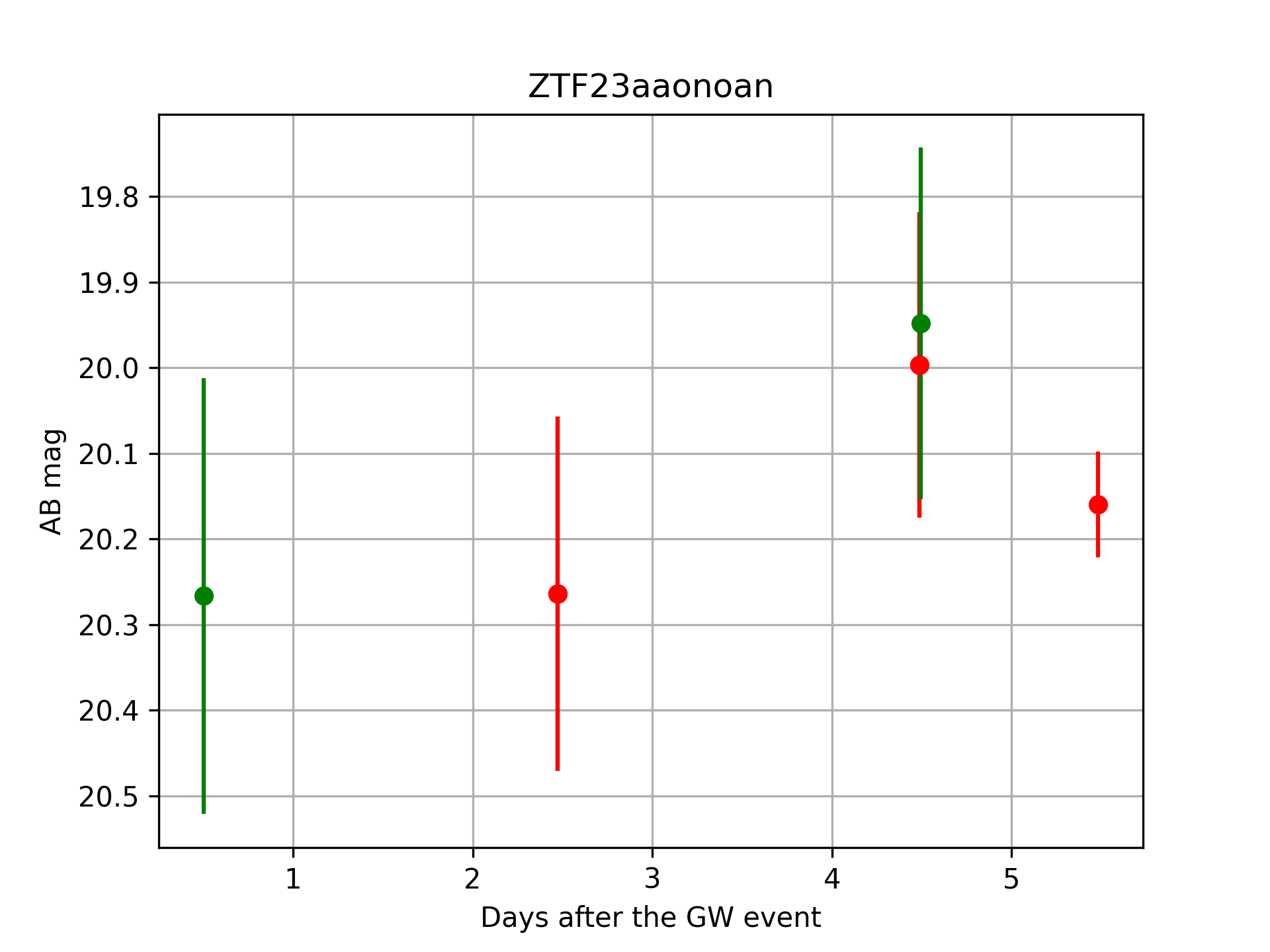}
\includegraphics[width=0.3\textwidth]{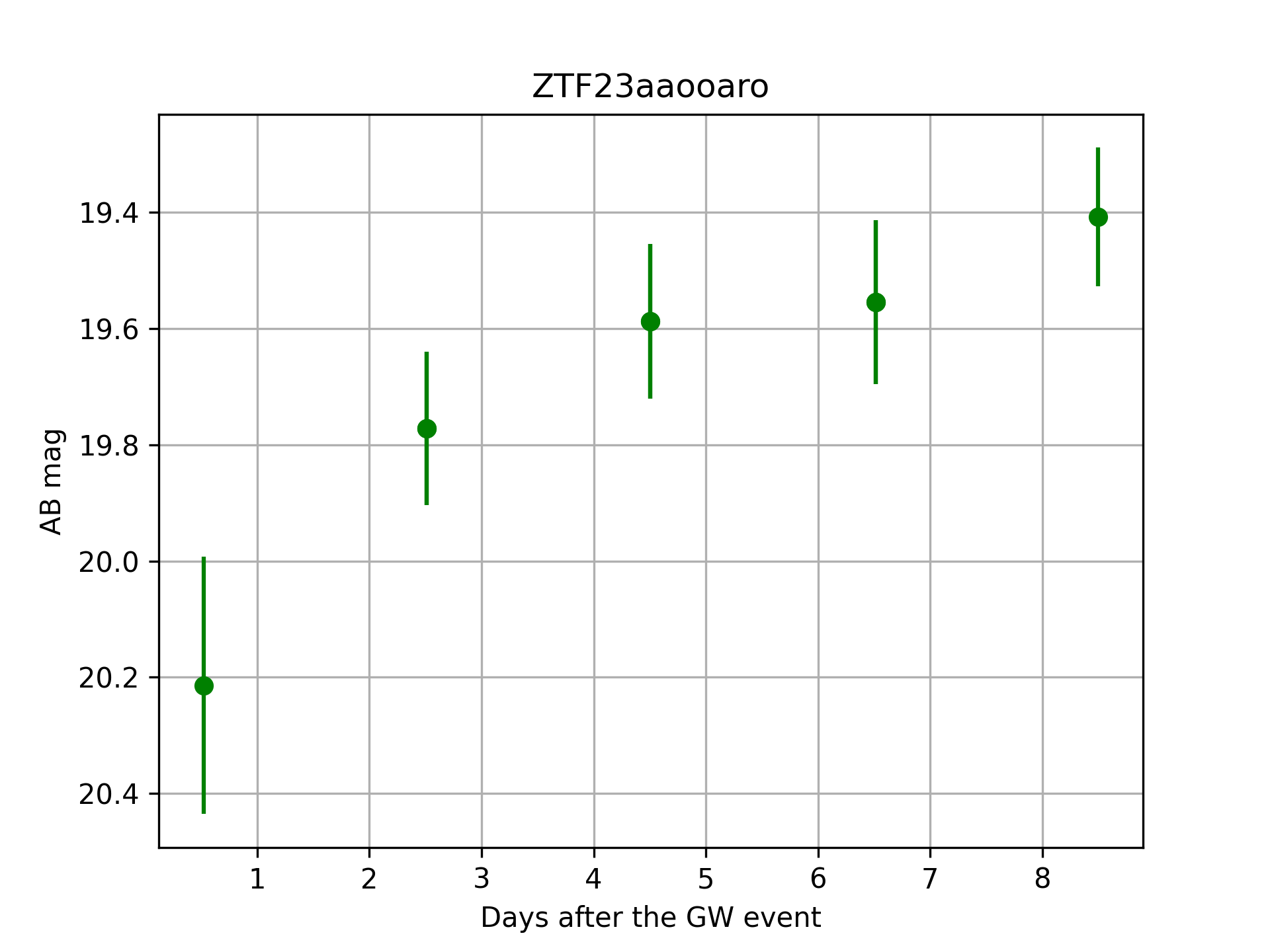}
    \caption{Light-curves for ZTF candidates found during O4a. These candidates correspond to the events S230521k and S230528a. }
    \label{fig:LC_2}
\end{figure*}

\begin{figure*}[h]
    \centering
\includegraphics[width=0.3\textwidth]{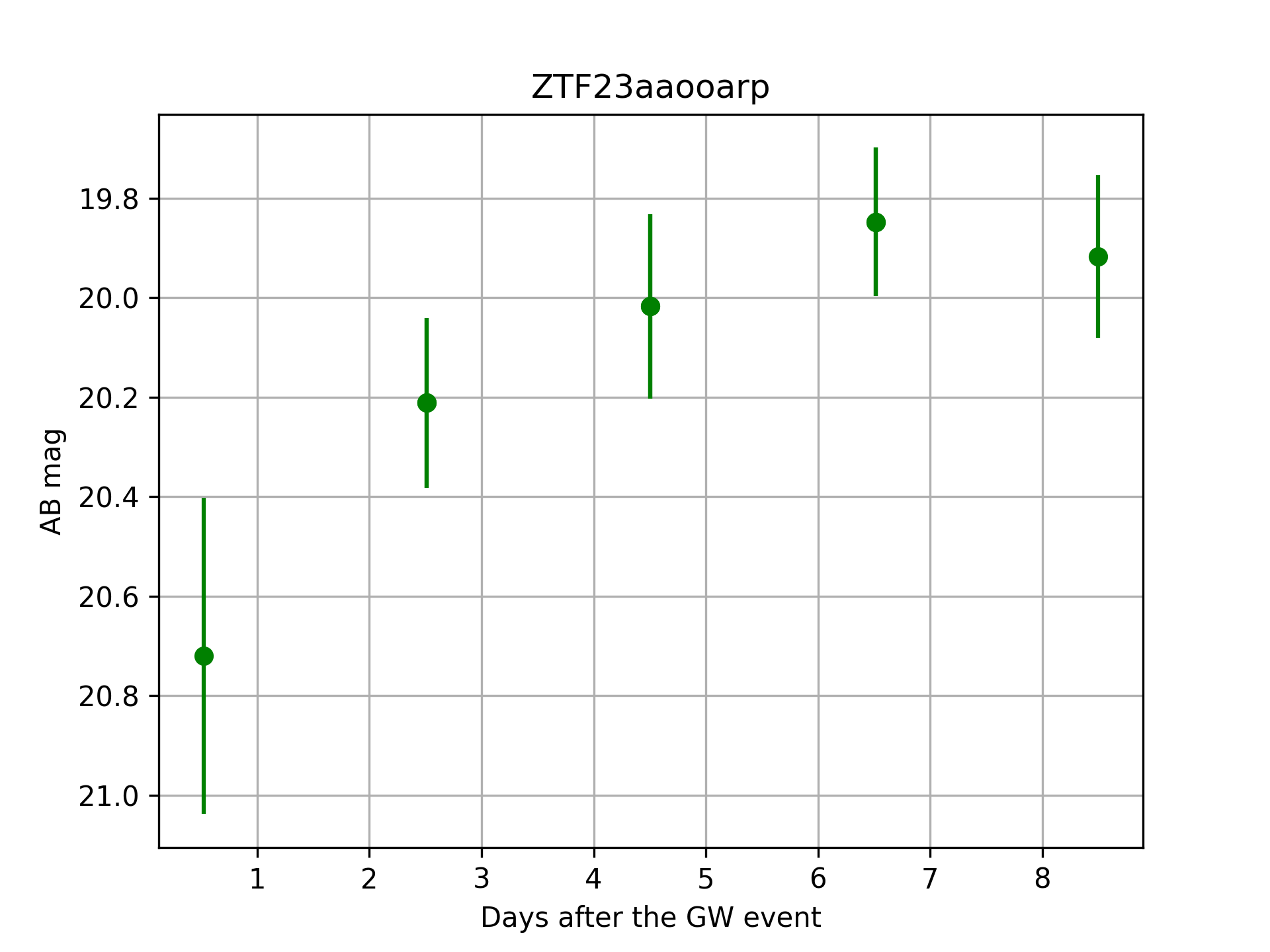}
\includegraphics[width=0.3\textwidth]{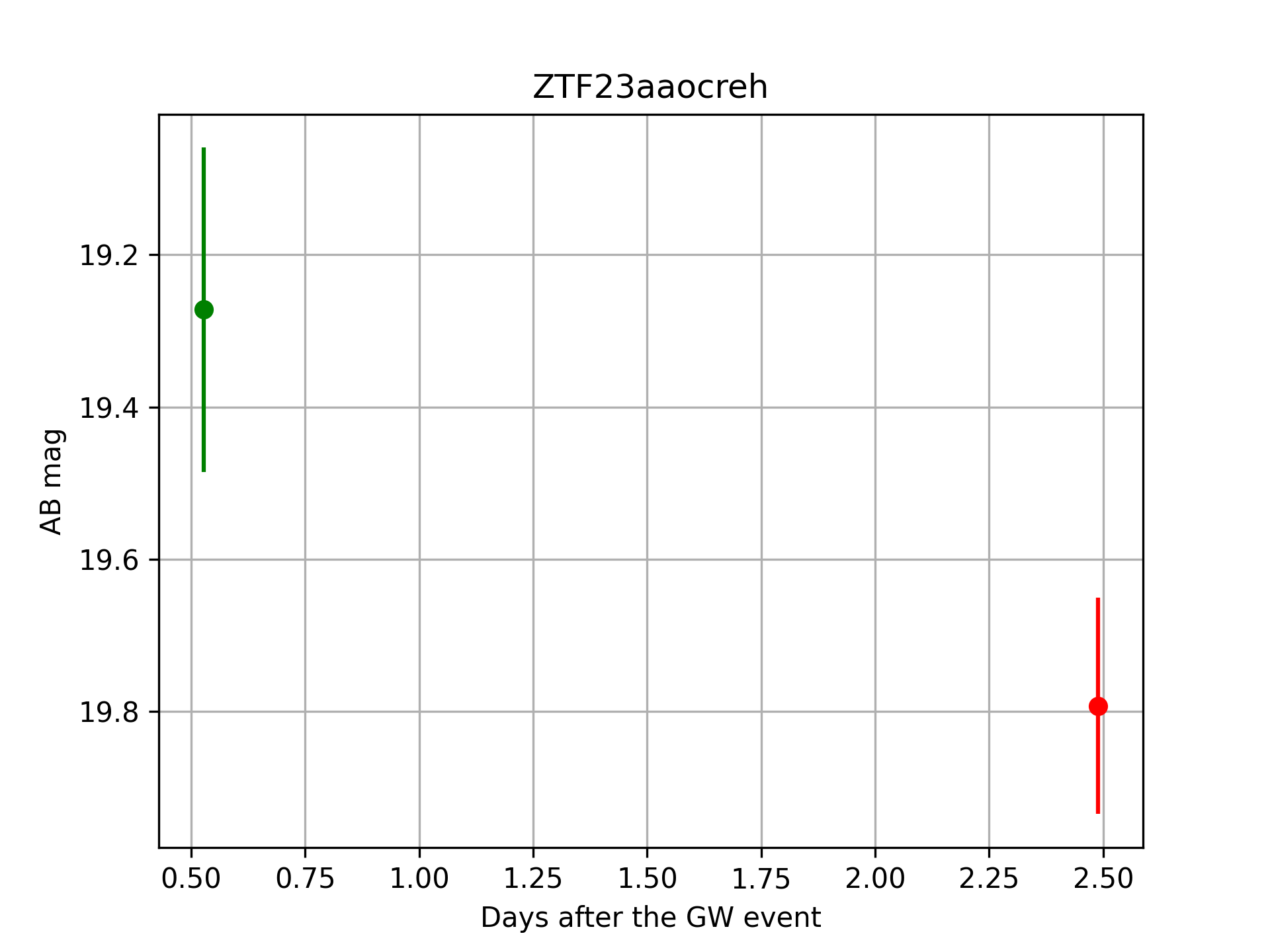}
\includegraphics[width=0.3\textwidth]{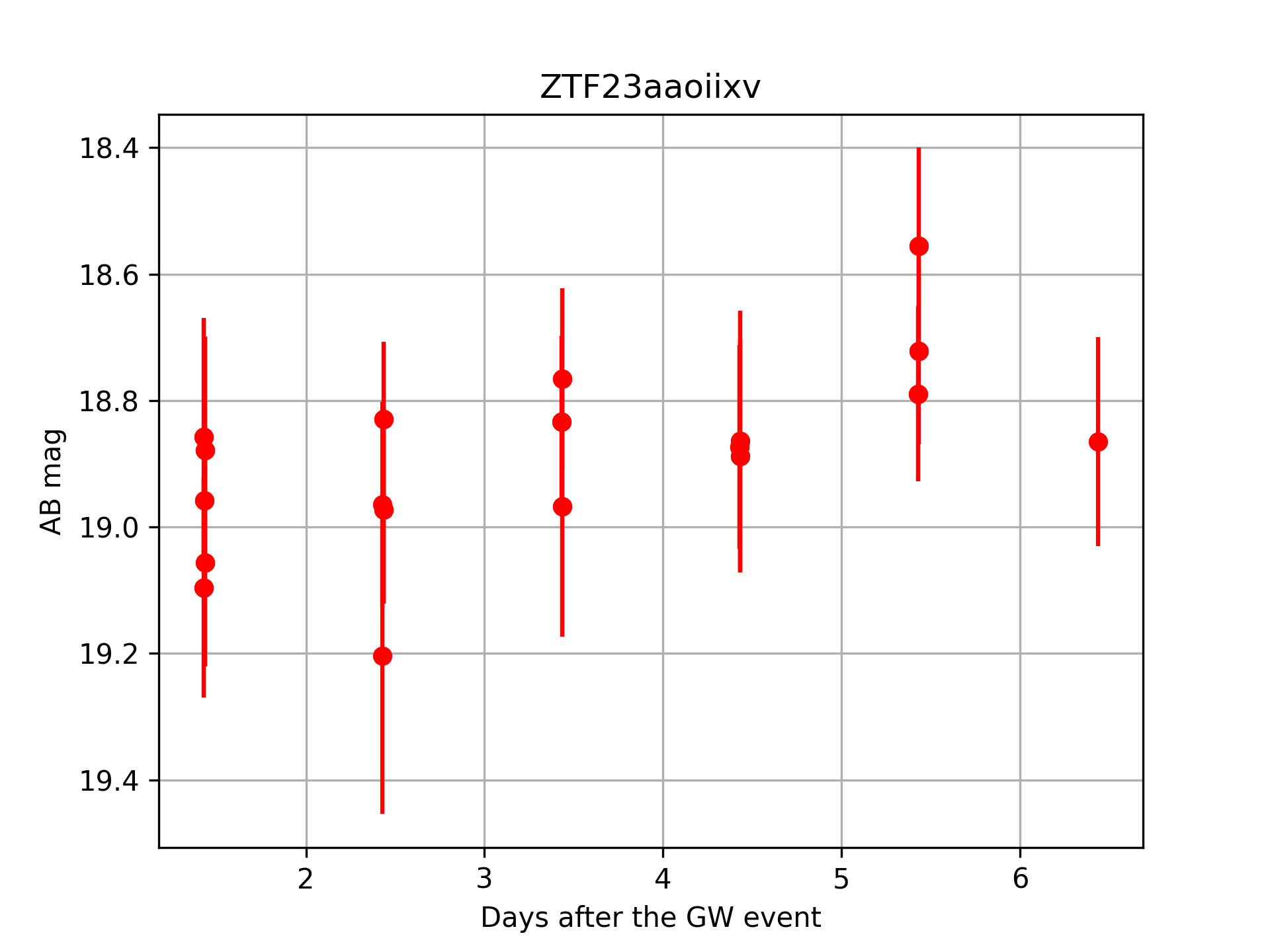}
\includegraphics[width=0.3\textwidth]{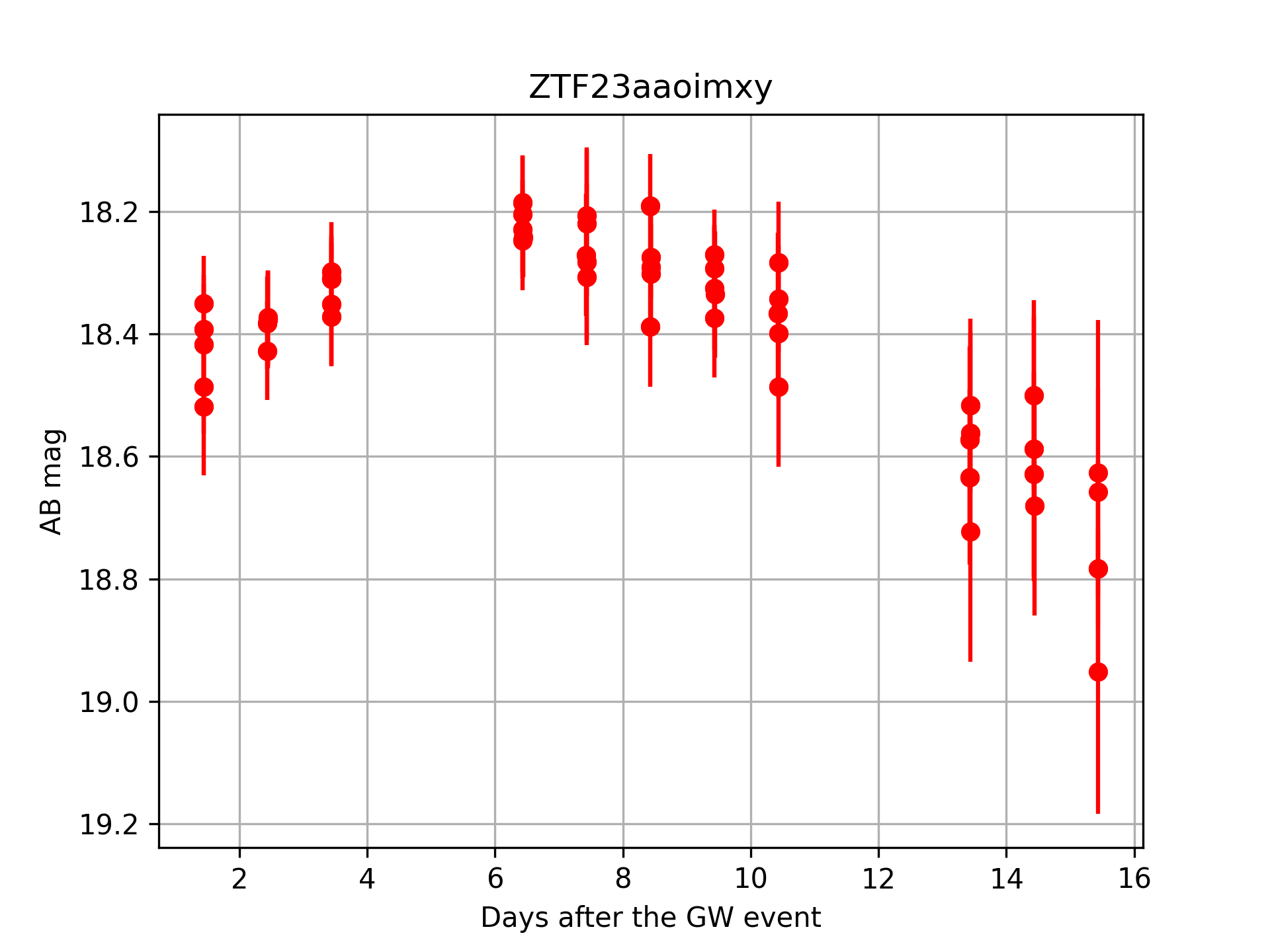}
\includegraphics[width=0.3\textwidth]{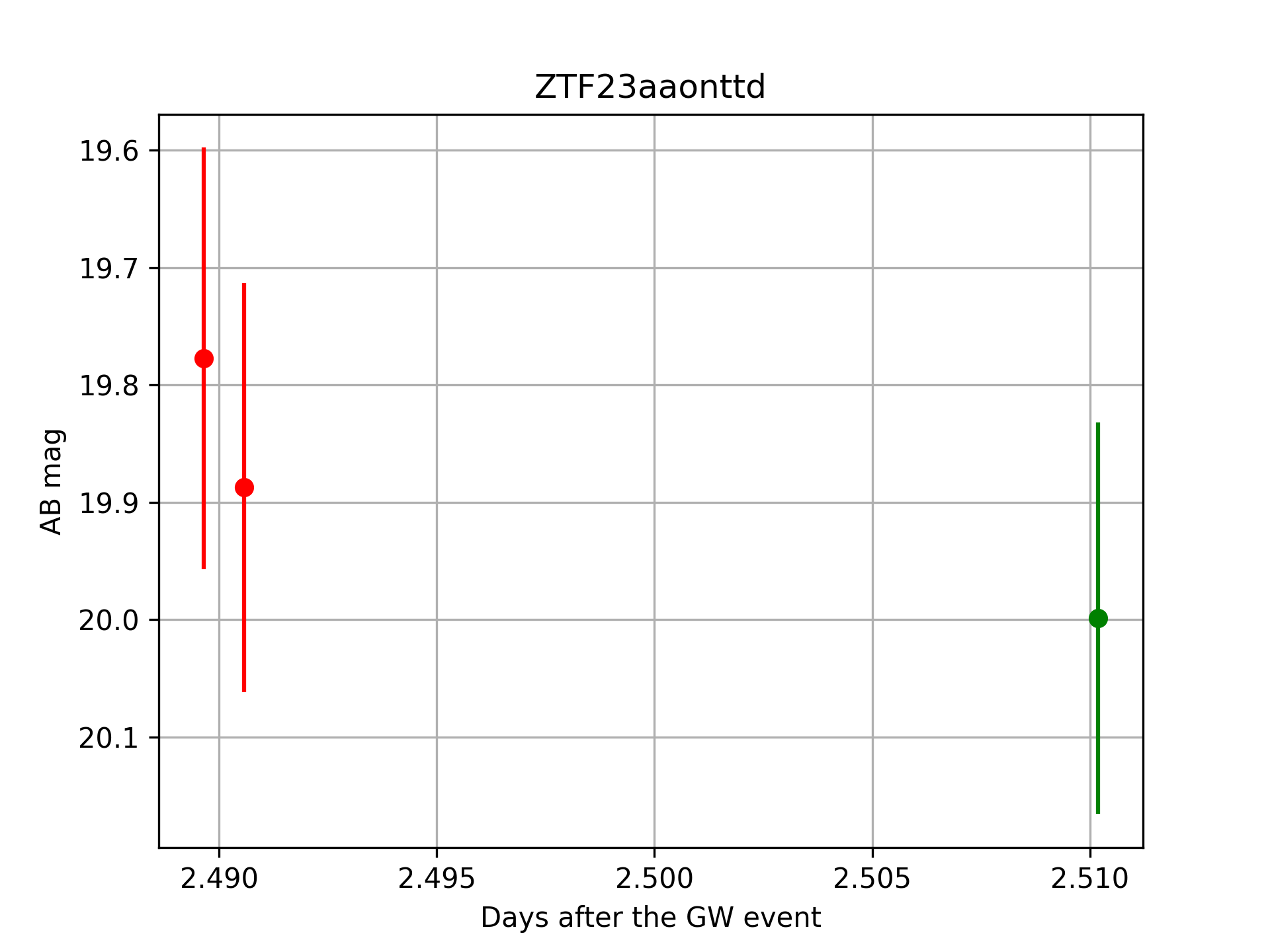}
\includegraphics[width=0.3\textwidth]{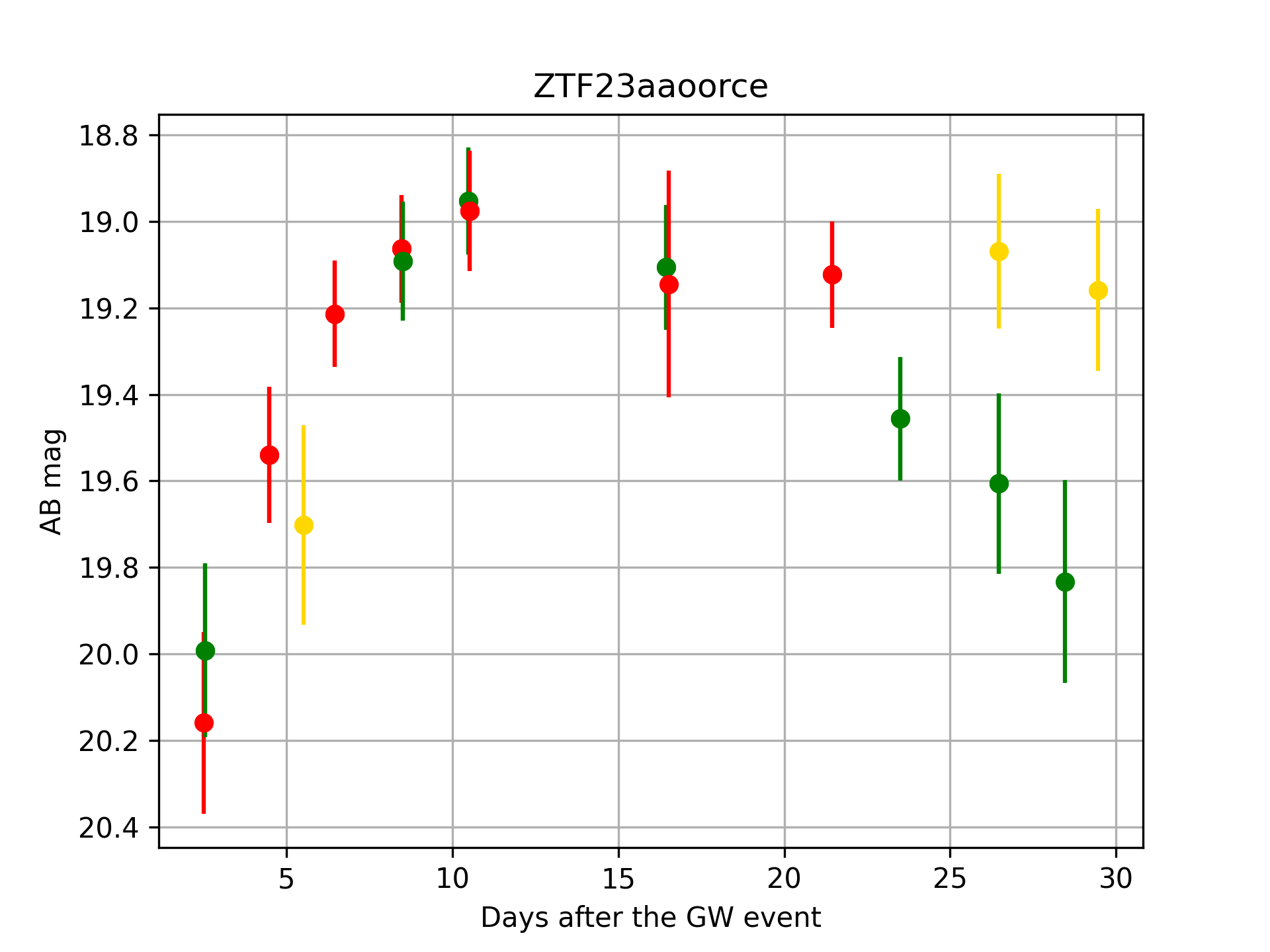}
\includegraphics[width=0.3\textwidth]{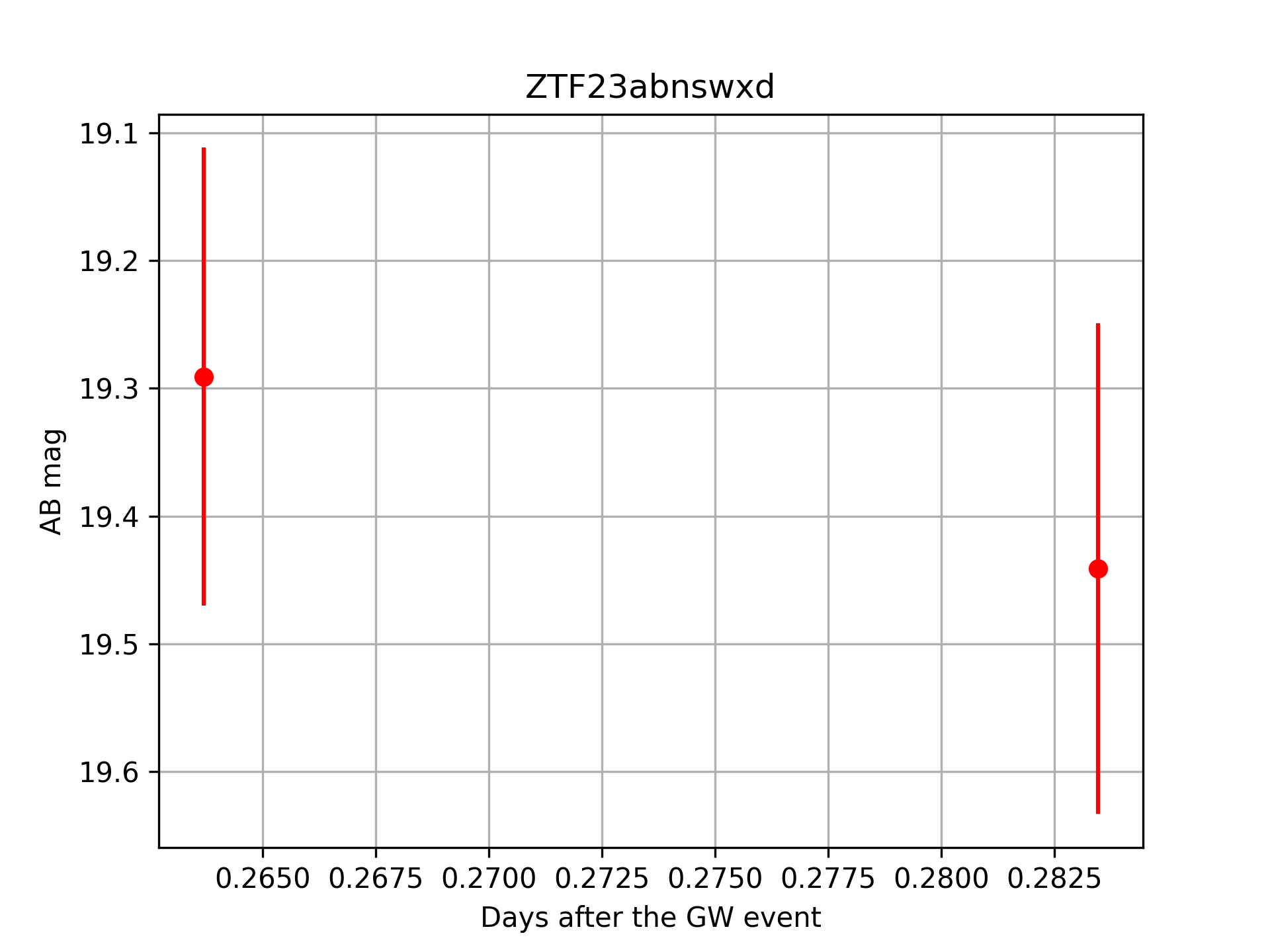}
\includegraphics[width=0.3\textwidth]{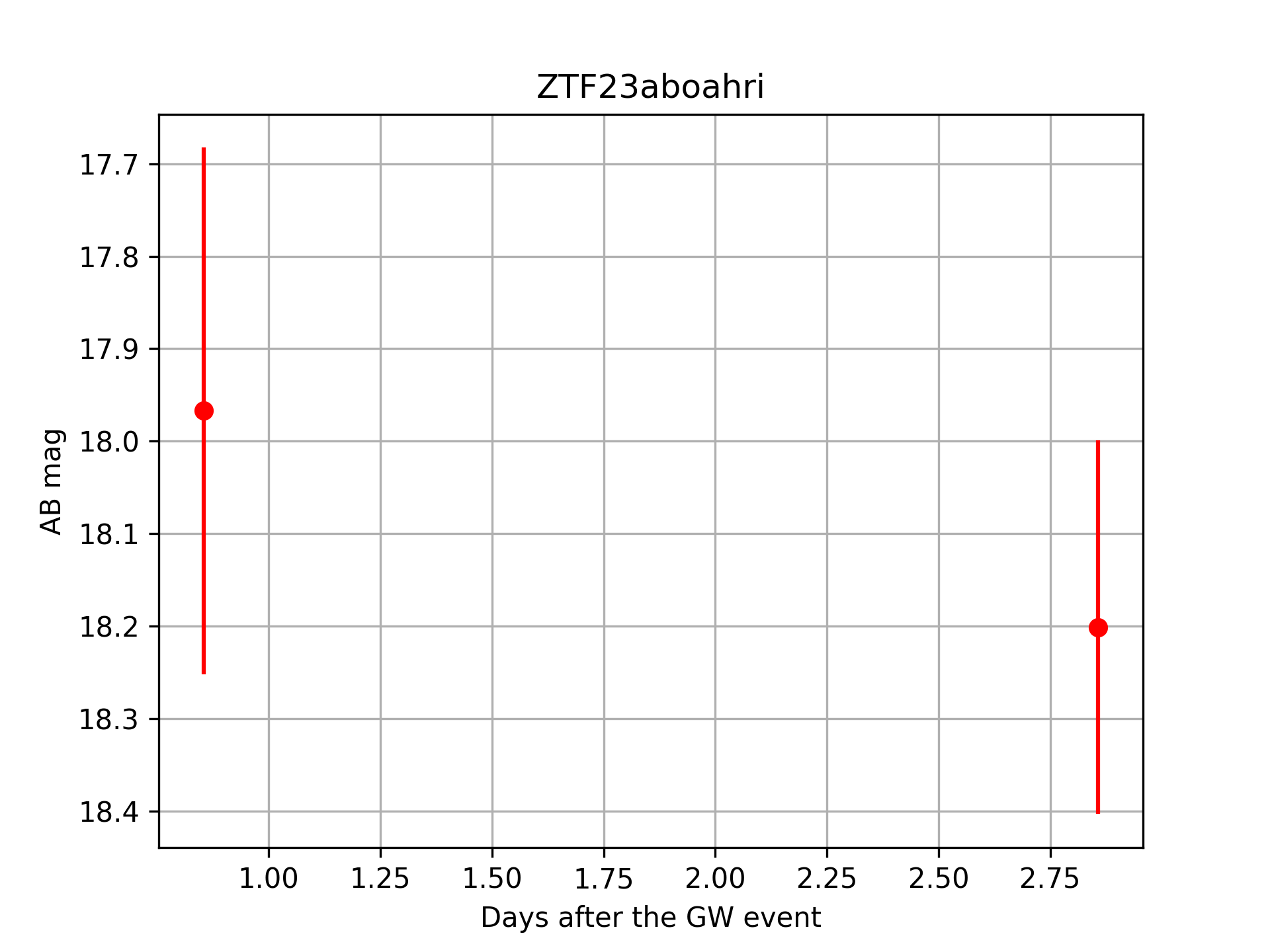}
\includegraphics[width=0.3\textwidth]{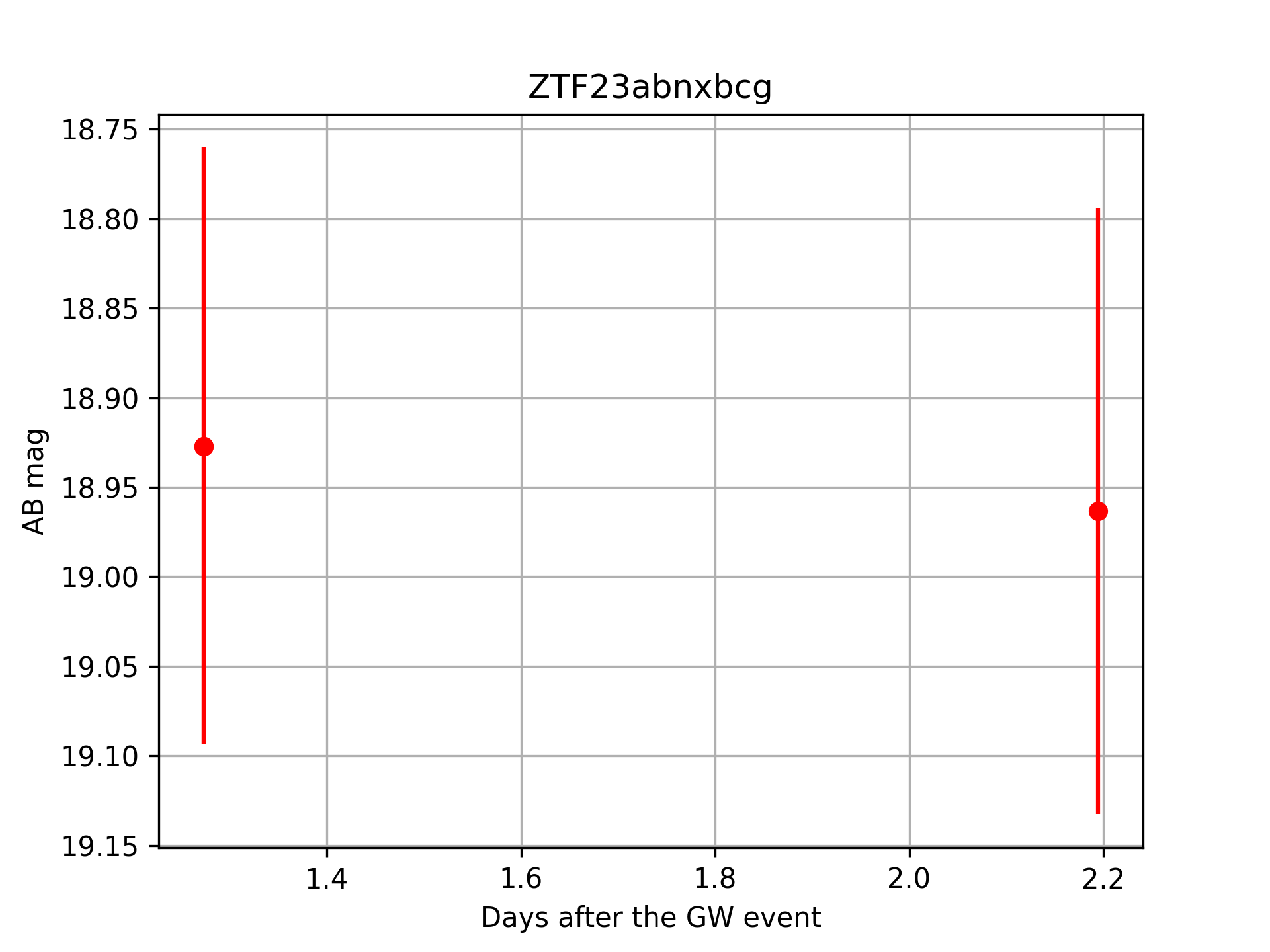}
\includegraphics[width=0.3\textwidth]{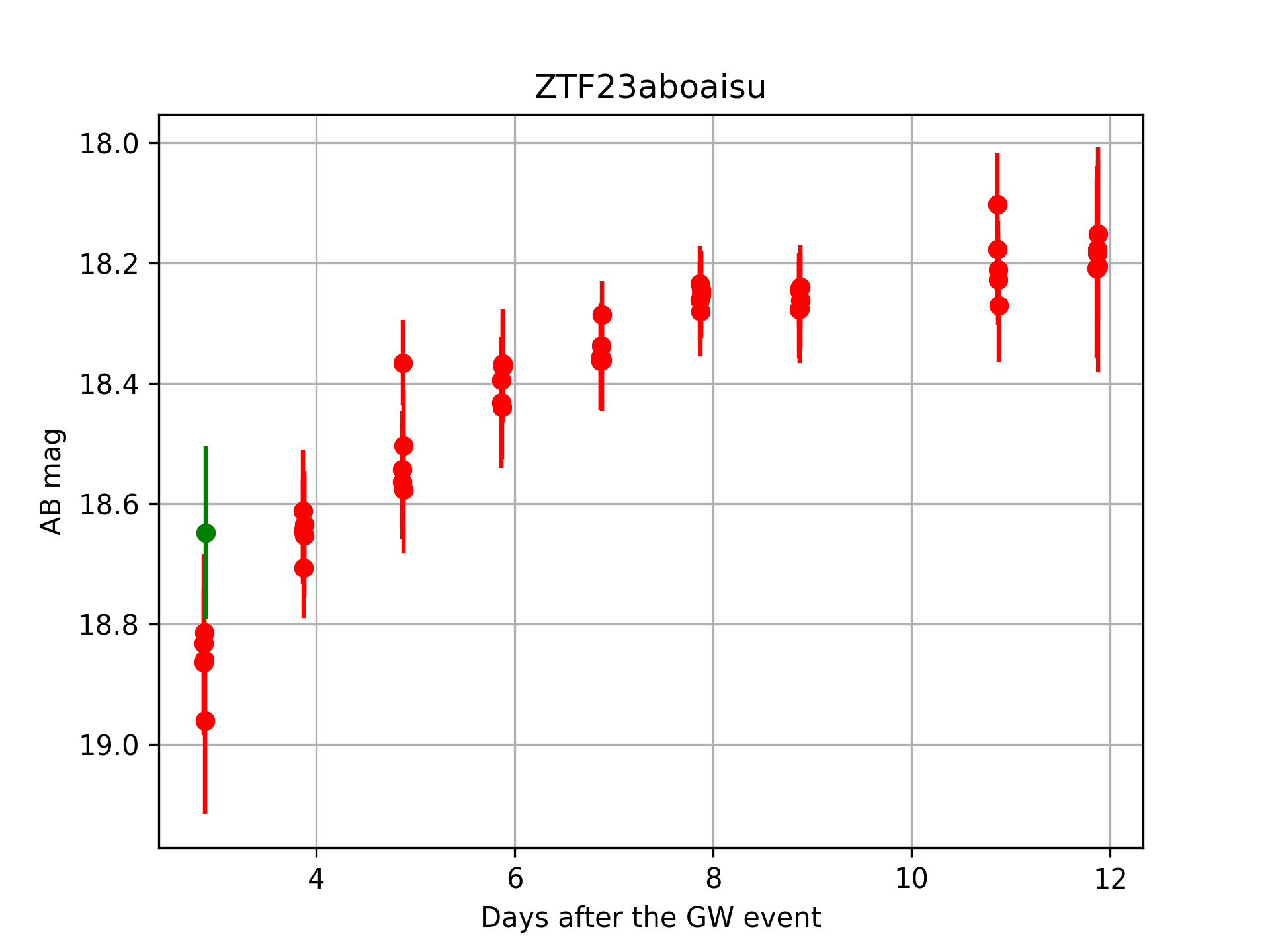}
\includegraphics[width=0.3\textwidth]{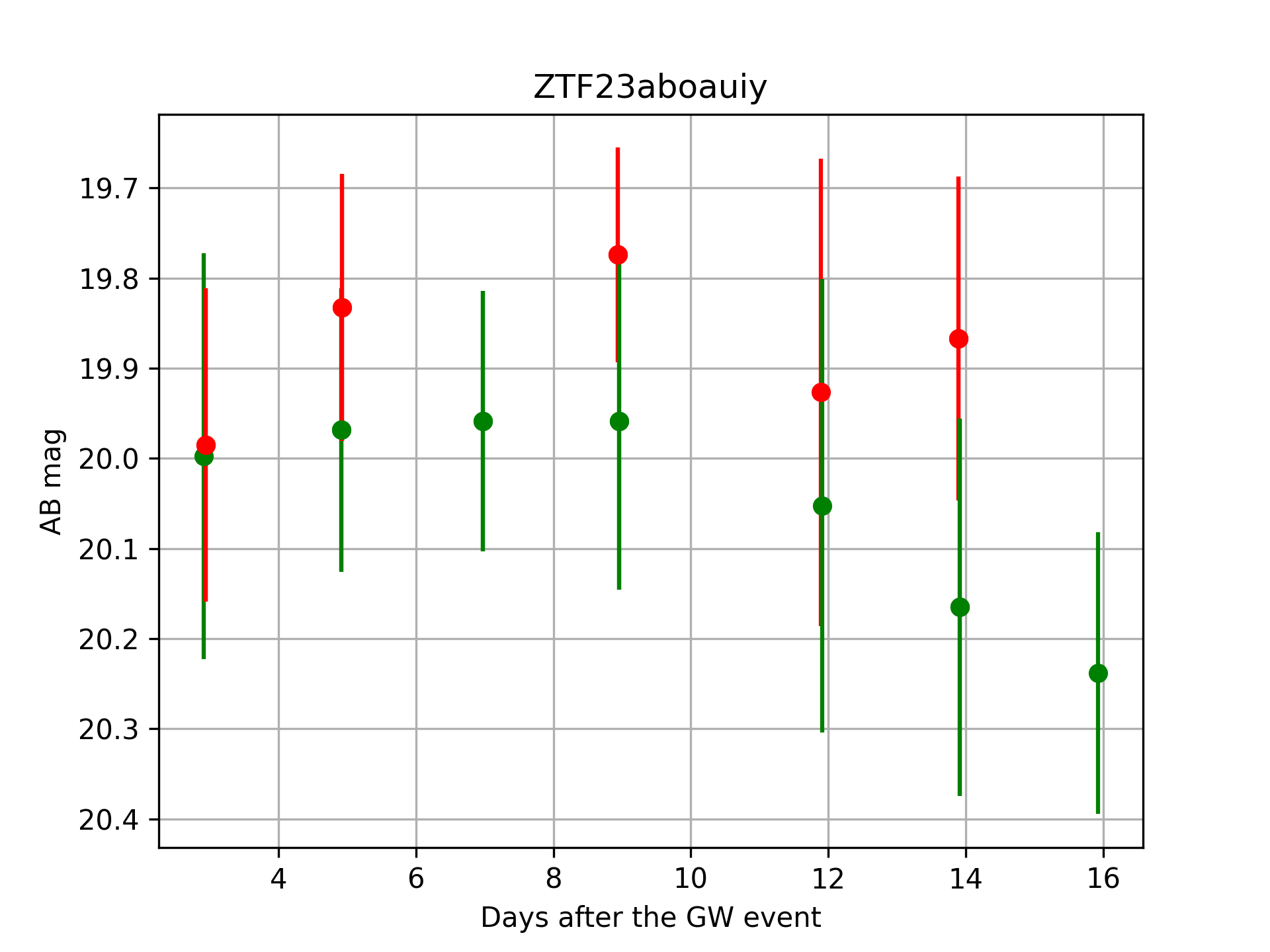}
\includegraphics[width=0.3\textwidth]{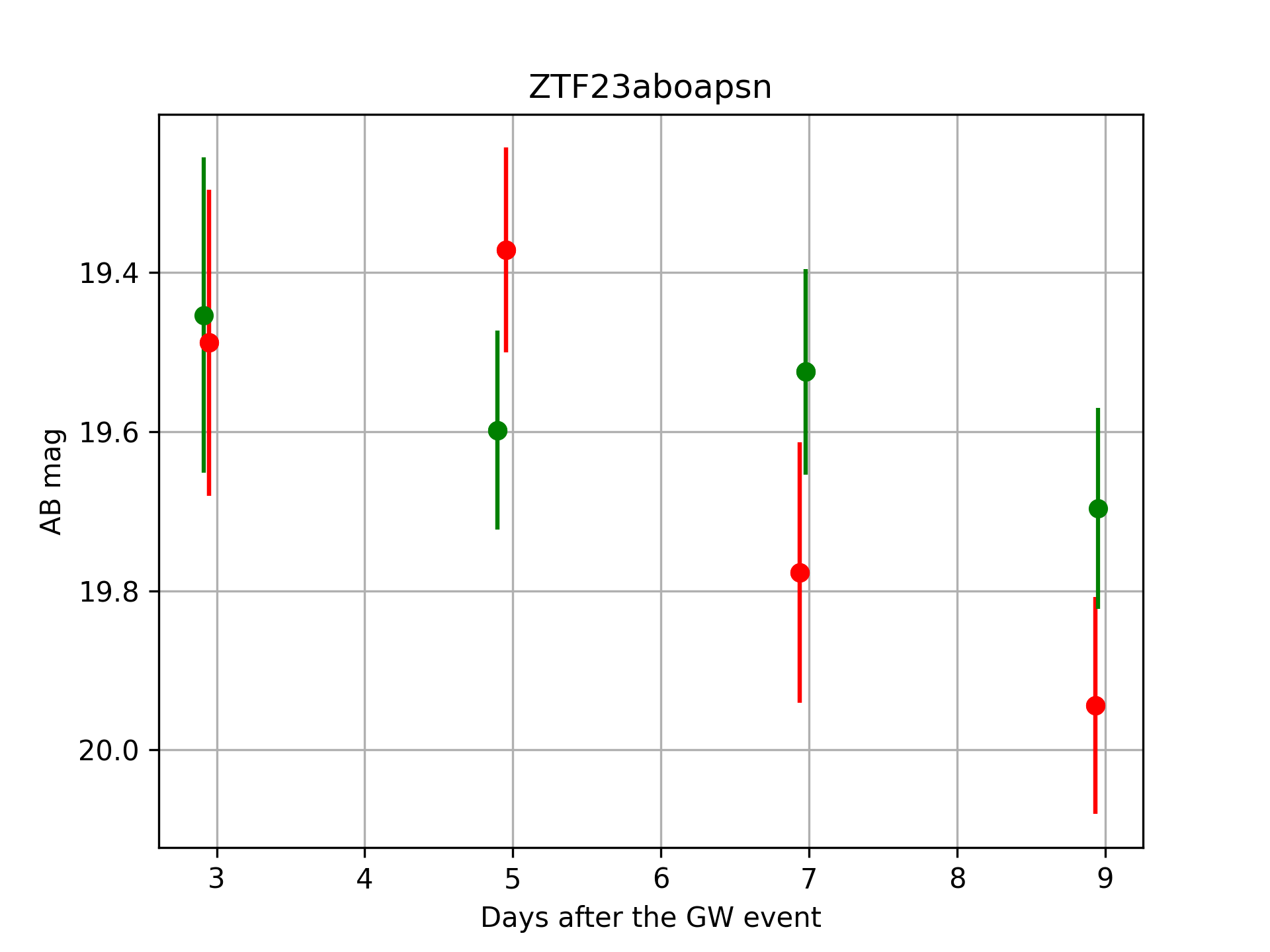}
    \caption{These candidates correspond to the events S230521k S230528a and S231029k. }
    \label{fig:LC_3}
\end{figure*}

We show the photometric light-curves of all the candidates in Figures \ref{fig:LC_1},\ref{fig:LC_2}, and \ref{fig:LC_3}.
    \paragraph{Palomar 60-inch} We acquired photometric data utilizing the Spectral Energy Distribution Machine (SEDM; \citealt{BlNe2018,Rigault2019}) mounted on the Palomar 60-inch telescope. The SEDM is a low resolution (R $\sim$ 100) integral field unit spectrometer with a multi- band ($ugri$) Rainbow Camera (RC). The follow-up request process is automated and can be initiated through \texttt{Fritz}. Standard requests typically involved 180 s exposures in the $g$-, $r$-, and $i$-bands, however it can be customized and for some transients we used 300 s exposures. The data undergoes reduction using a Python-based pipeline, which applies standard reduction techniques and incorporates a customized version of FPipe (Fremling Automated Pipeline; \citealt{FrSo2016}) for image subtraction.
    \paragraph{GROWTH-India Telescope}
We utilized the 0.7-meter robotic GROWTH-India Telescope (GIT) \citep{Kumar2022git}, located in Hanle, Ladakh. It is equipped with a 4k back-illuminated camera that results in a 0.82 deg$^2$ field of view. Data reduction is performed in real-time using the automated GIT pipeline. Photometric zero points were determined using the PanSTARRS catalogue, and PSF photometry was conducted with PSFEx \citep{Bertin2010}. In cases where sources exhibited a significant host background, we performed image subtraction using \texttt{pyzogy} \citep{Guevel2017Pyzogy}, based on the ZOGY algorithm \citep{Zackay_2016}.

   
    \paragraph{Liverpool Telescope} The images acquired with the Liverpool Telescope (LT) were taken using the IO:O \citep{steele2004liverpool} camera equipped with the Sloan \textit{griz} filterset. These images underwent reduction through an automated pipeline, including bias subtraction, trimming of overscan regions, and flat fielding. Image subtraction occurred after aligning with a PS1 template, and the final data resulted from the analysis of the subtracted image.

\subsection{Spectroscopic Follow-up} \label{subsec:spectroscopy}
\paragraph{Palomar 60-inch:} Through \texttt{Fritz}, we can assign transients for spectroscopic follow-up with SEDM. The low-resolution (R$\sim$100) integral field unit(IFU) spectrograph is used to charactherize sources brighter than 18.5 mag. The classification is done by running SNID \citep{Blondin2007snid} and NGSF \citep{Goldwasser2022ngsf} on the reduced spectra.

\paragraph{Palomar 200-inch:} We observed ZTF candidates using the Palomar 200-inch Double Spectrograph (DBSP; \citealt{Oke1982}). The setup configuration involved 1 arcsec, 1.5 arcsec, and 2 arcsec slitmasks, a D55 dichroic, a blue grating of 600/4000, and a red grating of 316/7500. We applied a custom PyRAF DBSP reduction pipeline \citep{BeSe2016} to process and reduce our data.

\begin{figure*}[h]
    \centering
    \includegraphics[width=\textwidth]{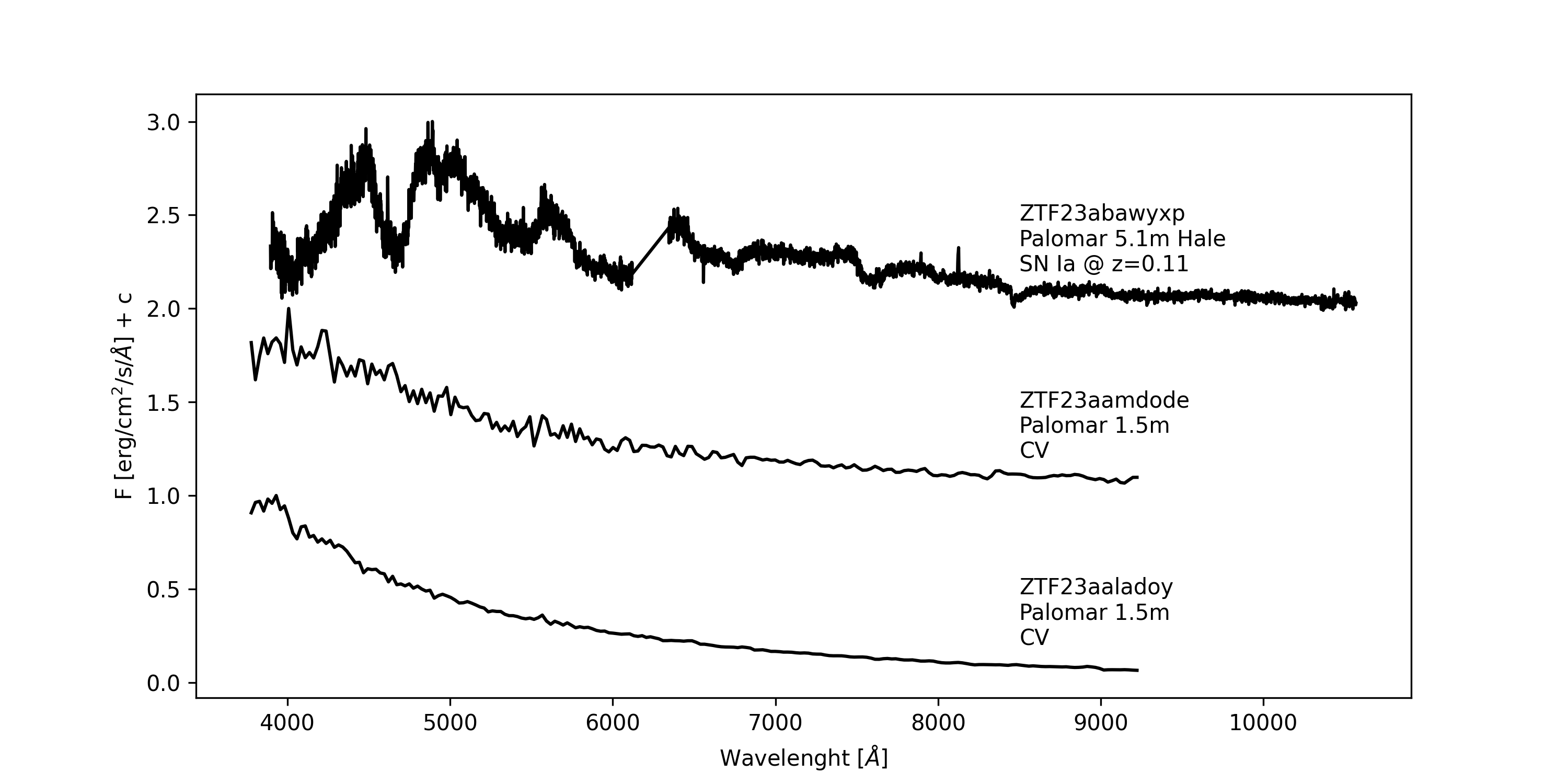}
    \caption{Spectra of the counterpart candidates taken during O4a. }
    \label{fig:spectra_O4}
\end{figure*}


\section{Additional ZTF triggers}\label{sec:ap_gwevents}

Throughout O4a, ZTF covered the region of events detected with a FAR $>$ 1 year$^{-1}$. We triggered observations for S230521k, S230528a, S230615az, S230729cj, and S231029k.

\begin{table*}[tb]
\centering
\resizebox{\textwidth}{!}{
\begin{tabular}{l|l|l|l|l|l|l|l|l|l|l|l|l}
{\bf Trigger }& {\bf Strategy }& {\bf  FAR } & {\texttt{$p_{BNS}$} }& {\texttt{$p_{NSBH}$} }& {\bf HasNS } & {\bf HasRemnant } & {\bf HasMassGap } & {\bf Distance } & {\bf Covered   }& {\bf Area covered }& {\bf g-band depth} & {\bf latency} \\
{\bf  }& {\bf  }& {\bf  [year$^{-1}$]} & {\bf  } & {\bf prob. } & {\bf prob. } & {\bf prob. } & {\bf prob. } & {\bf [Mpc]} & {\bf prob.  } &{\bf[deg$^2$]}& {\bf [AB mag]} & {\bf  [hr]} \\
\hline
S230521k  & Re-weighting  &  76    & 0.25 & 0.14 & 1.0  & 0.9  & 0.0  & 454  & 20\% & 1294 & 21.37 & 0.03 \\
S230528a  & Re-weighting  &  9     & 0.31 & 0.62 & 0.98 & 0.07 & 0.73 &  261 & 4\%  & 315  & 20.92 & 3 \\
S230615az & Re-weighting  & 4.7    & 0.85 & 0.0  & 1.0  & 1.0  & 0.01 & 260  & 31\% & 1063 & 21.25 & 11 \\
S230729cj & No coverage          & 3.82   & 0.0  & 0.39 & 0.0  & 0.61 & 1.0  & 344  & 0\%  & ---  & --    & -- \\
S231029k & Re-weighting   & 93     & 0.68 & 0.0  & 1.0  & 1.0  & 0.46 &  571 & 36\% & 6836 & 19.28 & 0.23 \\
\end{tabular}
}
\caption{Summary of ZTF observations and GW properties for 5 GW events additionally followed-up with ZTF, with FAR $>$ 1 year$^{-1}$. Similarly to Table \ref{tab:O4_summary} we quote other quantities intrinsic to the GW event, such as the mean distance to the merger, the \texttt{HasRemnant}, and the \texttt{HasMassGap} parameters. }
\label{tab:O4_summary_apendix}
\end{table*}

\subsection{S230521k}

S230521k had a source classification with 25\% probability of it being a BNS system and 14\% being a NSBH system but had a high FAR of 76 year$^{-1}$. S230521k properties did not merit a targeted search, thus we re-weighted the nominal ZTF schedule. The observations spanned a total area of 1294 deg$^2$, covering 20\% of the total probability. The first serendipitous observation was taken around $\sim$ 5 minutes after the GW event. The median seeing during the observations is $\sim 2$ arcsec, and limiting magnitudes of the first night are $g=21.37$ and $r=21.42$ mag. Based on the first two nights of observations, 13 candidates passed our automatic and manual inspection and upon further monitoring, none of them showed any promising nature \citep{GCN2023ztf1S230521k,GCN2023ztf2S230521k}. Details of the candidates along with the rejection criterion are presented in Table~\ref{tab:candid}.



\begin{figure*}[h]
    \centering
    \includegraphics[width=0.45\textwidth]{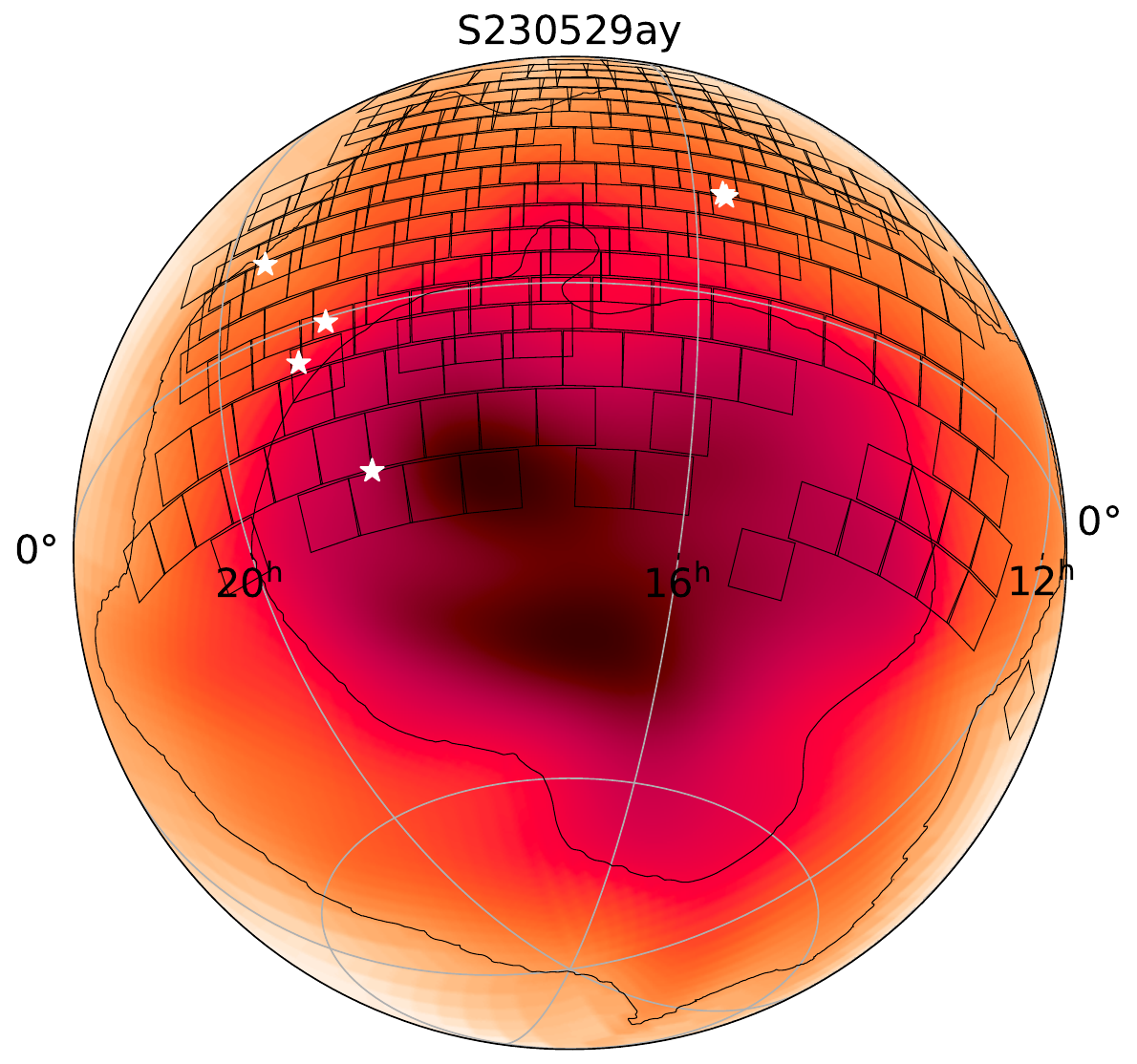}
    \includegraphics[width=0.45\textwidth]{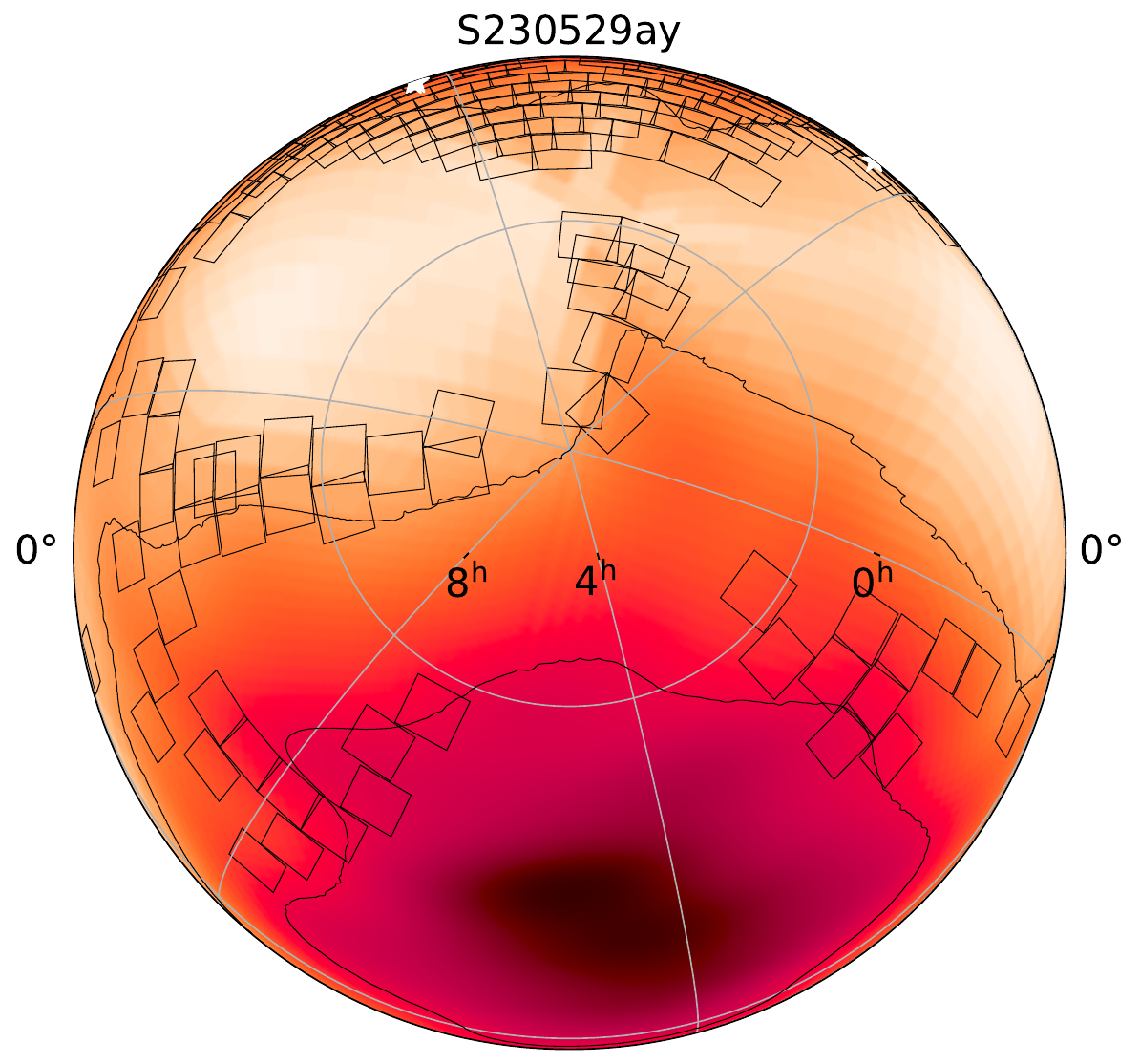}
    \caption{Localization of the high-significance event S230529ay, overplotted with the ZTF tiles and the 90\% probability contour.  }
    \label{fig:map_S230529ay}
\end{figure*}

\begin{figure*}[h]
    \centering
    \includegraphics[width=0.45\textwidth]{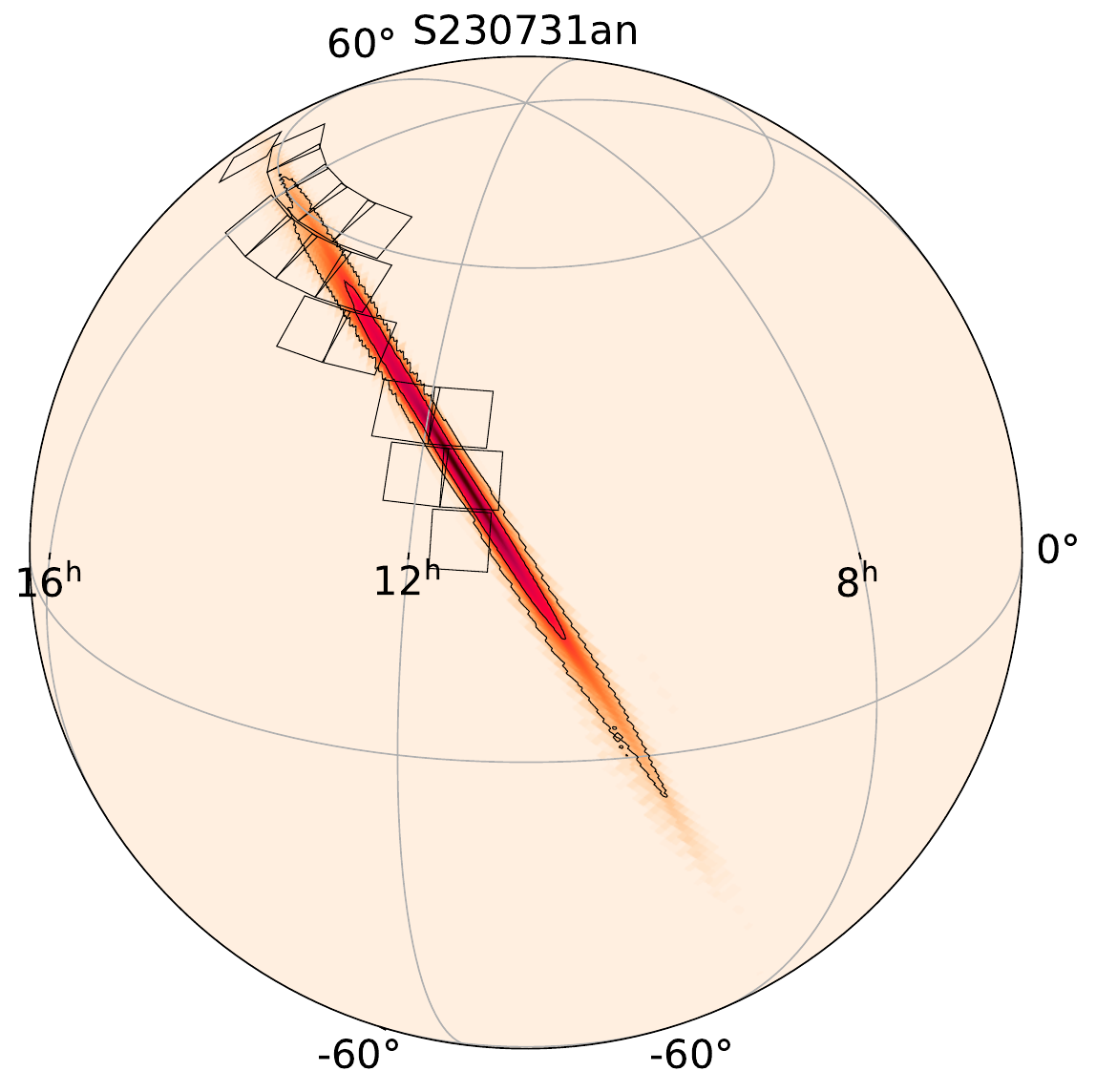}
    \includegraphics[width=0.45\textwidth]{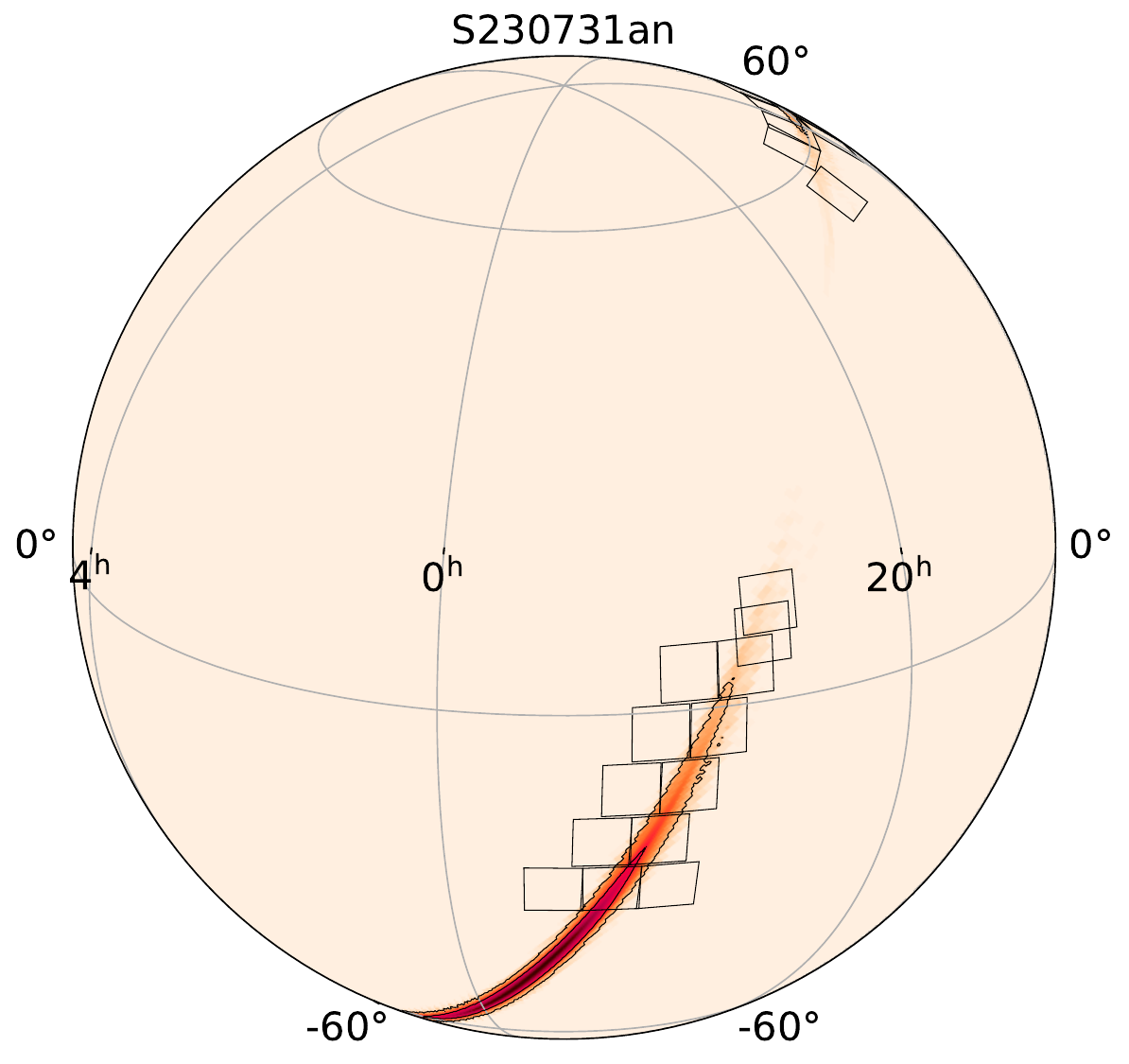}
    \caption{Localization of the high-significance event S230731an, overplotted with the ZTF tiles and the 90\% probability contour. We show the candidates discovered in the region as white stars. We note that even though we covered $\sim$ 2500 deg$^2$, the total enclosed probability is only 7\%. }
    \label{fig:map_S230731an}
\end{figure*}

\begin{figure*}[h]
    \centering
    \includegraphics[width=0.45\textwidth]{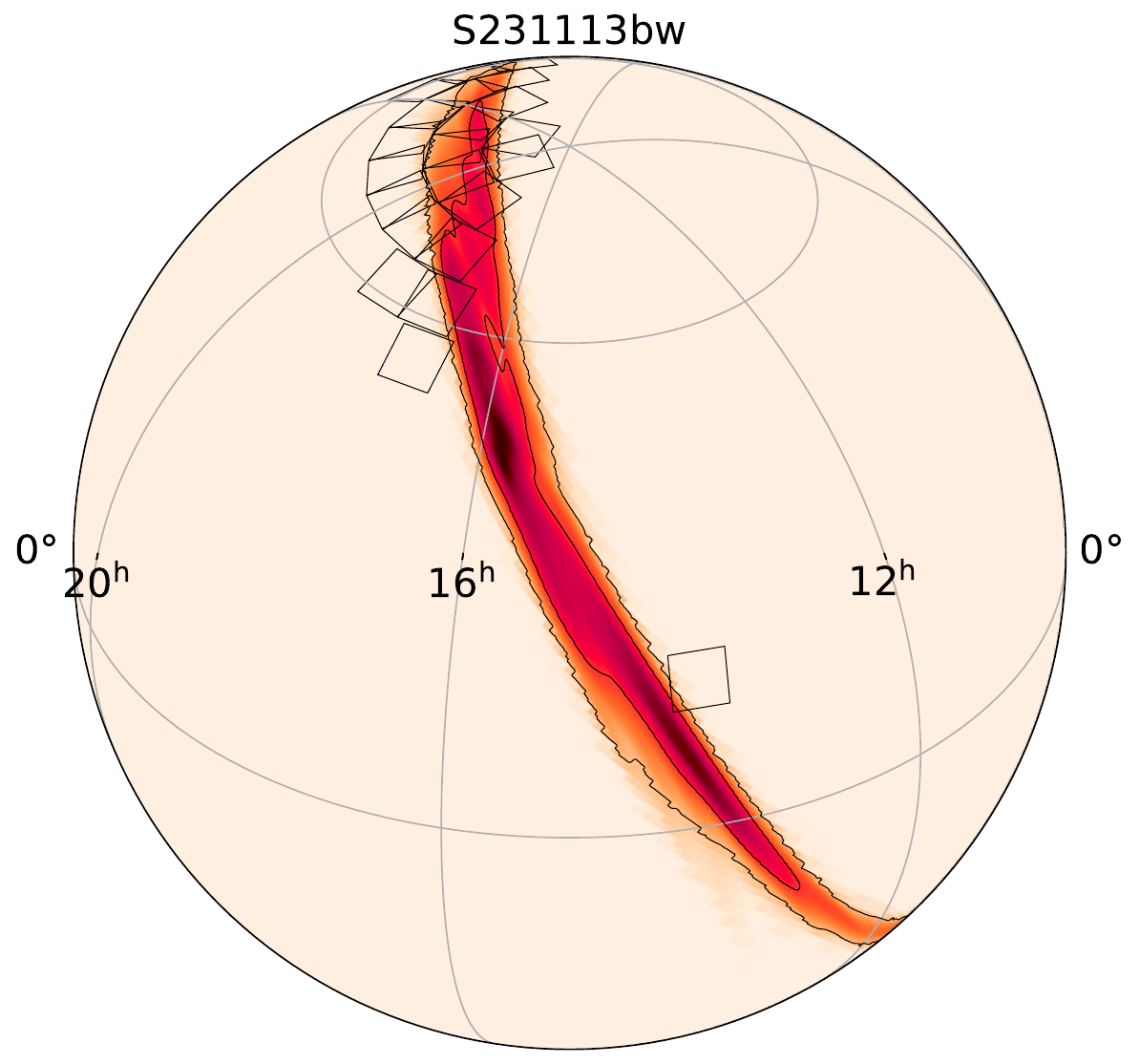}
    \includegraphics[width=0.45\textwidth]{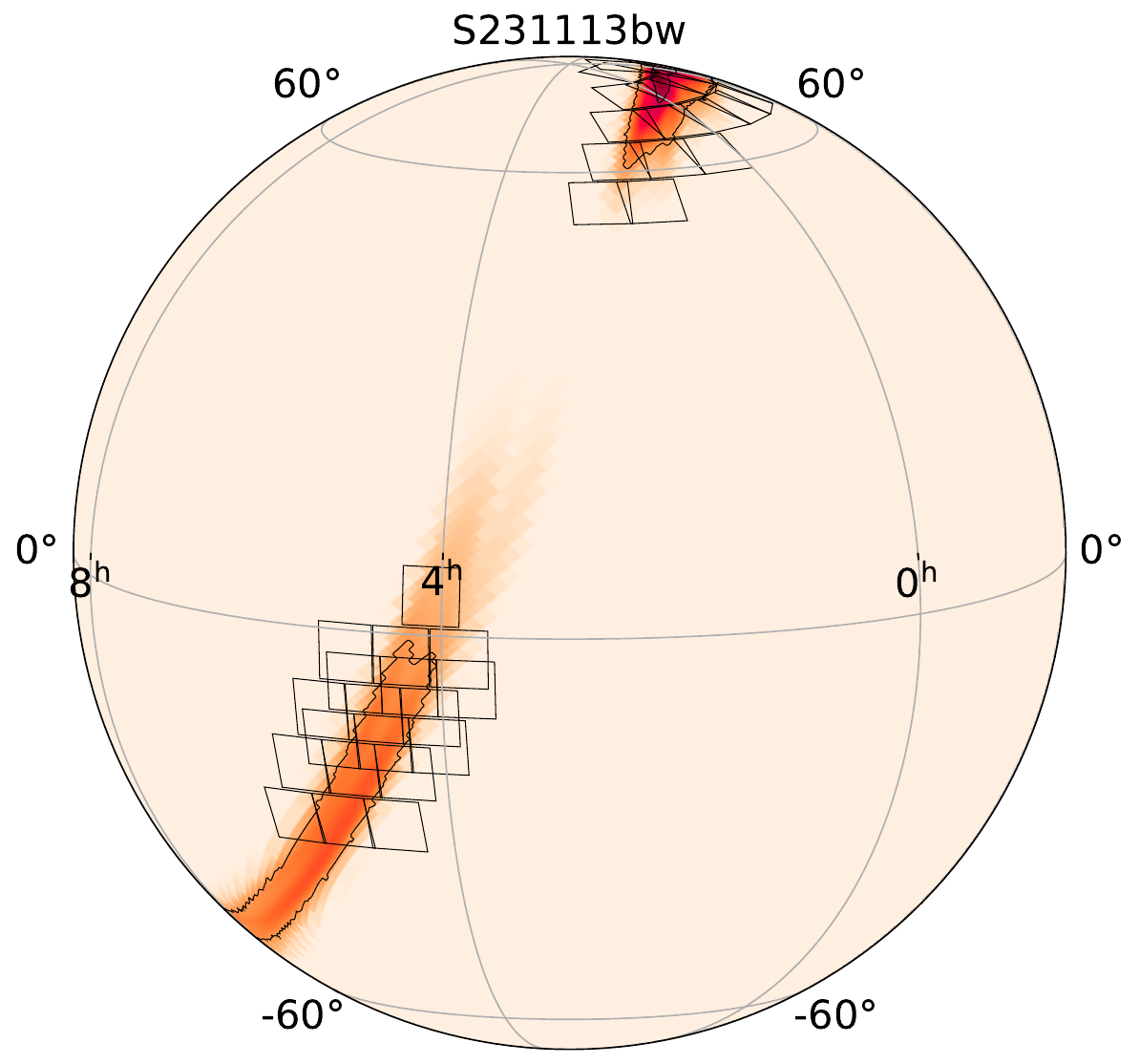}
    \caption{Localization of the high-significance event S231113bw, over plotted with the ZTF tiles and the 90\% probability contour. No candidates were found in this region. }
    \label{fig:map_S231113bw}
\end{figure*}

\begin{figure*}[h]
    \centering
    \includegraphics[width=0.45\textwidth]{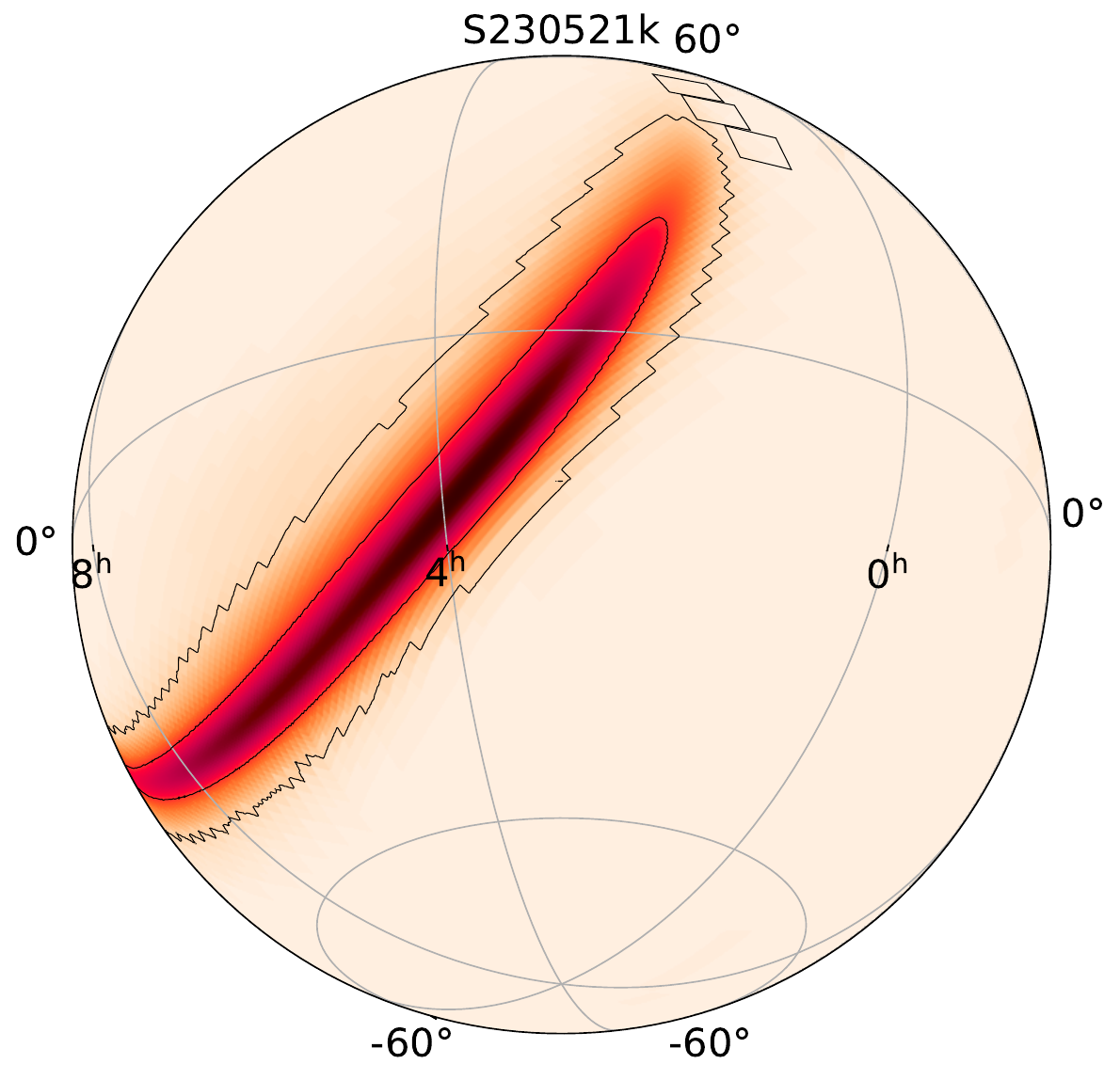}
    \includegraphics[width=0.45\textwidth]{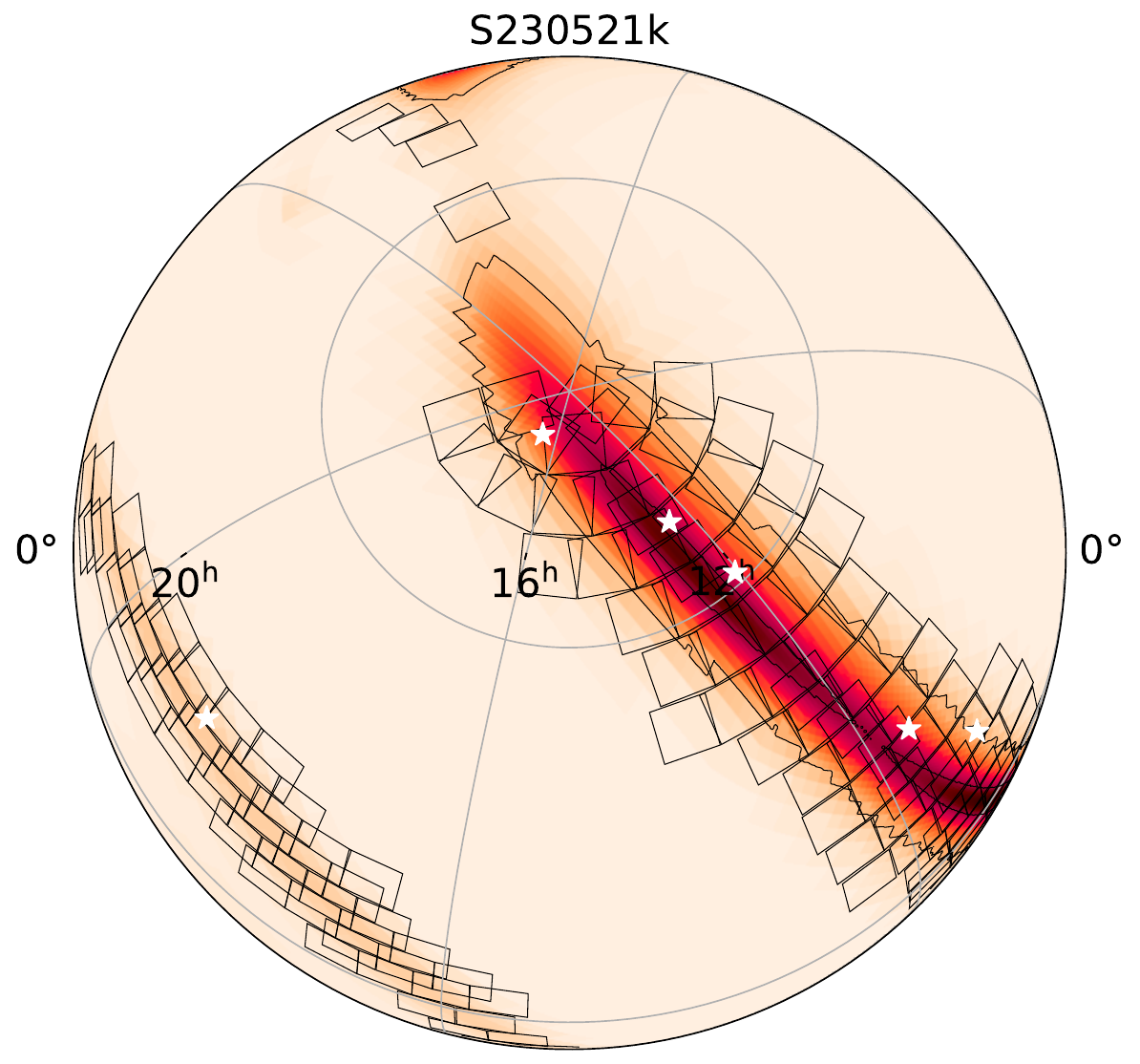}
    \caption{Localization of S230521k, overplotted with the ZTF tiles and the 90\% probability contour. We show the cnadidates in the region as white stars. }
    \label{fig:map_S230521k}
\end{figure*}

\subsection{S230528a}

S230528a was issued with a 40\% probability of it being an NSBH system and 20\% probability for a BNS system with a FAR of 9 year$^{-1}$. Observations included the re-weighting of the ZTF public fields for coverage and the first observation was taken $\sim$ 3\,hours after the GW alert. The observations during the first two days which covered 315 deg$^2$ and 4\% of the total probability. The median limiting magnitudes for the first night of observations was $g=20.92$ and $r=21.09$ mag. During the real-time search, we found four candidates \citep{GCN2023ztf1S230528a}. However, forced photometry on the archival ZTF data and ATLAS data revealed fainter detections in two candidates that predated the GW event and the other two showed flat evolution inconsistent with the expectations for KN emission, so none of the candidates survived for further follow-up (see Table~\ref{tab:candid}).



\subsection{S230615az}

S230615az was classified as a probable BNS event with 85\% probability and a FAR of $\sim$ four year$^{-1}$. The initial 90\% probability area covered $\sim 4400$ deg$^2$. The ZTF strategy for this event relied on the re-weighting of the nominal ZTF fields, covering in total 31\% of the region. While most of the probability lied in two southern lobes, ZTF was able to observe $\sim 1063$ deg$^2$. We found two candidates, but both of them had pre-detections $\sim 11$ days before the GW trigger. No candidates were selected for further follow-up. Additionally, GOTO found a candidate counterpart to the GW event with an L band magnitude of 19.43$\pm$0.08 \citep{GCN2023S230615azGOTO}, but forced photometry on ZTF data revealed that this candidate had a \textit{g}-band detections $36$~hours before the GW trigger and hence we ruled it out (see Table~\ref{tab:candid} for details). Observations with LBT classified the GOTO transient as a SN Ia \citep{GCN2023S230615azLBT}. GIT obtained multiple 300-sec exposures in the $r$ filter by starting to observe 6 min after the GW event, and was able to cover 0.4\% of the skymap. GIT found two interesting candidates that passed the cross-checks with Minor Planet Catalog (MPC) --- GIT230615aa and GIT230615ab \citep{GCN2023S230615azgit}. GIT230615aa was later rejected as an interesting candidate due to deep upper-limits reported \citep{2023GCN.34020....1S} soon after the first detection. 
\begin{figure*}[h]
    \centering
    \includegraphics[width=0.45\textwidth]{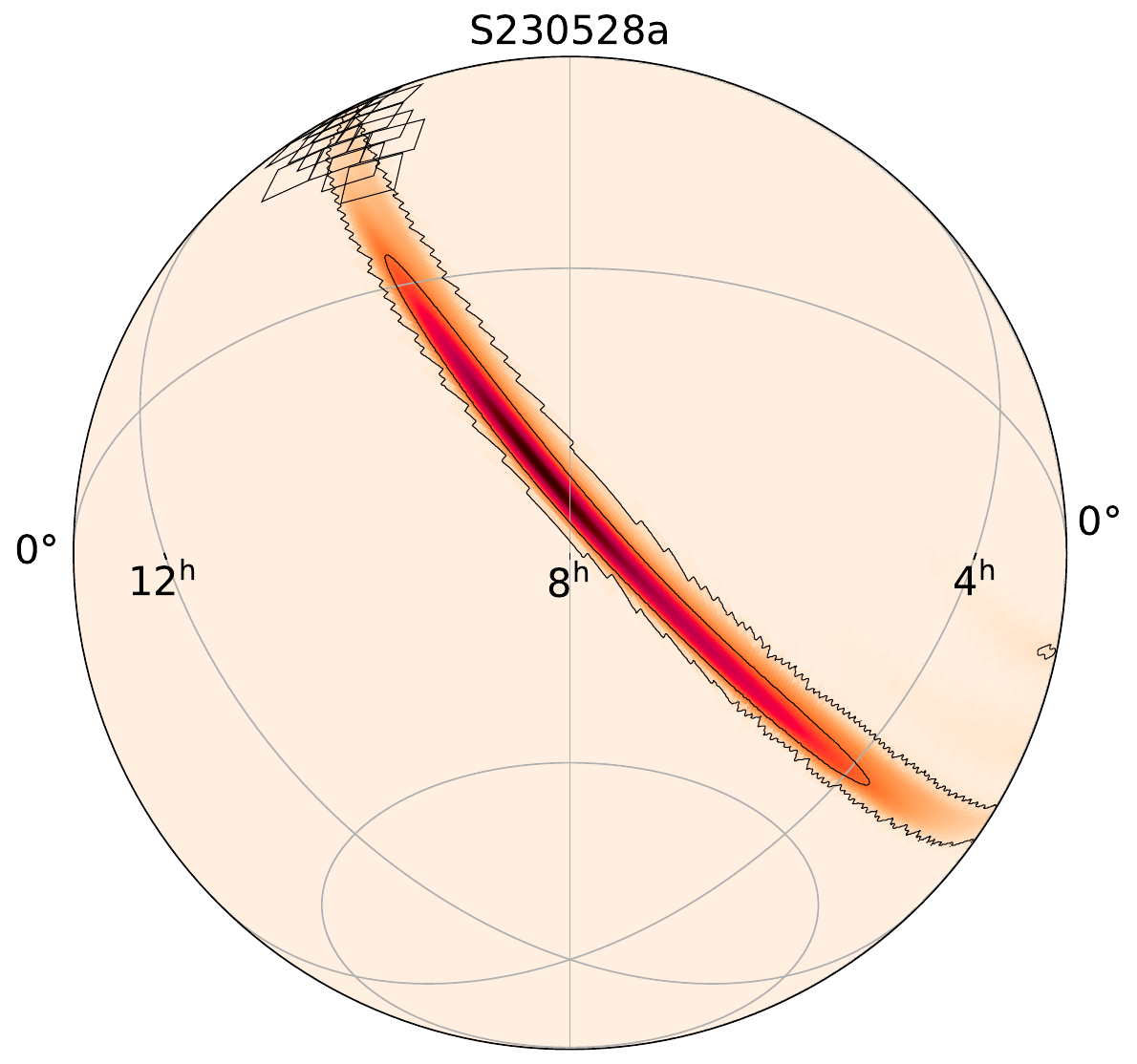}
    \includegraphics[width=0.45\textwidth]{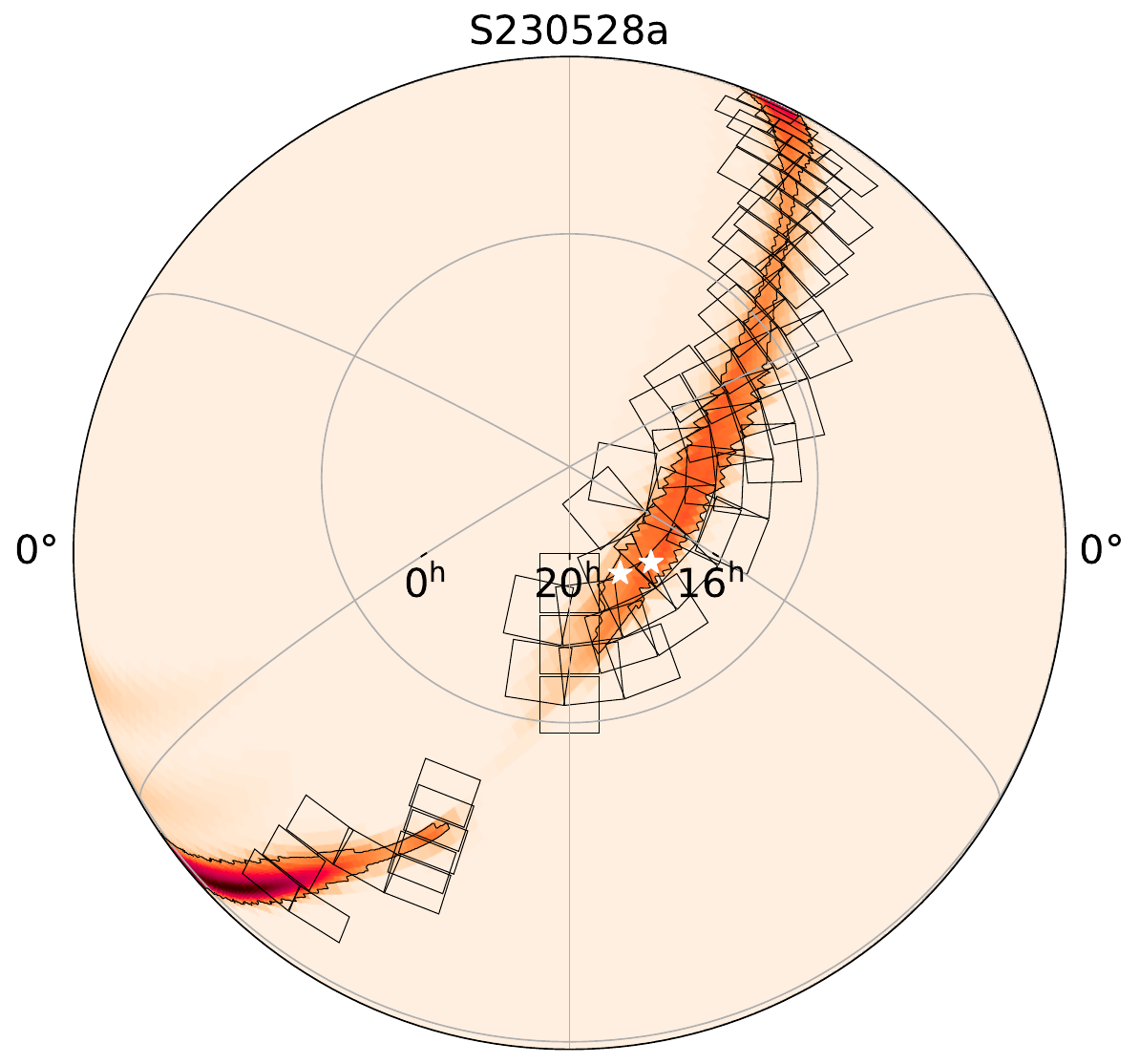}
    \caption{Localization of S230528a, overplotted with the ZTF tiles and the 90\% probability contour. We show the transients consistent with KNe candidates as white stars. }
    \label{fig:map_S230528a}
\end{figure*}

\subsection{S230729cj} 

This event had a FAR of 3.8 year$^{-1}$, however, the region was almost entirely behind the Sun and the ZTF coverage was of only 2\% of the skymap. Hence, we recovered no candidates.

\begin{figure*}[h]
    \centering
    \includegraphics[width=0.45\textwidth]{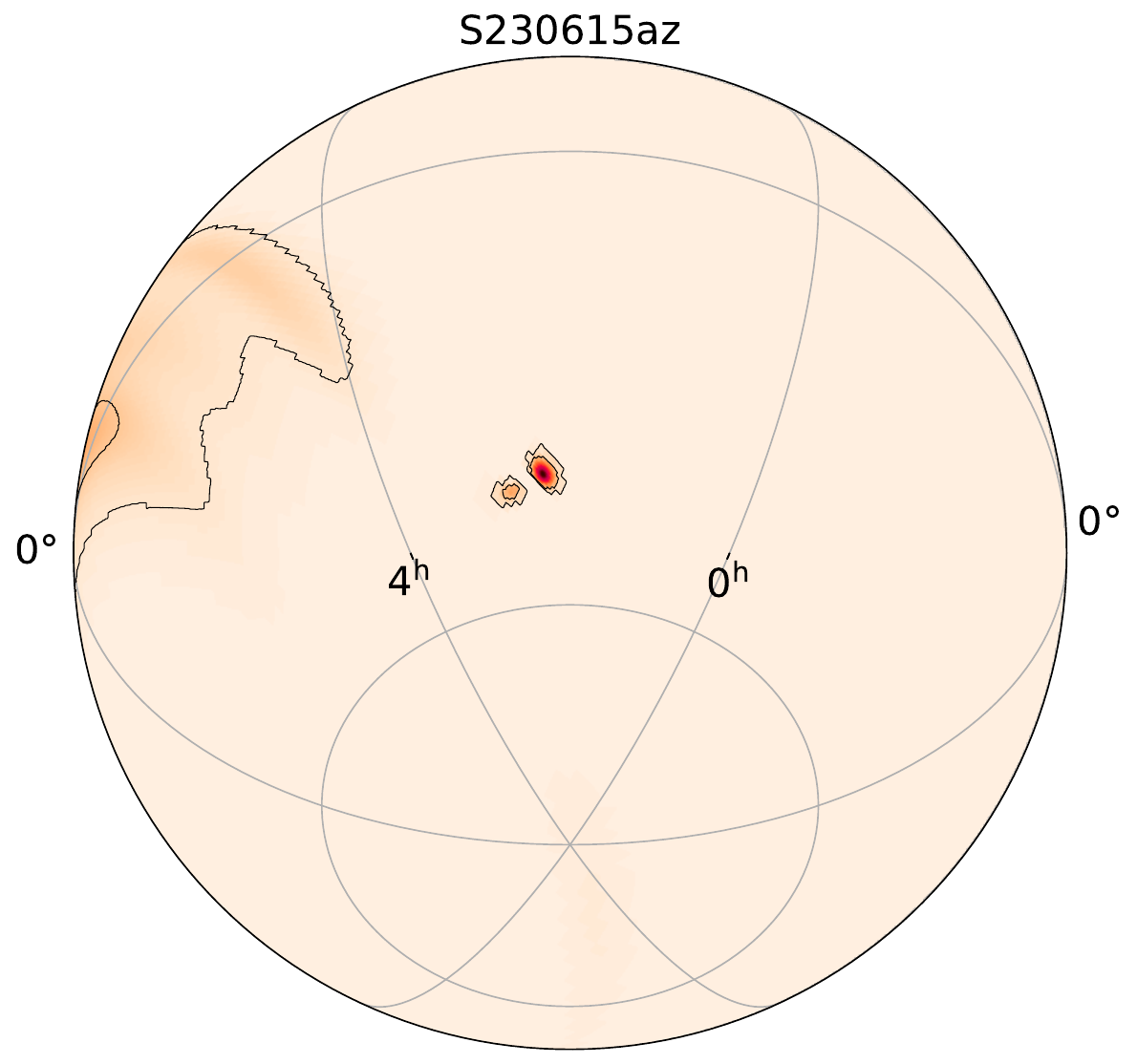}
    \includegraphics[width=0.45\textwidth]{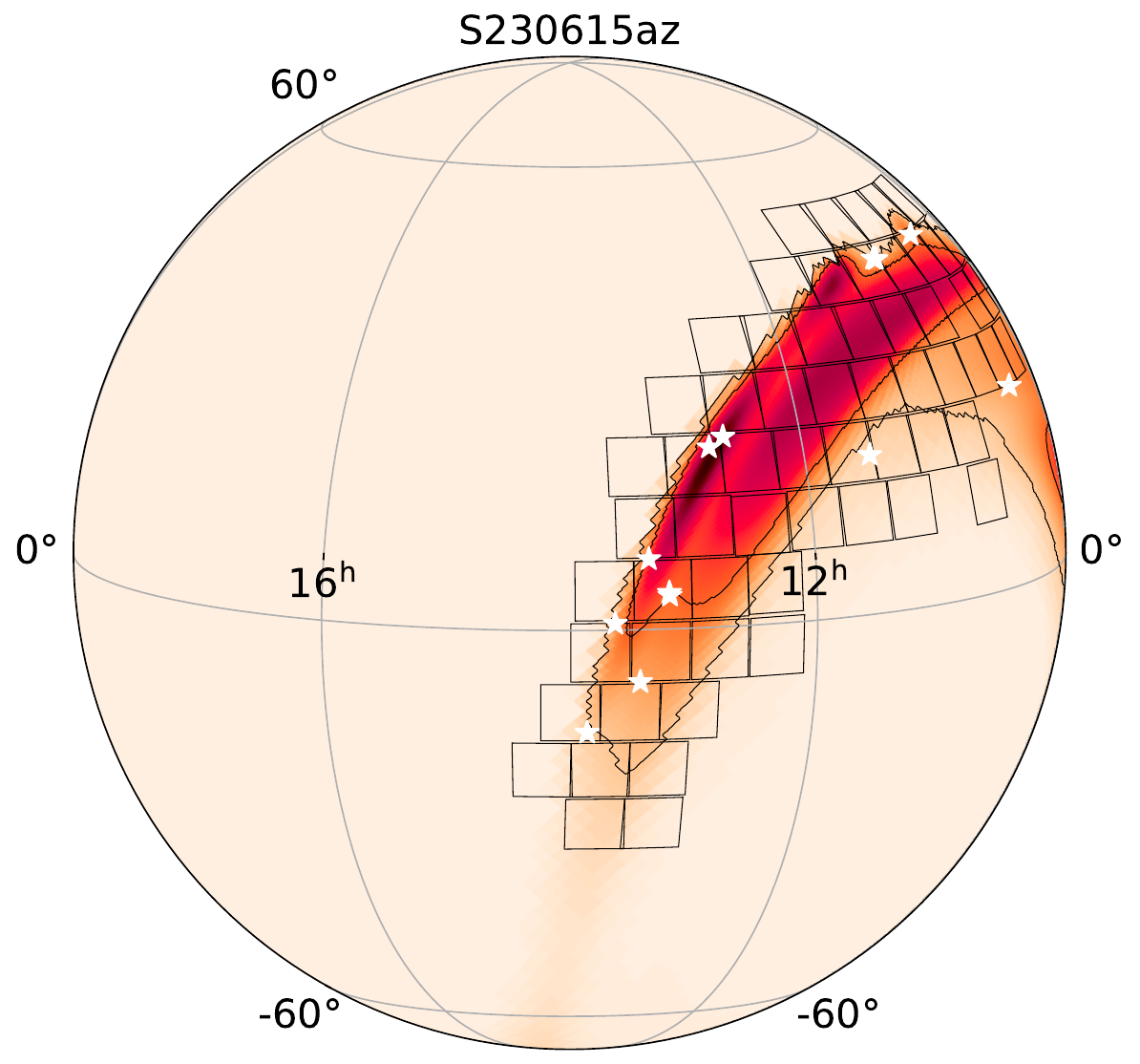}
    \caption{Localization of S230615az, overplotted with the ZTF tiles and the 90\% probability contour. We show the candidates as white stars. }
    \label{fig:map_S230615az}
\end{figure*}

\newpage
\subsection{S231029k}

S231029k, with a relatively high FAR of 93 year$^{-1}$, was detected by the \texttt{spiir} pipeline \citep{spiir} and was initially classified as a likely BNS (68\%), with a terrestrial probability of 32\%. The 90\% credible level of the skymap covered $\sim 14968$ deg$^2$, primarily in the southern hemisphere. Our serendipitous observations started about 15 min after the GW event and covered about 36\% ($\sim 6836$ deg$^2$) of the latest skymap. The first night of observations reached magnitude limits of 19.3 mag in the $g$ band and 19.5 in the $r$ band. No candidates passed our filters. 

\begin{figure*}[h]
    \centering
    \includegraphics[width=0.45\textwidth]{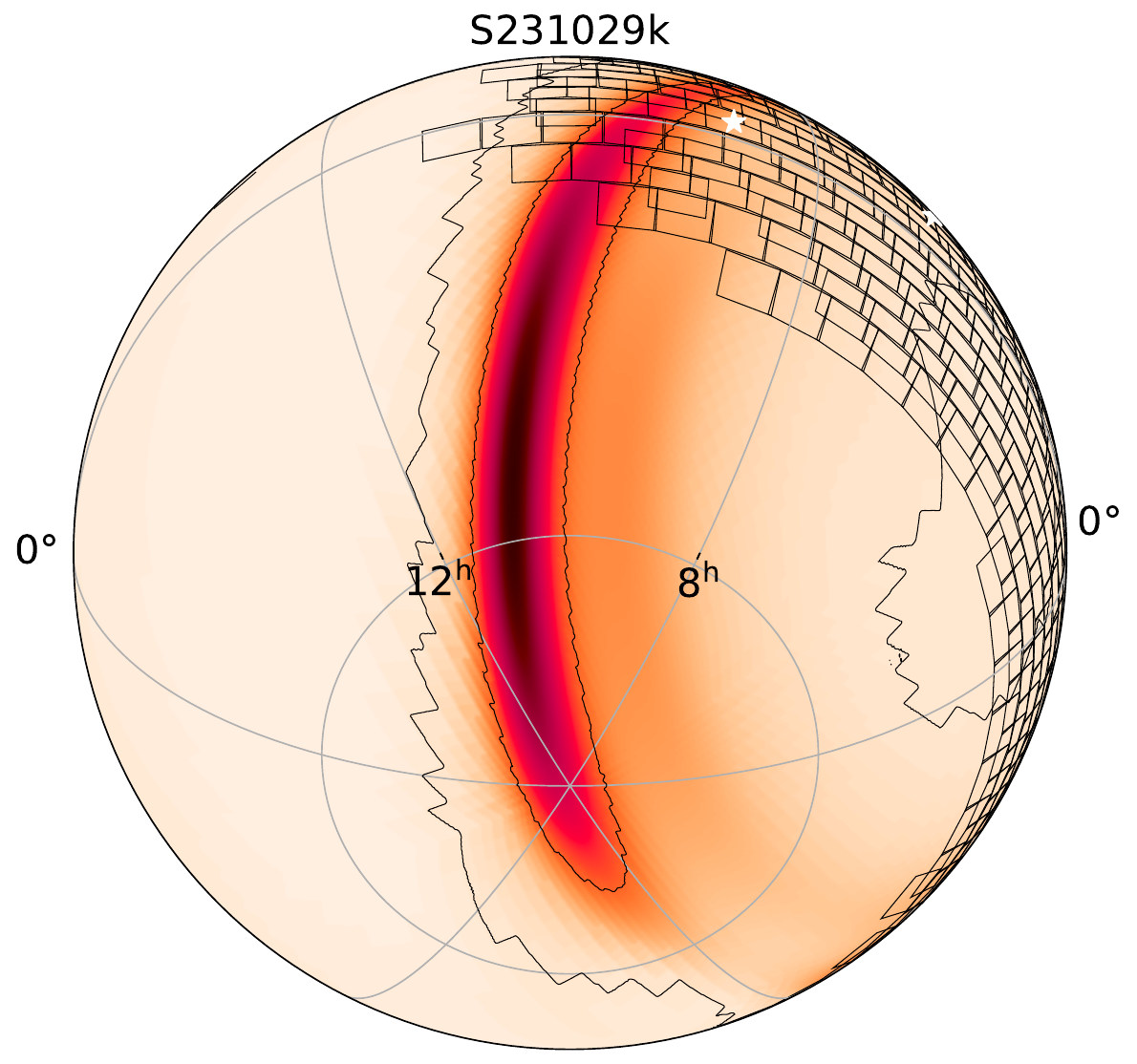}
    \includegraphics[width=0.45\textwidth]{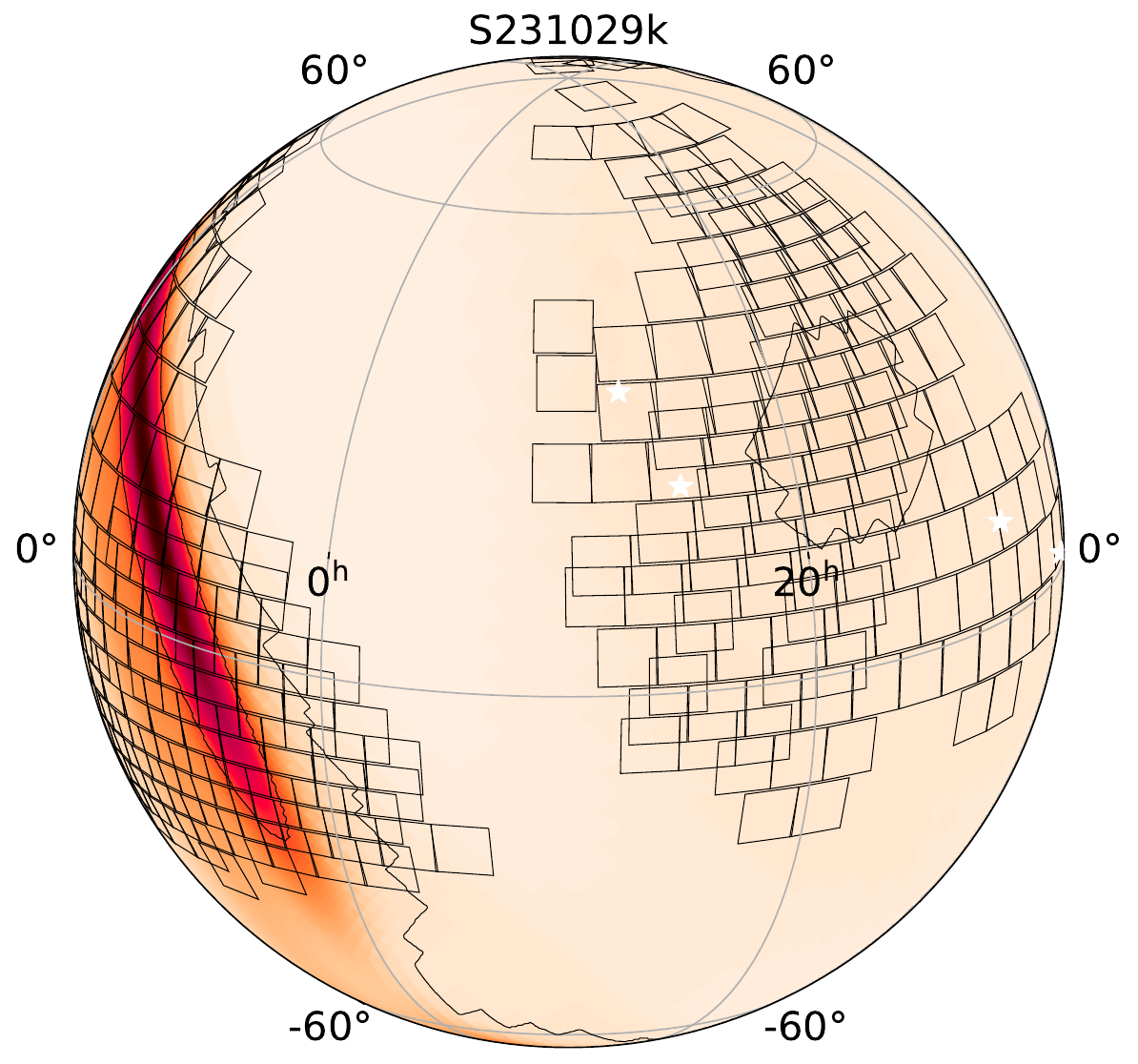}
    \caption{Localization of S231029k, overplotted with the ZTF tiles and the 90\% probability contour. We show the candidates as white stars.  }
    \label{fig:map_S231029k}
\end{figure*}

\clearpage

\section{Candidates from ZTF searches}

In this section we summarize all the candidates analyzed during the O4a searches. We include transients not originally detected with ZTF, but later ruled out by us.

\startlongtable
\begin{deluxetable*}{c|lcccccc}
\tablewidth{0.99\textwidth}
\tablecaption{Properties of the candidates that passed manual inspection and their rejection criteria for six of the followed-up LVK events. There are 15 candidates \\ from the follow-up of \textit{High-significance} events (FAR $<$ 1 year$^{-1}$), and 27 candidates for the \textit{Other ZTF triggers} with FAR $>$ 1 year$^{-1}$\label{tab:candid}.}
\tablehead{\colhead{Event} & \colhead{Candidate} & \colhead{RA} & \colhead{DEC} & \colhead{Discovery time\tablenotemark{a}} & \colhead{Discovery mag.} & \colhead{Redshift} & \colhead{Rejection criterion}\\
\colhead{} & \colhead{} & \colhead{[hhmmss]} & \colhead{[ddmmss]} & \colhead{(hours)} & \colhead{(AB magnitude)} & \colhead{} & \colhead{}}
\startdata
\multicolumn{8}{ l }{Candidates for the \textit{High-significance events}: FAR $<$ 1 year$^{-1}$}   \\
\hline
S230529ay & ZTF23aamnpce & 15h43m56.1s & +15d13m29.3s & 11.47 & \textit{r}=20.49$\pm$0.23 & 0.227 & Inconsistent with GW distance \\
\nodata & ZTF23aamnowb & 15h45m31.2s & +15d39m03.5s & 11.47 & \textit{r}=18.95$\pm$0.28 & \nodata & Slow evolution \\
\nodata & ZTF23aamnsjs & 18h40m49.0s & -20d39m35.6s & 14.77 & \textit{g}=19.20$\pm$0.14 & \nodata & Flat evolution \\
\nodata & ZTF23aamnycd & 19h34m57.5s & +11d15m58.8s & 15.21 & \textit{g}=19.28$\pm$0.15 & \nodata & Slow evolution \\
\nodata & ZTF23aamoeji & 18h57m55.6s & -0d37m42.5s & 15.25 & \textit{g}=20.15$\pm$0.20 & \nodata & Likely galactic \\
\nodata & ZTF23aamnwln & 19h13m03.8s & -4d53m51.1s & 15.34 & \textit{g}=19.93$\pm$0.13 & \nodata & Flat evolution \\
\hline
S230627c & ZTF23aaptuhp & 10h34m41.1s & +45d25m31.3s & 2.23 & \textit{g}=20.34$\pm$0.17 & \nodata & Slow evolution \\
\nodata & ZTF23aaptssn & 10h21m11.3s & +31d18m05.6s & 2.41 & \textit{g}=20.91$\pm$0.18 & 0.15 & Slow evolution \\
\nodata & ZTF23aapwrwg & 10h29m03.6s & +38d44m34.6s & 2.49 & \textit{g}=20.88$\pm$0.18 & 0.577 & Slow evolution \\
\nodata & ZTF23aaptsuy & 10h40m48.5s & +41d58m05.3s & 2.49 & \textit{g}=20.2$\pm$0.11 & \nodata & Slow evolution \\
\nodata & ZTF23aapttaw & 10h58m45.5s & +60d57m16.4s & 2.67 & \textit{g}=21.12$\pm$0.19 & 0.254 & Slow evolution \\
\nodata & ZTF23aaptudb & 11h06m13.5s & +78d33m34.7s & 2.75 & \textit{g}=21.03$\pm$0.27 & 0.188 & Slow evolution \\
\nodata & ZTF23aapdtga & 10h46m32.1s & +57d08m54.8s & 3.23 & \textit{r}=21.33$\pm$0.22 & 0.678 & Slow evolution \\
\nodata & ZTF23aaptusa & 10h48m10.6s & +71d50m29.1s & 3.89 & \textit{g}=21.32$\pm$0.2 & 0.175 & Slow evolution \\
\nodata & ZTF23aapwtcp & 10h49m43.9s & +71d24m34.0s & 4.03 & \textit{r}=21.22$\pm$0.25 & 0.918 & Slow evolution \\
\hline
\multicolumn{8}{ l }{Candidates for the \textit{Other ZTF triggers}: FAR $>$ 1 year$^{-1}$}   \\
\hline
S230521k & ZTF23aaladoy & 18h40m43.9s & +27d01m24.8s & 3.32 & \textit{g}=15.56$\pm$0.03 & \nodata & Featureless spectrum\\
 &  &  &  &  &  &  & long-lived ($\sim 200$days). \\
\nodata & ZTF23aalcvpw & 11h33m53.2s & +29d11m37.8s & 23.22 & \textit{r}=20.36$\pm$0.26 & 0.145 & Pre-detections \\
\nodata & ZTF23aalczjc & 12h29m02.8s & +70d51m01.8s & 23.26 & \textit{r}=20.19$\pm$0.29 & \nodata & Slow evolution \\
\nodata & ZTF23aakyfsk & 12h03m33.6s & +61d23m17.2s & 23.94 & \textit{g}=20.41$\pm$0.19 & 0.201 & Slow evolution \\
\nodata & ZTF23aaldkog & 17h03m21.3s & +83d56m32.9s & 24.24 & \textit{g}=20.60$\pm$0.25 & \nodata & Quasar \\
\hline
S230528a & ZTF23aamgkkz & 17h16m51.2s & +75d27m22.8s & 3.27 & \textit{g}=19.24$\pm$0.11 & 0.105 & Slow evolution \\
\nodata & ZTF23aamlhjz & 18h17m47.0s & +76d20m35.1s & 31.26 & \textit{r}=20.74$\pm$0.31 & \nodata & Slow evolution \\
\hline
S230615az &  GOTO23hu & 13h22m55.2s & +08d09m49.5s & -36.7\phn & \textit{g}=20.72$\pm$0.19 &  \nodata  &  Pre-detections\\
\nodata & GIT230615aa & 12h50m03.64s & +20d53m21.77s & 0.67 & \textit{r}=20.05$\pm$0.06 & \nodata & Asteroid \\
\nodata & GIT230615ab & 12h42m11.65s  & +22d03m25.09s & 1.05 & \textit{r}=19.81$\pm$0.05 & \nodata & Not rejected \\
\nodata & ZTF23aaoocrh & 10h19m00.1s & +41d53m02.8s & 11.98 & \textit{g}=20.12$\pm$0.31 & \nodata & Slow evolution \\
\nodata & ZTF23aaocgns & 10h17m01.7s & +41d44m38.9s & 11.98 & \textit{g}=20.05$\pm$0.22 & \nodata & Slow evolution \\
\nodata & ZTF23aaonoan & 11h21m35.9s & +18d24m26.8s & 12.06 & \textit{g}=20.27$\pm$0.25 & \nodata & Slow evolution \\
\nodata & ZTF23aaooaro & 13h13m14.5s & +3d53m10.4s & 12.53 & \textit{g}=20.21$\pm$0.22 & 0.095 & Slow evolution \\
\nodata & ZTF23aaooarp & 13h13m29.2s & +4d13m31.1s & 12.53 & \textit{g}=20.72$\pm$0.32 & \nodata & Slow evolution \\
\nodata & ZTF23aaocreh & 13h51m44.2s & -12d14m51.7s & 12.66 & \textit{g}=19.27$\pm$0.21 & \nodata & Slow evolution \\
\nodata & ZTF23aaoiixv & 9h06m17.9s & +22d29m45.7s & 34.23 & \textit{r}=18.86$\pm$0.19 & \nodata & Slow evolution \\
\nodata & ZTF23aaoimxy & 9h16m15.1s & +43d23m41.3s & 34.52 & \textit{r}=18.52$\pm$0.11 & \nodata & Slow evolution \\
\nodata & ZTF23aaonttd & 13h27m02.9s & -6d10m06.6s & 59.76 & \textit{r}=19.78$\pm$0.18 & \nodata & Slow evolution \\
\nodata & ZTF23aaoorce & 13h38m54.1s & +0d45m34.0s & 59.77 & \textit{r}=20.16$\pm$0.21 & \nodata & Slow evolution \\
\hline
S231029k & ZTF23abnswxd & 6h37m33.6s & +18d16m39.2s & 6.34 & \textit{r}=19.29$\pm$0.18 & \nodata & Slow evolution \\
\nodata & ZTF23aboahri & 16h24m19.9s & +1d43m55.8s & 20.51 & \textit{r}=17.97$\pm$0.28 & \nodata & Slow evolution \\
\nodata & ZTF23abnxbcg & 8h42m41.8s & +4d34m02.2s & 30.56 & \textit{r}=18.93$\pm$0.17 & 0.074 & Slow evolution \\
\nodata & ZTF23aboaisu & 17h48m33.8s & +11d34m10.8s & 68.79 & \textit{r}=18.86$\pm$0.12 & \nodata & Slow evolution \\
\nodata & ZTF23aboauiy & 21h02m41.4s & +24d07m28.7s & 69.85 & \textit{g}=20.0$\pm$0.22 & \nodata & Slow evolution \\
\nodata & ZTF23aboapsn & 21h31m33.2s & +35d42m43.5s & 69.87 & \textit{g}=19.45$\pm$0.2 & \nodata & Slow evolution 
\enddata
\tablenotetext{a}{Time relative to the GW event.}
\end{deluxetable*}

\section{Regression}

We develop a Random Forest (RF) regressor to predict kilonova properties using low-latency gravitational wave data.

\begin{itemize}
   \item  We adopt simulations from \citep{Kiendrebeogo23} as our training dataset. This includes 1189 simulated compact binary coalescences that passed detection criteria for O4. The simulations include binary distance, sky position, p-astro, FAR, and an area of 90\% sky localization that we include in our features.
   \item  We compute EM-bright\footnote{\url{https://pypi.org/project/ligo.em-bright/}} classifications (HasNS, HasRemnant, and HasMassGap) for the simulated data above to include as features.
  
   \item  We generate the light curves for each of the simulated events using the nuclear multi-messenger astronomy (NMMA)\footnote{\url{https://nuclear-multimessenger-astronomy.github.io/nmma/fitting.html}} framework, which relies on the POSSIS model (Bu2019lm; \citealt{Bulla2019,Diet2020}). We restrict our analysis to simulations with peak magnitudes $>$ 18 mag for r filter. We use the peak of the light curve in g and r filters as target.
   \item  We use the features and target (the information from the GW simulated events and the predicted peak magnitude) to train a RF regressor. To make sure that the scale and measurement units were consistent throughout the training dataset, we applied StandardScaler. The data is separated into two groups: an 80/20 ratio is used for training and testing, while a 70/30 ratio is used for validation. We obtain an MSE of 0.25 and an $R^2$ of 0.76 and an MSE of 0.14 and an $R^2$ of 0.82 in the $g$-band and $r$-band for our test data, respectively.
   
   \item  For the events included in this paper (see Table \ref{tab:O4_summary}), we collect the necessary features (FAR, area(90), distance, longitude, latitude, HasNS, HasRemnant, HasMassGap, and P-astro) and use these to predict the peak magnitude using our RF model. The analysis was conducted offline, after the manual candidate vetting was completed.

\end{itemize}

 Our main finding is the estimated peak magnitude for a KN associated with S230627c. Our model predicts a KN peaking at 21.61 mag in the $r$-band and 22.16 mag in the $g$-band. According to Table \ref{tab:candid}, ZTF23aapdtga is 21.80 mag in the $g$-band and 21.33 mag in the $r$-band, making this candidate consistent with our predictions within 3$\sigma$. No other candidate for any other GW event was within 3$\sigma$ of the predicted peak. 

We expect our RF model to have improved performance with larger and more representative training data, and we look forward to including our predictions to aid in real-time searches.

\startlongtable
\begin{deluxetable*}{l|l|l|l|l|l|l|l|l}
\tablecaption{Compilation of all the 150 IGWN events that had either a probability of BNS (\texttt{$p_{BNS}$}) greater than 0.1 or their probability of NSBH (\texttt{$p_{NSBH}$}) greater than 0.1. We divide the events into \textit{High-significance} (FAR $<$ 1 year$^{-1}$), \textit{Other ZTF trigger}s, and \textit{Not followed with ZTF}. The events we did not follow-up with ZTF all have FAR $>$ 1 year$^{-1}$, and were not used in the ZTF non-detection analysis. We additionally quote their FAR, their probability of BBH merger (\texttt{$p_{BBH}$}), their terrestrial probability (\texttt{$p_{Terrestrial}$}), and their publicly available properties \texttt{HasNS}, \texttt{HasRemnant}, and  \texttt{HasMassGap}. \label{tab:O4_summary_all}. }
\tablehead{{\bf Trigger }& {\bf FAR [year$^{-1}$] }& {\bf \texttt{$p_{BNS}$}} & {\bf \texttt{$p_{NSBH}$} } & {\bf \texttt{$p_{BBH}$} } & {\bf  \texttt{$p_{Terrestrial}$}} &\texttt{HasNS} & 	\texttt{HasRemnant} & 	\texttt{HasMassGap} }
\startdata
\hline
\multicolumn{6}{ l }{ \textit{High-significance}: FAR $<$ 1 year$^{-1}$}   \\
\hline
S230518h & 0.01 & 0.0 & 0.86 & 0.04 & 0.1 & 1.0 & 0.0 & 0.0   \\ 
S230529ay & 0.01 & 0.31 & 0.62 & 0.0 & 0.07 & 0.98 & 0.07 & 0.73   \\ 
S230627c & 0.01 & 0.0 & 0.49 & 0.48 & 0.03 & 0.0 & 0.0 & 0.14   \\ 
S230731an & 0.01 & 0.0 & 0.18 & 0.81 & 0.0 & 0.0 & 0.0 & 0.0 \\
S231113bw & 0.43 & 0.0 & 0.17 & 0.79 & 0.04 & 0.02 & 0.0 & 0.04  \\ 
\hline
\multicolumn{6}{ l }{ \textit{Other ZTF triggers}}   \\
\hline
S230521k & 76.34 & 0.25 & 0.14 & 0.0 & 0.6 & 1.0 & 0.9 & 0.0 \\ %
S230528a & 9.58 & 0.2 & 0.44 & 0.0 & 0.36 & 1.0 & 0.02 & 0.97 \\ 
S230615az & 4.7 & 0.85 & 0.0 & 0.0 & 0.15 & 1.0 & 1.0 & 0.01 \\ 
S230729cj & 3.82 & 0.0  & 0.39 & 0.0 & 0.61 & 1.0 & 0.0 & 0.86 \\
S231029k & 93.5 & 0.68 & 0.0 & 0.0 & 0.32 & 1.0 & 1.0 & 0.46 \\ %
\hline
\multicolumn{6}{ l }{\textit{Not followed with ZTF}}   \\
\hline
S230615i & 411.95 & 0.02 & 0.2 & 0.0 & 0.78 & 1.0 & 0.0 & 0.06 \\ %
S230617bc & 55.43 & 0.27 & 0.07 & 0.0 & 0.66 & 1.0 & 1.0 & 0.02 \\ %
S230618ba & 81.1 & 0.0 & 0.29 & 0.14 & 0.57 & 0.73 & 0.0 & 0.52 \\ %
S230619aa & 248.48 & 0.36 & 0.02 & 0.0 & 0.62 & 1.0 & 0.82 & 0.55 \\ %
S230619bd & 608.9 & 0.18 & 0.0 & 0.0 & 0.81 & 1.0 & 1.0 & 0.46 \\ %
S230620ad & 197.28 & 0.0 & 0.12 & 0.02 & 0.86 & 0.26 & 0.0 & 0.51 \\ %
S230621ap & 497.72 & 0.1 & 0.0 & 0.0 & 0.9 & 1.0 & 1.0 & 0.0 \\ %
S230622br & 101.94 & 0.0 & 0.36 & 0.1 & 0.54 & 1.0 & 0.0 & 0.06 \\ %
S230623ad & 80.08 & 0.49 & 0.0 & 0.0 & 0.51 & 1.0 & 1.0 & 0.0 \\ %
S230624s & 593.36 & 0.0 & 0.12 & 0.01 & 0.88 & 1.0 & 0.0 & 0.0 \\ %
S230627v & 339.99 & 0.17 & 0.0 & 0.0 & 0.83 & 1.0 & 1.0 & 0.0 \\ %
S230627ay & 213.68 & 0.0 & 0.22 & 0.03 & 0.75 & 1.0 & 0.0 & 0.58 \\ %
S230627bj & 245.72 & 0.0 & 0.12 & 0.04 & 0.84 & 1.0 & 0.0 & 0.0 \\ %
S230629y & 82.07 & 0.0 & 0.23 & 0.09 & 0.68 & 0.7 & 0.0 & 0.36 \\ %
S230701z & 509.73 & 0.0 & 0.12 & 0.02 & 0.86 & 0.82 & 0.0 & 0.13 \\ %
S230703aq & 596.07 & 0.14 & 0.0 & 0.0 & 0.86 & 1.0 & 1.0 & 0.0 \\ %
S230704bf & 312.88 & 0.12 & 0.0 & 0.0 & 0.88 & 1.0 & 1.0 & 0.01 \\ %
S230705bd & 528.13 & 0.19 & 0.0 & 0.0 & 0.81 & 1.0 & 1.0 & 0.0 \\ %
S230706al & 61.13 & 0.07 & 0.2 & 0.0 & 0.73 & 1.0 & 0.01 & 0.63 \\ %
S230706bv & 671.4 & 0.1 & 0.04 & 0.0 & 0.86 & 1.0 & 0.7 & 0.53 \\ %
S230708y & 317.51 & 0.18 & 0.0 & 0.0 & 0.82 & 1.0 & 1.0 & 0.46 \\ %
S230708ay & 694.89 & 0.14 & 0.0 & 0.0 & 0.86 & 1.0 & 1.0 & 0.52 \\ %
S230708bf & 314.04 & 0.1 & 0.01 & 0.0 & 0.88 & 1.0 & 1.0 & 0.56 \\ %
S230708bv & 167.08 & 0.0 & 0.29 & 0.05 & 0.66 & 0.54 & 0.0 & 0.12 \\ %
S230709aq & 127.88 & 0.0 & 0.16 & 0.16 & 0.68 & 0.13 & 0.0 & 0.23 \\ %
S230709bj & 358.29 & 0.0 & 0.22 & 0.13 & 0.64 & 0.82 & 0.0 & 0.06 \\ %
S230711j & 664.72 & 0.17 & 0.05 & 0.0 & 0.77 & 1.0 & 0.88 & 0.47 \\ %
S230711aj & 375.04 & 0.17 & 0.01 & 0.0 & 0.82 & 1.0 & 1.0 & 0.52 \\ %
S230712ab & 545.49 & 0.04 & 0.12 & 0.0 & 0.84 & 1.0 & 0.0 & 0.53 \\ %
S230713s & 674.74 & 0.0 & 0.13 & 0.0 & 0.86 & 1.0 & 0.0 & 0.0 \\ %
S230713x & 237.37 & 0.35 & 0.01 & 0.0 & 0.63 & 1.0 & 0.97 & 0.52 \\ %
S230714i & 89.78 & 0.2 & 0.39 & 0.0 & 0.41 & 1.0 & 0.0 & 0.31 \\ %
S230715z & 266.79 & 0.14 & 0.0 & 0.0 & 0.86 & 1.0 & 1.0 & 0.0 \\ %
S230720a & 402.07 & 0.04 & 0.1 & 0.0 & 0.85 & 1.0 & 0.0 & 0.67 \\ %
S230721x & 484.57 & 0.16 & 0.0 & 0.0 & 0.84 & 1.0 & 1.0 & 0.46 \\ %
S230723bl & 647.4 & 0.11 & 0.0 & 0.0 & 0.89 & 1.0 & 1.0 & 0.46 \\ %
S230726al & 355.31 & 0.0 & 0.14 & 0.09 & 0.76 & 0.62 & 0.0 & 0.02 \\ %
S230727am & 434.43 & 0.13 & 0.0 & 0.0 & 0.87 & 1.0 & 1.0 & 0.02 \\ %
S230729p & 66.91 & 0.41 & 0.0 & 0.0 & 0.59 & 1.0 & 1.0 & 0.46 \\ %
S230729ae & 484.22 & 0.11 & 0.0 & 0.0 & 0.89 & 1.0 & 1.0 & 0.02 \\ %
S230729bl & 719.23 & 0.03 & 0.17 & 0.0 & 0.8 & 1.0 & 0.0 & 0.06 \\ %
S230729bv & 385.66 & 0.2 & 0.0 & 0.0 & 0.8 & 1.0 & 1.0 & 0.46 \\ %
S230729cf & 439.08 & 0.04 & 0.17 & 0.0 & 0.79 & 1.0 & 0.0 & 0.38 \\ %
S230730av & 695.25 & 0.0 & 0.11 & 0.05 & 0.84 & 0.91 & 0.0 & 0.09 \\ %
S230805k & 594.66 & 0.12 & 0.0 & 0.0 & 0.88 & 1.0 & 1.0 & 0.0 \\ %
S230805at & 514.49 & 0.0 & 0.15 & 0.02 & 0.83 & 0.91 & 0.0 & 0.8 \\ %
S230805ax & 112.23 & 0.0 & 0.21 & 0.08 & 0.71 & 1.0 & 0.0 & 0.0 \\ %
S230806f & 130.82 & 0.16 & 0.0 & 0.0 & 0.84 & 1.0 & 1.0 & 0.0 \\ %
S230810r & 293.81 & 0.28 & 0.0 & 0.0 & 0.72 & 1.0 & 1.0 & 0.0 \\ %
S230812bu & 196.26 & 0.0 & 0.15 & 0.06 & 0.79 & 1.0 & 0.0 & 0.06 \\ %
S230812cd & 501.37 & 0.11 & 0.01 & 0.0 & 0.88 & 1.0 & 0.86 & 0.52 \\ %
S230819f & 495.6 & 0.05 & 0.13 & 0.0 & 0.82 & 1.0 & 0.0 & 0.4 \\ %
S230819h & 576.93 & 0.0 & 0.11 & 0.03 & 0.86 & 0.68 & 0.0 & 0.0 \\ %
S230820bj & 681.63 & 0.1 & 0.0 & 0.0 & 0.9 & 1.0 & 1.0 & 0.33 \\ %
S230820bn & 100.93 & 0.0 & 0.42 & 0.16 & 0.43 & 1.0 & 0.0 & 0.06 \\ %
S230821e & 413.82 & 0.01 & 0.14 & 0.0 & 0.85 & 1.0 & 0.0 & 0.15 \\ %
S230823ay & 158.96 & 0.0 & 0.31 & 0.03 & 0.66 & 1.0 & 0.04 & 0.0 \\ %
S230824av & 513.48 & 0.15 & 0.0 & 0.0 & 0.85 & 1.0 & 1.0 & 0.52 \\ %
S230824ay & 607.69 & 0.14 & 0.0 & 0.0 & 0.86 & 1.0 & 1.0 & 0.0 \\ %
S230825bf & 78.56 & 0.0 & 0.12 & 0.07 & 0.81 & 0.8 & 0.0 & 0.03 \\ %
S230826ac & 126.26 & 0.13 & 0.04 & 0.0 & 0.84 & 1.0 & 0.11 & 0.6 \\ %
S230826al & 61.85 & 0.43 & 0.0 & 0.0 & 0.57 & 1.0 & 1.0 & 0.46 \\ %
S230826ba & 615.77 & 0.21 & 0.0 & 0.0 & 0.79 & 1.0 & 1.0 & 0.07 \\ %
S230827au & 89.07 & 0.0 & 0.2 & 0.28 & 0.52 & 0.39 & 0.0 & 0.04 \\ %
S230827bj & 138.75 & 0.0 & 0.39 & 0.11 & 0.5 & 0.71 & 0.0 & 0.0 \\ %
S230827bl & 524.14 & 0.13 & 0.01 & 0.0 & 0.86 & 1.0 & 1.0 & 0.4 \\ %
S230828ah & 106.23 & 0.0 & 0.18 & 0.05 & 0.77 & 0.74 & 0.0 & 0.02 \\ %
S230830g & 105.68 & 0.0 & 0.12 & 0.08 & 0.8 & 0.18 & 0.0 & 0.41 \\ %
S230830an & 100.02 & 0.2 & 0.3 & 0.0 & 0.5 & 1.0 & 0.0 & 0.77 \\ %
S230901h & 322.3 & 0.2 & 0.0 & 0.0 & 0.8 & 1.0 & 1.0 & 0.46 \\ %
S230902ak & 589.69 & 0.0 & 0.14 & 0.03 & 0.82 & 0.65 & 0.0 & 0.13 \\ %
S230903aw & 353.71 & 0.21 & 0.01 & 0.0 & 0.79 & 1.0 & 1.0 & 0.33 \\ %
S230903bk & 297.01 & 0.0 & 0.15 & 0.21 & 0.64 & 0.27 & 0.0 & 0.03 \\ %
S230904i & 138.6 & 0.0 & 0.34 & 0.01 & 0.64 & 1.0 & 0.0 & 0.0 \\ %
S230906al & 202.08 & 0.0 & 0.26 & 0.09 & 0.65 & 0.79 & 0.0 & 0.35 \\ %
S230907ap & 512.92 & 0.0 & 0.16 & 0.01 & 0.83 & 0.97 & 0.14 & 0.0 \\ %
S230907az & 728.0 & 0.1 & 0.0 & 0.0 & 0.9 & 1.0 & 1.0 & 0.02 \\ %
S230909an & 277.05 & 0.15 & 0.0 & 0.0 & 0.85 & 1.0 & 1.0 & 0.29 \\ %
S230910p & 378.32 & 0.14 & 0.01 & 0.0 & 0.85 & 1.0 & 1.0 & 0.46 \\ %
S230910ay & 249.17 & 0.03 & 0.29 & 0.0 & 0.69 & 1.0 & 0.06 & 0.04 \\ %
S230911am & 162.54 & 0.25 & 0.05 & 0.0 & 0.7 & 1.0 & 1.0 & 0.02 \\ %
S230912g & 259.7 & 0.13 & 0.0 & 0.0 & 0.87 & 1.0 & 1.0 & 0.0 \\ %
S230912y & 153.36 & 0.16 & 0.02 & 0.0 & 0.82 & 1.0 & 1.0 & 0.07 \\ %
S230918bq & 253.07 & 0.0 & 0.13 & 0.09 & 0.78 & 0.61 & 0.0 & 0.06 \\ %
S230918bu & 201.7 & 0.11 & 0.01 & 0.0 & 0.88 & 1.0 & 1.0 & 0.0 \\ %
S230919j & 69.33 & 0.25 & 0.0 & 0.0 & 0.75 & 1.0 & 1.0 & 0.02 \\ %
S230919m & 94.0 & 0.0 & 0.16 & 0.01 & 0.84 & 1.0 & 0.08 & 0.01 \\ %
S230920p & 377.03 & 0.22 & 0.0 & 0.0 & 0.77 & 1.0 & 1.0 & 0.0 \\ %
S230920bc & 85.93 & 0.38 & 0.0 & 0.0 & 0.62 & 1.0 & 1.0 & 0.46 \\ %
S230923f & 566.72 & 0.11 & 0.0 & 0.0 & 0.89 & 1.0 & 1.0 & 0.0 \\ %
S230924v & 167.54 & 0.24 & 0.2 & 0.0 & 0.56 & 1.0 & 0.05 & 0.62 \\ %
S230924ah & 485.38 & 0.0 & 0.13 & 0.06 & 0.81 & 0.91 & 0.0 & 0.0 \\ %
S230925ac & 269.34 & 0.12 & 0.0 & 0.0 & 0.88 & 1.0 & 1.0 & 0.46 \\ %
S230925au & 89.14 & 0.23 & 0.19 & 0.0 & 0.57 & 1.0 & 0.2 & 0.53 \\ %
S230925bx & 96.48 & 0.44 & 0.0 & 0.0 & 0.56 & 1.0 & 1.0 & 0.56 \\ %
S230928q & 400.2 & 0.0 & 0.11 & 0.05 & 0.84 & 0.73 & 0.0 & 0.2 \\ %
S230928am & 249.42 & 0.0 & 0.25 & 0.06 & 0.69 & 0.7 & 0.0 & 0.03 \\ %
S230928cc & 239.25 & 0.0 & 0.13 & 0.03 & 0.84 & 0.65 & 0.0 & 0.09 \\ %
S230930bt & 536.97 & 0.1 & 0.0 & 0.0 & 0.9 & 1.0 & 1.0 & 0.33 \\ %
S231003ab & 574.58 & 0.0 & 0.21 & 0.07 & 0.72 & 0.94 & 0.0 & 0.06 \\ %
S231003bg & 97.4 & 0.0 & 0.15 & 0.12 & 0.74 & 0.82 & 0.0 & 0.03 \\ %
S231004f & 92.6 & 0.47 & 0.0 & 0.0 & 0.53 & 1.0 & 1.0 & 0.0 \\ %
S231005bt & 387.57 & 0.24 & 0.0 & 0.0 & 0.76 & 1.0 & 1.0 & 0.46 \\ %
S231006c & 498.88 & 0.0 & 0.22 & 0.07 & 0.71 & 0.91 & 0.0 & 0.1 \\ %
S231006ac & 341.4 & 0.0 & 0.11 & 0.15 & 0.74 & 0.23 & 0.0 & 0.03 \\ %
S231010ak & 364.29 & 0.11 & 0.01 & 0.0 & 0.88 & 1.0 & 1.0 & 0.52 \\ %
S231013ai & 114.44 & 0.03 & 0.42 & 0.0 & 0.55 & 1.0 & 0.0 & 0.05 \\ %
S231013bo & 502.53 & 0.21 & 0.0 & 0.0 & 0.79 & 1.0 & 1.0 & 0.0 \\ %
S231014g & 374.43 & 0.14 & 0.0 & 0.0 & 0.86 & 1.0 & 1.0 & 0.46 \\ %
S231014w & 374.3 & 0.13 & 0.03 & 0.0 & 0.84 & 1.0 & 1.0 & 0.02 \\ %
S231014be & 163.23 & 0.1 & 0.19 & 0.0 & 0.71 & 1.0 & 0.01 & 0.85 \\ %
S231015g & 290.21 & 0.25 & 0.01 & 0.0 & 0.74 & 1.0 & 1.0 & 0.0 \\ %
S231015by & 211.49 & 0.14 & 0.0 & 0.0 & 0.86 & 1.0 & 1.0 & 0.0 \\ %
S231016br & 220.42 & 0.0 & 0.16 & 0.06 & 0.78 & 0.82 & 0.0 & 0.01 \\ %
S231017t & 200.55 & 0.14 & 0.07 & 0.0 & 0.8 & 1.0 & 0.0 & 0.55 \\ %
S231017z & 259.76 & 0.27 & 0.05 & 0.0 & 0.68 & 1.0 & 1.0 & 0.0 \\ %
S231018v & 203.66 & 0.2 & 0.0 & 0.0 & 0.79 & 1.0 & 1.0 & 0.0 \\ %
S231018ax & 213.21 & 0.36 & 0.01 & 0.0 & 0.63 & 1.0 & 1.0 & 0.33 \\ %
S231019ak & 228.54 & 0.01 & 0.41 & 0.02 & 0.56 & 1.0 & 0.0 & 0.0 \\ %
S231020br & 519.02 & 0.13 & 0.05 & 0.0 & 0.82 & 1.0 & 0.72 & 0.53 \\ %
S231021az & 7.38 & 0.0 & 0.26 & 0.01 & 0.73 & 0.96 & 0.0 & 0.02 \\ 
S231022bk & 468.37 & 0.23 & 0.0 & 0.0 & 0.76 & 1.0 & 1.0 & 0.0 \\ %
S231022bl & 373.51 & 0.06 & 0.11 & 0.0 & 0.83 & 1.0 & 0.0 & 0.72 \\ %
S231025a & 29.42 & 0.59 & 0.0 & 0.0 & 0.41 & 1.0 & 1.0 & 0.0 \\ 
S231025c & 578.54 & 0.0 & 0.1 & 0.04 & 0.85 & 1.0 & 0.0 & 0.09 \\ %
S231025r & 498.2 & 0.11 & 0.0 & 0.0 & 0.89 & 1.0 & 1.0 & 0.46 \\ %
S231025t & 146.73 & 0.16 & 0.33 & 0.0 & 0.51 & 1.0 & 0.0 & 0.35 \\ %
S231025az & 551.51 & 0.17 & 0.0 & 0.0 & 0.83 & 1.0 & 1.0 & 0.02 \\ %
S231026n & 415.8 & 0.15 & 0.06 & 0.0 & 0.79 & 1.0 & 0.3 & 0.62 \\ %
S231026z & 361.62 & 0.16 & 0.01 & 0.0 & 0.83 & 1.0 & 0.85 & 0.52 \\ %
S231027bk & 498.81 & 0.2 & 0.0 & 0.0 & 0.8 & 1.0 & 1.0 & 0.0 \\ %
S231028r & 679.89 & 0.16 & 0.0 & 0.0 & 0.84 & 1.0 & 1.0 & 0.02 \\ %
S231028ai & 274.01 & 0.13 & 0.0 & 0.0 & 0.87 & 1.0 & 1.0 & 0.02 \\ %
S231028aw & 306.48 & 0.3 & 0.0 & 0.0 & 0.7 & 1.0 & 1.0 & 0.0 \\ %
S231029e & 444.94 & 0.22 & 0.0 & 0.0 & 0.78 & 1.0 & 1.0 & 0.27 \\ %
S231029ai & 160.69 & 0.0 & 0.46 & 0.04 & 0.5 & 1.0 & 0.0 & 0.0 \\ %
S231029bd & 281.51 & 0.22 & 0.0 & 0.0 & 0.78 & 1.0 & 1.0 & 0.0 \\ %
S231030t & 707.41 & 0.16 & 0.01 & 0.0 & 0.82 & 1.0 & 1.0 & 0.0 \\ %
S231102i & 185.19 & 0.29 & 0.16 & 0.0 & 0.55 & 1.0 & 0.98 & 0.03 \\ %
S231104s & 256.82 & 0.0 & 0.21 & 0.09 & 0.69 & 0.91 & 0.0 & 0.06 \\ %
S231107a & 410.49 & 0.21 & 0.0 & 0.0 & 0.78 & 1.0 & 1.0 & 0.0 \\ %
\enddata

\end{deluxetable*}

\newpage

\clearpage


\end{document}